# The Hidden Subgroup Problem

## Master's Project

Frédéric Wang
20090566

Supervisor: Ivan Damgård

July 2010

Datalogisk Institut
Det Naturvidenskabelige Fakultet
Aarhus Universitet
Danmark

École Nationale Supérieure
d'Informatique pour l'Industrie et l'Entreprise
Evry
France

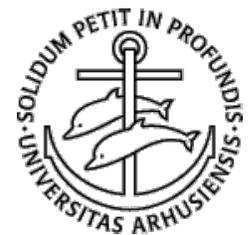

ii

# Abstract

We review the Hidden Subgroup Problem (HSP), one of the most prominent topic in quantum computing. We start with some "historical" quantum algorithms which are exponentially faster than their classical counterparts, including the famous Shor's factoring algorithm. Then, we define HSP and show how the previous algorithms can be interpreted as solutions to this problem.

In a second part, we describe in detail the "standard method". We recall the solution to the abelian case as well as some representation theory background that allows to use Fourier Transform over Finite Groups. We also define Fourier Sampling and mention how the weak version extends the abelian case to find normal hidden subgroups and solve the dedekindian HSP.

In a third part, we discuss all the results known so far for the cases of the Dihedral and Symmetric groups, which are expected to provide new efficient algorithms. Solving the dihedral HSP in a particular way gives an algorithm for $\text{poly}(n)$-uniqueSVP while a solution to the symmetric HSP gives an algorithm for the graph isomorphism problem.

Finally, we conclude our study with the general HSP. We make some proposals for a general approach and discuss the solutions and techniques known so far. Next, we indicate how HSP relates to various efficient algorithms. We also give an overview of miscellaneous variants and extensions to the hidden subgroup problem that have been proposed.

In addition to gathering in a single document an introduction to the Hidden Subgroup problem as well as an overview of the state-of-the-art for this research topic, we also bring some contributions:

- We provide a detailed and corrected computation of the fact that, in the Strong Fourier Sampling, measuring the row yields no information. In the sketchy proof of Grigni et al., a family of vectors was claimed to be orthonormal whereas it is only orthogonal.
- In appendix B, we compute the exact expression of Strong Fourier Sampling over the dihedral group. When the hidden subgroup is reduced to $\{(0, 0), (d, 1)\}$, it has already been mentioned that Strong Fourier Sampling over $D_N$ is similar to using the Quantum Fourier Transform over $\mathbb{Z}_N \times \mathbb{Z}_2$ and hence a particular case of the dihedral coset problem. For a general hidden subgroup, we prove that the expression of the Strong Fourier Sampling is the same as if we were directly working in the quotient group $D_N / (H \cap (\mathbb{Z}_N \times \{0\}))$.
- In appendix G, we propose an entirely new approach to try to find the slope $d$ of the dihedral HSP. Rather than using coset sampling, we consider uniform superpositions over large subsets of $f(D_N)$. In particular, we can solve the case $N = 2^n$ if we have an efficient process to create for $b = 0, 1$ the states $\frac{1}{\sqrt{2^{n-1}}} \sum_{i=0}^{2^{n-1}-1} |f(2i, b)\rangle$.
- We give more detail on the relation between HSP and lattice problems. In particular, we show that if Regev's algorithm is based on a solution to the dihedral coset problem with a query complexity $O(n^D)$ then it gives a solution to $\Theta\left(n^{\frac{1}{2}+2D}\right)$-uniqueSVP. We indicate that the relation still holds if we use a solution over $D_{2^n}$ or $\text{Dih}\left(\mathbb{Z}_{2^{4n}}^n\right) = \mathbb{Z}_{2^{4n}}^n \rtimes \mathbb{Z}_2$ using a "recursive DCP algorithm". However, we also provide an overview of lattice-based problems in appendix E and warn that Regev's algorithm would only have

- small impact on lattice-based cryptosystems. Hence, we propose to consider also HSP algorithms for stronger lattice problems.
- In appendix D, we give a reduction of Monotone 1-in-3 3SAT to GapCVP$^\infty$ where one step uses the abelian HSP algorithm to find the kernel of group homomorphism. Although the quantum part is not actually needed here, this provides another example where a hidden subgroup problem can be used.
- In appendix F, we give an alternative algorithm for the cyclic and abelian hidden subgroup problems, based on a change of the underlying group. Even if it does not generalize, it is a good example of subgroup reduction.
- We present a general approach to the Hidden Subgroup Problem and show that in theory, we can solve it if we have a solution to HSP over simple groups and a way to build efficient oracles for some reduction techniques. We also propose iterative subgroup and quotient reductions: using a maximal supergroup of $H$ for the former and the normal subgroup obtained by Weak Fourier Sampling for the latter. Maximal subgroup reduction is a possible approach for HSP over simple groups.
- We indicate how to reduce the rigid graph isomorphism problem to HSP over the alternating group $A_n$, which is simple for $n > 4$.
- We notice how the Hidden Polynomial Problem is a natural extension of the abelian HSP over $G = F_q^m$, when the subgroup $H$ is promised to be a hyperplane.

# Table of Contents

# Acknowledgments

*Jeg vil gerne sige tak til Ivan Damgård, for at have lært mig Quantum Information Processing på først semester og have accepteret at blive min vejleder. Mange tak for at have fulgt mit arbejde og have gjort alt for at jeg kunne have en mundtlig eksamen.*

*Jeg vil også give tak til alle folk, som har givet mig til at have et dejligt ophold i Danmark. Specialt tak til Mikkel Gravgaard, for sin hjælp med administrative ting da jeg kom. Mange tak til alle beboere på Skejbygårdkollegiet for alle hyggelige aftener samt sjove fodboldkampe.*

*Je tiens aussi à exprimer ma gratitude envers toutes les personnes de l'ENSIIE qui m'ont permis de passer ma troisième année à Århus dans les meilleurs conditions.*

*Enfin, je souhaite remercier ma famille pour son soutien tout le long de mon séjour au Danemark. Merci notamment à mon frère et ma sœur pour être venus me rendre visite quelques jours à Århus.*

# 1. Introduction

The Hidden Subgroup Problem is one of the most prominent topic in quantum computing. Most quantum algorithms running exponentially faster than their classical counterparts fall into this framework. It is expected that finding more solutions to Hidden Subgroup Problems will provide new efficient quantum algorithms.

The first example of a black-box algorithm of this kind was given by Simon [Sim1994]. Classical algorithms can not give the right answer to that black-box problem unless they use an exponential number of oracle calls. This result strengthened the earlier Deutsch–Jozsa algorithm, which is only exponentially more efficient than *deterministic* classical algorithm. Despite its theoretical importance for demonstrating the power of quantum computer, Simon's algorithm does not lead to practical applications.

However, shortly after Simon published his algorithm, Shor [Sho1995] extended it to find periods of functions over $\mathbb{Z}_N$ and $\mathbb{Z}_{p-1}^2$. He used these period finding algorithms as key steps to solve two difficult problems: factoring integer and computing discrete logarithms. The hardness of both problems is the assumption on which some cryptographic systems such that the ubiquitous RSA system are based. Hence Shor's algorithms had a strong impact since they proved how quantum computers can break cryptographic systems and it has been one of the great success of quantum computing.

Simon's problem and Shor's algorithms can be better understood in the framework of the Hidden Subgroup Problem. We work on a group $G$ and try to find an unknown subgroup $H$ using calls to a function $f$. This function is constant on cosets of $H$ and takes different values on distinct cosets. Ettinger, Høyer and Knill proved that the subgroup $H$ can always be determined using only $O(\log|G|)$ calls to the oracle $f$ but the whole procedure is not necessarily efficient [EttHøyKni1999].

We recall the "historical" results as well as the definition of the Hidden Subgroup Problem in the first part of this report. Although they are well known by researchers working on this topic, they remain a good introduction and motivate the subsequent work.

Once the framework of the Hidden Subgroup Problem set, the special case of abelian group was addressed. Indeed, the algorithms given by Simon and Shor work on abelian groups and the main ingredients such that entanglement over a coset and Quantum Fourier Transform could easily be extended to solve the abelian Hidden Subgroup Problem. The algorithm requires the knowledge of a decomposition $G \cong \mathbb{Z}_{t_1} \times \mathbb{Z}_{t_2} \times \mathbb{Z}_{t_3} \times \ldots \times \mathbb{Z}_{t_k}$ but if the group is given in a black box form, we can still determine that decomposition. Because the algorithm is fundamental, we start the second part of this report by recalling how the abelian HSP algorithm works. We also sketch the idea of Cheung and Mosca [CheMos2001] to get the decomposition for a black-box group.

The success of the Fourier Sampling to solve the abelian HSP naturally made the researchers study its generalization over non-abelian groups. This is quite technical, so we take care to define precisely Quantum Fourier Transform and Fourier Sampling. We study how they apply to non-abelian HSP in the remainder of the second part of this report.

We first extract the essential points from the reference book on group representation theory by Serre [Ser1971] to show that it is always possible to define a general Quantum Fourier Transform over finite groups as a unitary operation. We indicate how it extends the classical Quantum Fourier Transform used for abelian groups. In the non-abelian case, it turns out that each choice of basis for the set of irreducible representations gives rise to a specific Quantum Fourier Transform. We use the Schur's lemma to show that they are all described modulo basis changes in the space of each irreducible representation.

Next, we turn our attention to the Fourier Sampling over finite groups using the studies of [HalRusTa-2000] and [GriSchVaz2000]. It is the occasion to recall the standard method of coset sampling: we gather information on the hidden subgroup from coset states i.e. uniform superposition over a random coset of $H$. Then, we apply several Fourier Samplings to these cosets states: a Quantum Fourier Transform followed by a measurement of an irreducible representation. We define the Weak and Strong versions and recall classical expressions on distribution probabilities. In particular, we give a detailed proof of the fact that measuring the row is not relevant for Strong Fourier Sampling.

We then see how the Weak Fourier Sampling is enough to solve the Dedekindian Hidden Subgroup Problem i.e. for groups that have only *normal* subgroups. In the new framework, the algorithm of the abelian case can be seen as computing the hidden subgroup as the intersection of the kernels of the representations measured by Fourier Sampling. Independently, [HalRusTa-2000] and [GriSchVaz2000] showed that this technique still works when the hidden subgroup is normal, thus solving the dedekindian in theory. [HalRusTa-2000] contains an explicit algorithm for the class of hamiltonian groups which are the only non-abelian dedekindian groups. We also mention briefly more results of [GriSchVaz2000] and [Gav2004] such that extensions to groups that are in some sense not too far from the dedekindian case as well as a black box result to find normal subgroups.

The third part of this report is dedicated to the two most important open non-abelian HSPs: those over the dihedral and symmetric groups. Regev showed that a solution to the Hidden Subgroup Problem over the dihedral group using coset samplings on a particular reduced case would provide an efficient algorithm for the $\text{poly}(n)$-uniqueSVP [Reg2004]. uniqueSVP is an instance of lattice problems, which are involved in many tasks believed to be computationally hard. More specifically, some cryptographic systems are based on lattice problems and some of them use the $\text{poly}(n)$-uniqueSVP. A solution to the Hidden Subgroup Problem over the symmetric group gives a polynomial algorithm to determine whether two given graphs are isomorphic, one of the very rare problems whose exact complexity has remained unknown for many decades.

The dihedral Hidden Subgroup Problem was first considered by Ettinger and Høyer as a first study of the non-abelian case [EttHøy1998]. They gave an interesting reduction of the problem to the case where the hidden subgroup is $\{(0, 0), (d, 1)\}$ and showed that we can determine $d$ with efficient query and measurement but exponential post-processing. We describe the structure of the dihedral group and explain the techniques of Ettinger and Høyer. We also discuss whether the algorithms of the previous section work. In particular, we give a very detailed description of Fourier Sampling over the dihedral group in appendix B and present some attempts to solve DHSP using this method in appendix C. We introduce the Dihedral Coset Problem which is essentially asking whether we can solve the dihedral HSP using a black box that outputs coset states. This problem encapsulates all the previous approaches. However, we propose a totally new method in appendix G.

In another section, we study the results obtained from this dihedral coset black box. In [Kup2003], Kuperberg gave a subexponential time algorithm to determine the parity of $d$ from a dihedral coset black box, using a new combine-and-measure technique. This algorithm can be easily extended to get a solution to the dihedral HSP with subexponential time. Actually, Kuperberg's algorithm uses exponential space but Regev gave a modification that makes the space requirement polynomial [Reg2004]. More generally, Bacon, Childs and van Dam used tools from quantum-information theory to characterize the measurement on $k$ outputs of the black box that gives the optimal information [BacChiDam2005]. They showed that $k$ must be at least linear even for determining the parity of $d$ as in Kuperberg's algorithm. They also demonstrated a relation between the implementation of the optimal measurement and the subset sum problem. Actually, Regev had already shown such a relation in [Reg2003].

Next we look more carefully to Regev's algorithm [Reg2003] that establishes a connection between the polynomial uniqueSVP and a solution to the dihedral coset problem. We show how the degree of the polynomial complexity of a solution to the dihedral coset problem is related to the degree of the approximation in the uniqueSVP. We notice that *mutatis mutandis* his algorithm can be applied using more HSP problems over the family of generalized dihedral groups. Similarly, we indicate that the algorithm still holds for some kinds of recursive dihedral coset problem as in the case of Kuperberg's algorithm. We sum up in appendix E the overview of [MicReg2008] and notice that the problem considered is however not of major importance. We suggest some research directions to investigate.

Then we turn our attention to the case of the symmetric group. We recall the definition of the graph isomorphism problem as well as related problems. We describe the straightforward reduction of the equivalent graph automorphism problem to the symmetric hidden subgroup problem, or more specifically to the case where we want to determine whether the hidden subgroup is trivial or not. We also talk about another reduction for the case of rigid graphs given in [MooRusSch2005]: the underlying group is $S_n \wr \mathbb{Z}_2 \subseteq S_{2n}$ and the hidden subgroup $H = \{\mathrm{Id}, \sigma\}$. Moreover, we prove that the rigid graph isomorphism problem can actually be set in the reduced case of the simple group $A_{2n}$. Unfortunately, even these simpler cases have remained unsolved. We recall the negative results on the symmetric group and suggest to split the problem in more separate cases.

In a final part, we study the general hidden subgroup problem. We recall the two fundamental theorems describing how finite groups are built from simple groups using composition series. We state conditions to solve the general HSP: find a way to solve the HSP over simple group and to build efficient oracles when breaking down the group. For the second conditions, we isolate two particular reduction methods: subgroup reduction and quotient reduction. We propose iterative quotient reduction to make the hidden subgroup not contain any non-trivial normal subgroup. We also suggest iterative maximal subgroup reduction as a possible way to solve the HSP over the simple groups. We sum up our ideas in a general schema for a possible HSP algorithm. We notice how these ideas relate to an alternative abelian HSP algorithm proposed in appendix F and to dihedral and symmetric HSP. We explain how the classical attempts for dihedral HSP are using quotient and subgroup reductions. For the symmetric HSP, the reductions give two difficult problems: either solving HSP over the large simple group $A_n$ or building a oracle over $\mathbb{Z}_2$ from the big initial oracle over $S_n$.

In the next section, we review the solutions and techniques obtained for non-abelian groups. We recall the Fourier Sampling techniques as well as all the methods that have been discovered in order to find a solution to the dihedral HSP. We mention other techniques relying on the blackbox paradigm. We give an overview of efficient HSP algorithms based on these methods. They can essentially be classified into three groups: the extensions to the Dedekindian HSP which apply to groups with a large amount of normal subgroups, the semi-direct products of two abelian groups $A \rtimes B$ which are broken down into the abelian groups $A$ and $B$ with respect to the description of the previous section and another category of groups whose normal series structure is simple enough to apply blackbox techniques. In some sense, they are all close to the abelian case. We mention a partial result for an exception which includes the simple group $\mathrm{PGL}(2, p^m)$, namely finding 1-point stabilizer of some Lie groups.

The following section is devoted to relation between HSP and efficient algorithms with concrete applications. We recall the results for dihedral and symmetric HSP. We mention Hallgren's generalization to infinite groups $\mathbb{R}^r$ which allowed him to solve Pell's equation and other number fields problems. We also mention more relationship between HSP an cryptography. Considering that the general HSPs are actually difficult, Moore, Russell, Vazirani constructed a one-way function which is as hard to invert as the graph isomorphism problem. [Dam1988] contains a proposal for a hard problem with application to cryptography: Given $l + 1 = O(\log p)$ successive evaluations of the Legendre symbol $\left(\frac{s}{p}\right)$, $\left(\frac{s+1}{p}\right)$, ..., $\left(\frac{s+l}{p}\right)$ predict the next value $\left(\frac{s+l+1}{p}\right)$. We mention how a weaker problem can be set in the HSP framework and has been solved [DamHallp2002]. The authors say that the solution to this weaker problem allows attacks to some cryptosystems, already broken by Shor's algorithm though. We also discuss the comparaison given in [LomKau2006] between Shor's algorithm and Grover's algorithm. We compare their method with the algorithms for the 1-point stabilizer given in [DenMooRus2008] and wonder whether they can be interpreted as exponentially fast solutions to some concrete search problems.

We conclude our study with some variants and extensions to the hidden subgroup problem which are also expected to yield new efficient quantum algorithms. We mention the generalization to hypergroups and infinite groups. Other extensions include the Hidden Symmetries Problem and the Hidden Covering Space Problem. We describe the quite popular variants of the Hidden Shift Problem which is related to the dihedral and symmetric HSP, as well as its generalizations the Generalized Hidden Shift Problem and the Hidden Coset Problem. We mention three problems introduced in [FriEtAl2002] to solve some instances of HSP: Hidden Stabilizer Problem, Orbit Coset Problem and Orbit Superposition Problem. We also talk about some decision/search versions of hidden subgroup problems. Next, we show how finding hyperplane subgroups of the vector space $G = F_q^m$ naturally generalizes to the Hidden Polynomial Problem. Finally, we discuss the category of Hidden Shifted Subset Problems.

## 2. Prerequisites

Unless otherwise specified, we use the notations described in this section as well as other classical ones not mentioned below. We assume basic knowledge of quantum computing: an excellent introduction is the book of Nielsen and Chuang [NieChu2007]. The paper of Lomont [Lom2004] also contains a presentation of the main ideas as well as a good overview of the Hidden Subgroup Problem and its status as of 2004. Some advanced mathematical concepts will also be required and sometimes recalled as needed.

- log (binary logarithm), $\varphi(n)$ (Euler's totient function), $|S|$ (cardinality).
- numbers: $\mathbb{i}$ (imaginary unit), $\mathrm{e}$ (euler's number), $c^*$ (conjugate), $\lfloor c \rfloor$ and $\lceil f \rceil$ (ceiling and floor).
- limit and asymptotic approximation: $f \to c$ ($f$ has limit $c$), $f = O(g)$ ($f$ bounded above by a constant times $g$), $f = \Omega(g)$ ($f$ bounded below by a constant times $g$), $f = \Theta(g)$ (the two previous equalities hold), $f = o(g)$ ($f$ bounded above by $\varepsilon g$ with $\varepsilon \to 0$), $f \sim g$ ($f$ and $g$ are equivalent), $\mathrm{comp}(f)$ means the worse-case complexity of $f$, $\mathrm{poly}(x)$ means $O(P(x))$ for polynomial $P$ of $x$. Recall that $n! \sim \sqrt{2\pi n}\left(\frac{n}{\mathrm{e}}\right)^n$ (Stirling formula) and $\frac{\varphi(n)}{n} = \Omega\left(\frac{1}{\log(\log(n))}\right)$.
- Group laws are noted + for abelian groups and by a multiplication for general groups. $\mathbb{Z}_2^n$ may be seen as $n$-bits numbers, so in that case $\oplus$ is sometimes used instead of +.
- algebraic structures: $G_1 \oplus G_2$ (direct sum), $G_1 \times G_2$ (direct product), $G_1 \rtimes_\varphi G_2$ (semi-direct product), $G_1 \wr G_2$ (wreath product), $G/H$ (quotient), $H \triangleleft G$ (normality), $\mathrm{Ker}\, f$ (kernel), $\mathrm{Im}\, f$ (image), $G_1 \cong G_2$ (isomorphy), $gH$ and $Hg$ (left and right coset), $\langle S \rangle$ (group generated by $S$), $N(S) = \{g \in G, S = gSg^{-1}\}$ (normalizer of $S$), $[G_1, G_2] = \langle \{g_1^{-1} g_2^{-1} g_1 g_2 | (g_1, g_2) \in G_1 \times G_2\} \rangle$, $[G, G]$ (derived or commutator subgroup), $\Phi(G) = \bigcap_{H \text{ proper maximal subgroup } G} H$ (Frattini subgroup), $Z(G)$ (center of $G$). We also simply use the term "Coset" for left coset. We also define the normal subgroup $M_G = \bigcap_{H \text{ subgroup of } G} N(H)$.
- some algebraic structures: $\mathbb{Z}_k$ (cyclic group $\mathbb{Z}/k\mathbb{Z}$), $\mathbb{Z}_p^*$ with $p$ prime (multiplicative group of invertible elements of $\mathbb{Z}_p$), $\mathbb{Q}_8$ (quaternion group), $D_N$ (dihedral group), $S_N$ (symmetric group), $A_N$ (alternating group), $\mathrm{GL}_n(K)$ (general linear group of degree $n$ over a field $K$). $F_q$ is the finite field of order $q$. $\mathrm{GL}(n, q) = \mathrm{GL}_n(F_q)$.
- vectors and matrices: $x.y = \sum_{i=1}^n x_i y_i$, $0^n$ (zero vector of length $n$), $A^\dagger = (A^*)^t = (A^t)^*$ (hermitian conjugate), $\langle A, B \rangle_F$ (Frobenius hermitian product), $\|A\|_F = \sqrt{\mathrm{tr}(A^\dagger A)}$ (Frobenius norm). Recall that if $U$, $V$ are unitary, $\|UAV\|_F = \|A\|_F$.
- quantum mechanics: $|x\rangle$, $\langle x|$, $\langle x|y\rangle$, $|x\rangle|y\rangle$ (bra-ket notations).
- quantum circuit: ⌐⌙⌐ (measurement), $H$ (hadamard gate), $U \otimes V$ and $U^{\otimes n}$ (tensor product of unitary transforms). For any function $f : \mathbb{Z}_2^n \to \mathbb{Z}_2^m$, we define the gate $U_f$ by $\forall (x, y) \in \mathbb{Z}_2^n \times \mathbb{Z}_2^m, U_f(|x\rangle|y\rangle) = |x\rangle|f(x) \oplus y\rangle$. If $f$ can be implemented using a $\mathrm{poly}(n)$ complexity then it is also the case for the quantum version $U_f$.
- probability theory: $P(X)$ (probability of an event $X$), $P(X/Y)$ (conditional probability of $X$ given $Y$), $E(X)$ (mean). To bound probabilities, we use variants of Hoeffding's inequality: if we have independent bounded variables $X_k \in [0, 1]$ then their sum $S = \sum_{k=1}^{n} X_k$ satisfies $P(S - E(S) \geq t) \leq \exp\left(-\frac{2t^2}{n}\right)$.

- number theory: let $p$ be an odd prime, the Legendre symbol $\left(\frac{a}{p}\right)$ is 0 if $a$ is a multiple of $p$ and $a^{(p-1)/2} \mod (p) = \pm 1$ otherwise.

# 3. Efficient Quantum Algorithms and The Hidden Subgroup Problem

## 3.1. Simon's Problem

This problem imagined by Simon [Sim1994] demonstrates a quantum algorithm solving a black-box problem exponentially faster than any probabilistic classical algorithm. Contrary to Deutsch–Jozsa algorithm [DeuJoz1992], it is even exponentially faster than *probabilistic* classical algorithms.

**Definition 3.1 (Simon's problem)**

Let $m \geq n$ be natural numbers and $f : \mathbb{Z}_2^n \to \mathbb{Z}_2^m$ a function. Assume that there is a string $s \in \mathbb{Z}_2^n$ such that two distinct $x, x'$ have the same image iff $x' = x \oplus s$. *Simon's problem* is to determine $s$. ◇

The quantum solution to the problem is to repeat enough time the procedure given by the following circuit:

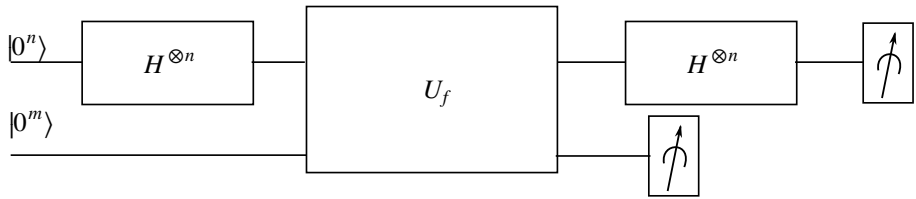

Figure 3.1

After applying $U_f$, we are in the state:

$$\frac{1}{\sqrt{N}} \sum_{x=0}^{N-1} |x\rangle |f(x)\rangle \quad (3.1)$$

Measuring the second register gives a value $y$ and makes the first register collapse to the superposition of its two preimages $x, x \oplus s$:

$$\frac{1}{\sqrt{2}} (|x\rangle + |x \oplus s\rangle)|y\rangle \quad (3.2)$$

After the second Hadamard gate, the state in the first register is

$$\frac{1}{\sqrt{2}} \frac{1}{\sqrt{N}} \sum_{z=0}^{N-1} ((-1)^{z.x} + (-1)^{z.(x \oplus s)})|z\rangle \quad (3.3)$$

Performing a standard measurement on the first register yields a random $n$-bits vector $z$ such that $z.s = 0$ i.e. in the subspace orthogonal to $s$. Repeating the procedure $O(n)$ times gives a set of generators $z^1, \ldots z^{O(n)}$ for this subspace with probability exponentially close to 1. Said otherwise, we have a system of equations $\sum_{i=1}^{n} z_i^j s_i \equiv 0 \mod (2)$ of rank the dimension of the subspace and $n$ unknowns $s_1, \ldots s_n$. If $s = 0$, then the rank of the system is $n$, so we get the unique solution $s = 0$. Otherwise $s \neq 0$ and the rank of the system is $n - 1$, so solving the system gives a subspace of dimension $n - (n - 1) = 1$, which is simply $\{0, s\}$. In either case, we have obtained $s$ as expected. The procedure is repeated $O(n)$ times, so a polynomial number of queries to the oracle $f$ are used.

What about (probabilistic) classical algorithm? Consider the special case $m = n$. Let's choose randomly $f$ to be a permutation ($s = 0$) or a two-to-one function ($s \neq 0$). In the former case, we randomly choose a permutation $f$ and in the latter case a nonzero string $s$. Simon proved that a classical algorithm calling $f$ no more than $2^{n/4}$ times can not determine whether $s$ is zero or not with success probability greater than $\frac{1}{2} + 2^{-n/2}$. Hence for a fixed minimal probability of success $\frac{1}{2} + \delta$, no classical algorithm using a polynomial number of queries can solve Simon's problem.

## 3.2. Shor's Factoring Algorithm

Shor's algorithm [Sho1995] allows to factor an integer $N_0$ in $\text{poly}(\log N_0)$ time. No classical algorithm are currently known to run this task in polynomial time. Actually, some cryptographic protocols are based on the assumption that no efficient algorithm exists for integer factoring and hence Shor's algorithm shows that they can be broken by quantum computation.

Shor's algorithm contains classical parts that can be executed in polynomial time and quantum computation is actually only used in a sub-procedure. More precisely, one step is to choose a random integer $a < N_0$ (which can be assumed to be coprime with $N_0$). The function defined on $\mathbb{Z}$ by $f(x) = a^x \mod (N_0)$ is $r$-periodic and the problem reduces to find the period $r$. We can restrict our study to the domain $\mathbb{Z}_N \subseteq \mathbb{Z}$ for some multiple $N$ of the period. Of course, we do not know how to choose $N$ precisely but it is possible to find some value $N = O(N_0^2)$ for which the approximation does not change the final result. For simplicity, let's assume $N = rM$ is an exact multiple of the period. The period-finding submodule can be executed in $\text{poly}(\log N) = \text{poly}(\log N_0)$ using the circuit of figure 3.2, which is very similar to the one of Simon's problem. We define the Fourier transform $F_N$ to be the symmetric matrix

$$F_N = \frac{1}{\sqrt{N}} \sum_{i=0}^{N-1} \sum_{j=0}^{N-1} e^{\frac{2\pi i}{N} i j} |j\rangle\langle i| \quad (3.4)$$

It can easily be shown to be unitary and hence a valid quantum operation. We assume that there exists an efficient implementation of $\text{poly}(\log N)$ elementary gates computing $F_N$. This implementation uses $n = \lceil \log N \rceil$ qubits, so we are working in an Hilbert Space of dimension $2^n \geq N$. Hence some basis states may not be used, they are just unchanged by the gate implementing $F_N$.

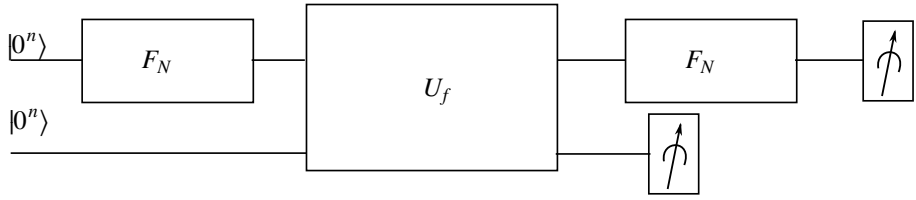

Figure 3.2

The first call to $F_N$ on $0^n$ returns the first column of $F_N$ i.e. a uniform superposition of basis vectors as in Simon's algorithm. Hence the state after $U_f$ is also

$$\frac{1}{\sqrt{N}} \sum_{x=0}^{N-1} |x\rangle|f(x)\rangle \quad (3.5)$$

Measuring the second register gives a value $y$ and makes the first register collapse to the superposition of the preimages of $y$. If $x_0$ is such an image all the others are obtained by $x_0 + ir$ for $i$ going from $0$ to $M-1$:

$$\frac{1}{\sqrt{M}} \sum_{i=0}^{M-1} |x_0 + ir\rangle|y\rangle \quad (3.6)$$

After the second Fourier Transform gate, the state in the first register is

$$\frac{1}{\sqrt{MN}} \sum_{j=0}^{N-1} \left( \sum_{i=0}^{M-1} e^{\frac{2\pi i}{N}(x_0+ir)j} \right) |j\rangle \quad (3.7)$$

Each sum over $i$ is a geometric series of initial term $e^{\frac{2\pi i}{N} x_0 j}$ and ratio $e^{\frac{2\pi i}{N} rj}$. So the sum is nonzero (and equal to $M e^{\frac{2\pi i}{N} x_0 j}$) iff $e^{\frac{2\pi i}{N} rj} = 1$ iff $j \equiv 0 \mod \left(\frac{N}{r}\right)$ iff $j$ is multiple of $M$. Hence the state can be rewritten:

$$\frac{1}{\sqrt{r}} \sum_{j \text{ multiple of } M} e^{\frac{2\pi i x_0 j}{N}} |j\rangle \quad (3.8)$$

All the vectors in the sum have same amplitude, so measuring the first register yields a uniformly random multiple of $M$ between $0$ and $N-1$. Using $k$ trials gives $j_1, \ldots j_k$ multiples of $M$. One can show that $\gcd(j_1, \ldots j_k) = M$ with success probability $\geq 1 - 2^{-k/2}$ (see Appendix E of [Lom2004] applied to $t_i = \frac{j_i}{M}$) and thus we get $r = \frac{N}{M}$.

### 3.3. Shor's Discrete Logarithm Algorithm

**Definition 3.2 (discrete logarithm)**

Let $p$ be a prime number and $g$ a generator of $\mathbb{Z}_p^*$. Any $x \in \mathbb{Z}_p^*$ can be written uniquely as $x = g^y$ for some $y \in \mathbb{Z}_{p-1}$. $y = \log_g x$ is called the *discrete logarithm* of $x$ (with respect to $g$). ◇

Similarly to Shor's algorithm, some cryptographic protocols are based on the difficulty of computing discrete logarithms. In [Sho1995], Shor describes a quantum algorithm to compute these logarithms in polynomial time. So suppose $x, g$ are given and that we want to find $y$.

We first define the function $f(a, b) = g^a x^b \mod (p)$ going from $\mathbb{Z}_{p-1} \times \mathbb{Z}_{p-1}$ to $\mathbb{Z}_p$. Each call to $f$ is clearly done in time $\text{poly}(\log(p))$. Note that we can rewrite $f$ using the discrete logarithm $y$: $f(a, b) = g^{a+yb} \mod (p)$. Hence $(a_1, b_1) \equiv (a_2, b_2) + \lambda(y, -1) \mod (p-1)$ implies $f(a_1, b_1) = f(a_2, b_2)$. Conversely, if this equality is true, we have $a_2 - a_1 \equiv y(b_1 - b_2) \mod (p-1)$ and we can take $\lambda \in \mathbb{Z}_{p-1}$ to be the unique element congruent with $b_2 - b_1$ modulo $p-1$ to recover the previous equality. As a consequence, we can say that $(y, -1)$ is the period of $f$ and all the values are obtained when $\lambda$ varies from $0$ to $p-2$.

We now define a circuit which is really similar to those we saw earlier. Here, the first register is the tensor product of two $\lceil \log(p-1) \rceil$-qubits state and the second a $\lceil \log(p) \rceil$-qubits state. On the first register we apply $F_{p-1} \otimes F_{p-1}$ i.e. a Fourier transform on each state of the tensor product:

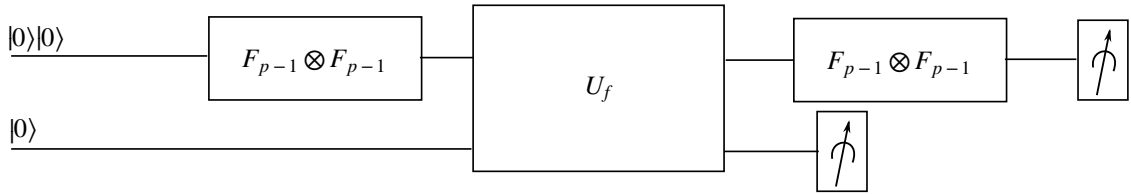

Figure 3.3

On the first register applying the first $F_{p-1} \otimes F_{p-1}$ gives, as in Shor's algorithm, a superposition over the whole group

$$\left(\frac{1}{\sqrt{p-1}} \sum_{a=0}^{p-2} |a\rangle\right)\left(\frac{1}{\sqrt{p-1}} \sum_{b=0}^{p-2} |b\rangle\right) = \frac{1}{p-1} \sum_{a,b=0}^{p-2} |a\rangle|b\rangle \quad (3.9)$$

Then, applying the $U_f$ operator gives the superposition

$$\frac{1}{p-1} \sum_{a,b=0}^{p-2} |a\rangle|b\rangle|f(a,b)\rangle \quad (3.10)$$

Measuring the second register yields a $f(a_0, b_0)$ and makes the first register collapse to the superposition of the preimages. By the discussion above it is:

$$\frac{1}{\sqrt{p-1}} \sum_{\lambda=0}^{p-2} |a_0 + \lambda y\rangle|b_0 - \lambda\rangle|f(a_0, b_0)\rangle \quad (3.11)$$

Now, we apply $F_{p-1} \otimes F_{p-1}$ on the first register, giving the state

$$\frac{1}{\sqrt{p-1}} \sum_{\lambda=0}^{p-2} \left( \frac{1}{\sqrt{p-1}} \sum_{u=0}^{p-2} e^{\frac{2\pi i (a_0 + \lambda y)u}{p-1}} |u\rangle \right) \left( \frac{1}{\sqrt{p-1}} \sum_{v=0}^{p-2} e^{\frac{2\pi i (b_0 - \lambda)v}{p-1}} |v\rangle \right) \quad (3.12)$$

which we can further simplify to

$$\frac{1}{(\sqrt{p-1})^3} \sum_{\lambda, u, v=0}^{p-2} e^{\frac{2\pi i ((a_0 u + b_0 v) + \lambda(yu - v))}{p-1}} |u\rangle|v\rangle = \frac{1}{(\sqrt{p-1})^3} \sum_{u,v=0}^{p-2} e^{\frac{2\pi i (a_0 u + b_0 v)}{p-1}} \left( \sum_{\lambda=0}^{p-2} \left[ e^{\frac{2\pi i (yu - v)}{p-1}} \right]^\lambda \right) |u\rangle|v\rangle$$

$$(3.13)$$

The sum over lambda is a geometric series, and its value is nonzero (and of value $p - 1$) iff $yu - v \equiv 0 \mod (p - 1)$. Replacing $v$ by $yu$ in the expression gives the final state:

$$\frac{1}{\sqrt{p-1}} \sum_{u=0}^{p-1} e^{\frac{2\pi i (a_0 + b_0 y)u}{p-1}} |u\rangle|yu\rangle \quad (3.14)$$

A measurement in the standard basis gives $u$ and $yu$ modulo $p - 1$. If $u$ and $p - 1$ are coprime, using the euclidean algorithm we can find $v$ such that $uv = 1$ modulo $p - 1$ from the first state. Then, we get $y = yuv$ modulo $p - 1$ from the second state. So we have obtained $y = \log_g x$ as expected. Note that u and $p - 1$ are coprime with probability $\frac{\varphi(p-1)}{p} = \Omega\left(\frac{1}{\log(\log(p))}\right)$ so we only need to repeat the procedure about $\log(\log(p))$ times to ensure a success with high probability.

### 3.4. The Hidden Subgroup Problem

In this section, we interpret the algorithms seen in previous sections in a common framework. The hope is to find new algorithms exponentially faster than their classical counterparts.

**Definition 3.3 (hidden subgroup problem)**

Let $G$ be a group and $H \subseteq G$ one of its subgroup. Let $S$ be any set and $f : G \to S$ a function that distinguishes cosets of $H$ i.e. $\forall g_1, g_2 \in G, f(g_1) = f(g_2) \Leftrightarrow g_1 H = g_2 H$. The *hidden subgroup problem* (HSP) is to determine the subgroup $H$ using calls to $f$. ◇

**Example 3.4 (interpretation of previous algorithms)**

| Problem | $G$ | $H$ | $S$ | $f$ | Comment |
|---|---|---|---|---|---|
| Simon | $\mathbb{Z}_2^n$ | $\{0, s\}$ | $\mathbb{Z}_2^m$ | Such that two distinct elements $x$, $x'$ have the same image iff $x' = x \oplus s$. | $s$ is any $n$-bits strings. |
| Shor's algorithm (order finding subroutine) | $\mathbb{Z}_N$ | $r\mathbb{Z}_N$ | $\mathbb{Z}_N$ | $f(x) = a^x \bmod (N_0)$ | $N_0$ is the number to factor, $a < N_0$ a random integer coprime with $N_0$, $r$ is the order of $a$ modulo $N_0$. Ideally, $N$ should be a multiple of $r$ but we can choose some $N = O(N_0^2)$ giving a good approximation. |
| Discrete logarithm | $\mathbb{Z}_{p-1} \times \mathbb{Z}_{p-1}$ | $(y, -1)\mathbb{Z}_{p-1}$ | $\mathbb{Z}_p^*$ | $f(a, b) = g^a x^b \bmod (p)$ | $p$ is prime, $g$ a generator of $\mathbb{Z}_p^*$, $x$ an element of $\mathbb{Z}_p^*$, $y$ the discrete logarithm of $x$. |

Figure 3.4

In the remaining of this document, we are only considering the case of a *finite* group $G$. In particular, the set $S$ may also be taken to be finite and we may even assume $|S| \leq |G|$. Consequently, we can formulate the problem in terms of circuits: we encode the elements of the two sets with at most $n = \lceil \log|G| \rceil$ bits and hence $f$ can be viewed as a function $f : \mathbb{Z}_2^n \to \mathbb{Z}_2^n$ and represented by a quantum logic gate $U_f$. A naive algorithm for the hidden subgroup problem is the following:

1. compute $s_0 = f(0)$
2. compute $f(g)$ for all $g \in G$ and return those for which $f(g) = s_0$ i.e. the elements of $H$.

The runtime complexity of this algorithm is $O(|G|\text{comp}(f))$ so at least $O(\exp(n))$. This is really bad compared with the efficient quantum algorithms previously seen. For instance Shor's algorithm run in $\text{poly}(\log N) = \text{poly}(n)$ i.e. exponentially faster. Hence we are lead to the following definition:

**Definition 3.5 (efficient)**

An algorithm for the hidden subgroup problem is said to be *efficient* iff it returns a generating set of elements of $H$ using a complexity polynomial in $n = \lceil \log |G| \rceil$. ◇

In particular, we have the following requirements for an efficient solution:

1. $f$ has a polynomial complexity.
2. $f$ is called a polynomial number of times.
3. the set generating $H$ has a polynomial size.

Note that not all functions $f : \mathbb{Z}_2^n \to \mathbb{Z}_2^n$ have a polynomial complexity, as indicated in exercise 3.16 of [NieChu2007]. The idea is that there are $(2^n)^{(2^n)}$ such functions so one can note encode all of them using only a polynomial number of elementary gates. As a consequence, $f$ has to be chosen carefully to satisfy the first property.

The argument given in appendix A.2.1.1 of [NieChu2007] shows that there is a set of size at most $\log(|H|) \leq \log(|G|)$ that generates $H$. Actually, for any group $K$, if we pick uniformly at random $t + \log(|K|)$ elements of $K$ (for some integer $t \geq 0$) then the set obtained generates $K$ with probability $\geq 1 - 2^{-t}$ (see Appendix D of [Lom2004]).

Ettinger, Høyer and Knill [EttHøyKni1999] proved that the second requirement can always be satisfied i.e. only a polynomial number of calls to $f$ is needed to identify $H$. The idea is to prepare the state

$$\frac{1}{\sqrt{|G|^m}} \sum_{g_1, \ldots, g_m \in G} |g_1, \ldots, g_m\rangle |f(g_1), \ldots, f(g_m)\rangle \quad (3.15)$$

and measure the second register to get a tensor product of $m$ coset states (see definition 4.8). Then they apply measurements for each $g \in G$, that check whether $g \in H$. They prove that the answers given by the algorithm are correct with high probability for some $m = O(n)$. However, the whole algorithm is not $\text{poly}(n)$ since we test $|G| = \Omega(\exp(n))$ elements.

**Example 3.6**

In Simon's problem an efficient algorithm means a $\text{poly}(n)$ (i.e. the number of bits in the strings $x$, $s$...) complexity and $n = \log(|\mathbb{Z}_2^n|)$. Here, we measure the complexity using the numbers of queries to the oracle rather than elementary operations (because $f$ may not be computed efficiently per comment above).

In Shor's algorithm and the discrete logarithm problem, it means polynomial in the number of bits needed to encode the numbers we work on, i.e. $\lceil \log(N) \rceil$ and $\lceil \log(p) \rceil$ respectively. $f$ uses basic operations on such numbers so has a polynomial complexity. The sizes of the groups we work with are $N$ and $(p-1)^2$ respectively, so their $\log G$'s are equivalent to the values we previously used to measure efficiency.

In all cases, there is only one generator to the group $H$, namely $s$, $r$ and $(y, -1)$. ◻

In the particular examples of previous sections, there are natural encoding of the elements and algorithm to compute the product of two elements. For instance, for the cyclic HSP $G = \mathbb{Z}_N$ the elements are encoded as binary number of length $\lceil \log(N) \rceil$ and there are well-known algorithms of complexity polynomial in $\lceil \log(N) \rceil$ for the group operations. To study the general HSP, it is often useful to see the group $G$ as a black-box where the operations are performed by a group oracle. As in the case of the HSP oracle computing $f$, replacing the group oracles by polynomial-time operations yields an efficient algorithm.

**Definition 3.7 (black-box group, unique encoding, encoding length, input size)**

Let $n = O(\log(N))$ and $m \geq 0$. $n$ is called the *encoding length* and $mn$ the *input size*. A *black-box group with unique encoding* has the following properties:

1. $G$ is given by generators $g_1, \ldots, g_m$.
2. there are oracles to perform multiplication and inversion.
3. each element can be represented by a binary string of length $n$.
4. the previous representation is unique.

For a *black-box group without unique encoding* we replace the point 4 by an oracle performing identity testing (hence equality). ◇

# 4. The "Standard Method" for The Hidden Subgroup Problem

## 4.1. The Abelian Hidden Subgroup Problem

This section deals with the abelian HSP and is mostly based on [Dam2001]. We just give the big picture of the general method and show how it generalizes the three previous examples. A more detailed presentation is given in [Lom2004].

In the previous chapter, we have seen three efficient quantum algorithms that fit into the framework of the hidden subgroup problem. Actually, these problems are also sharing the same kind of algorithm. First, they are all based on a particular case of HSP where the group is abelian. As it is well-known, such a group is isomorphic to a product of cyclic groups $\mathbb{Z}_{t_1} \times \mathbb{Z}_{t_2} \times \mathbb{Z}_{t_3} \times \ldots \times \mathbb{Z}_{t_k}$ and this is clearly the case in the three previous algorithms. We can then describe the quantum circuit in a general way:

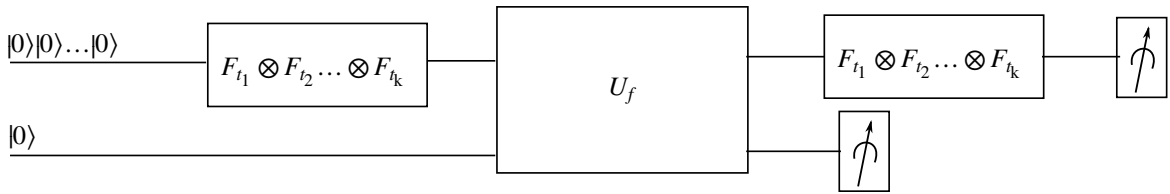

Figure 4.1

The first register is composed of a product of $k$ states of $\lceil \log(t_i) \rceil$ qubits and we apply to this register the operation $F_G = F_{t_1} \otimes F_{t_2} \ldots \otimes F_{t_k}$ i.e. the Fourier transform $F_{t_i}$ to the i-th state (note that for Simon's problem, $H = F_2$). First, let's consider the effect of $F_G$ on any element $g \in G$:

$$F_G |g\rangle = F_G\left(\bigotimes_{i=1}^{k} |g_i\rangle\right)$$
$$= \bigotimes_{i=1}^{k} \left(F_{t_i} |g_i\rangle\right)$$
$$= \bigotimes_{i=1}^{k} \left(\frac{1}{\sqrt{t_i}} \sum_{g_i'=0}^{t_i-1} e^{\frac{2i\pi g_i g_i'}{t_i}} |g_i'\rangle\right)$$
$$= \frac{1}{\sqrt{\prod_{i=1}^{k} t_i}} \sum_{g' \in G} \left(\prod_{i=1}^{k} e^{\frac{2i\pi g_i g_i'}{t_i}}\right)\left(\bigotimes_{i=1}^{k} |g_i'\rangle\right)$$
$$= \frac{1}{\sqrt{|G|}} \sum_{g' \in G} \chi_g(g') |g'\rangle$$

(4.1)

Where we have introduced the characters $\chi_g(g')$. Note that $g \mapsto \chi_g$ is a group homomorphism i.e. $\chi_{g+g'} = \chi_g \chi_{g'}$. We also have $\chi_g(g') = \chi_{g'}(g)$ and so $\chi_g(g' + g'') = \chi_g(g') \chi_g(g'')$. Hence we can define the group $H^\perp =$ $\{g \in G | \chi_g(H) = 1\}$. By Theorem 3.10 of [Lom2004], $H^\perp$ is isomorphic to $G/H$ and moreover $(H^\perp)^\perp = H$.

As we have already seen many times, the state after the gate $U_f$ is the following superposition:

$$\frac{1}{\sqrt{|G|}} \sum_{g \in G} |g\rangle |f(g)\rangle \text{ where } |g\rangle = |g_1\rangle \ldots |g_k\rangle \quad (4.2)$$

Measuring the second register gives a value $y = f(g^0)$ and leaves the first register in the state

$$\frac{1}{\sqrt{|H|}} \sum_{h \in H} |g^0 + h\rangle \quad (4.3)$$

Then we apply the Fourier transform $F_G$. Using the expression of formula 4.1 and the properties of the characters, we get

$$\begin{aligned}
F_G \left( \frac{1}{\sqrt{|H|}} \sum_{h \in H} |g^0 + h\rangle \right) &= \frac{1}{\sqrt{|H|}} \sum_{h \in H} \left( F_G |g^0 + h\rangle \right) \\
&= \frac{1}{\sqrt{|H||G|}} \sum_{h \in H} \sum_{g \in G} \chi_{(g^0 + h)}(g) |g\rangle \\
&= \frac{1}{\sqrt{|H||G|}} \sum_{h \in H} \sum_{g \in G} \chi_g(g^0) \chi_g(h) |g\rangle \\
&= \frac{1}{\sqrt{|H||G|}} \sum_{g \in G} \left( \chi_g(g^0) \sum_{h \in H} \chi_g(h) \right) |g\rangle
\end{aligned} \quad (4.4)$$

If we measure the first register, the probability to get an element of $H^\perp$ is:

$$\begin{aligned}
\frac{1}{|H||G|} \sum_{g \in H^\perp} \left| \chi_g(g^0) \sum_{h \in H} \chi_g(h) \right|^2 &= \frac{1}{|H||G|} \sum_{g \in H^\perp} \left( |\chi_g(g^0)|^2 \left| \sum_{h \in H} 1 \right|^2 \right) \\
&= \frac{1}{|H||G|} \sum_{g \in H^\perp} \left( 1^2 |H|^2 \right) \\
&= \frac{|H^\perp||H|^2}{|H||G|} \\
&= \frac{|G||H|^2}{|H|^2|G|} \\
&= 1
\end{aligned} \quad (4.5)$$

Moreover, the previous calculation shows that each element of $H^\perp$ has the same probability to be measured. Hence, measuring the first register yields a uniformly random element of $H^\perp$. Repeating the previous circuit about $n = \lceil \log |G| \rceil$ times gives with high probability a set of generators $g^1, \ldots, g^{O(n)}$ of $H^\perp$. We have $h \in H = (H^\perp)^\perp$ iff for all $j$, $\chi_h(g^j) = 1 = \prod_{i=1}^{k} e^{\frac{2\mathrm{i}\pi h_i g_i^j}{t_i}}$. If we denote $e$ the lower common multiple of the $t_i$'s and $\alpha_i^{g^j} = eg_i^j/t_i$, $h \in H$ iff $\sum_{i=1}^{k} \alpha_i^{g^j} h_i \equiv 0 \bmod (e)$. We can put this system in Smith normal form (i.e. write as a diagonal system) in time polynomial of $n$. Then get $O(n)$ linear congruences i.e. of the form $ax \equiv b \bmod (e)$. The method to enumerate the list of solution to a linear congruence is well-known. We randomly pick one solution to each linear congruence to get a uniformly random element of $H$. Repeating this about $n$ times gives with high probability a set of generators of $H$.

As previously stated, we also need an efficient quantum circuit to compute the Fourier transform $F_G$. Note that such a circuit is known for $F_{2^x}$ and an approximate one (i.e. measuring the result gives a probability distribution very close to the exact version) for other $F_x$'s. See for instance [Lom2004]. This allows to have an approximate implementation of $F_G = F_{t_1} \otimes F_{t_2} \ldots \otimes F_{t_k}$ that does not change too much the measurement of the circuit of figure 4.1.

As a consequence, we have an efficient solution to the abelian HSP if we know a decomposition $\mathbb{Z}_{t_1} \times \mathbb{Z}_{t_2} \times \mathbb{Z}_{t_3} \times \ldots \times \mathbb{Z}_{t_k}$ of $G$. Otherwise, we can get it if we are working in the black-box model with unique encoding. We describe the idea of [CheMos2001], without trying to optimize the computation. We choose randomly $k' = O(n)$ elements $a_1, \ldots, a_{k'}$ of $G$, so that they generate the group with high probability. Then we define $f : \mathbb{Z}_{|G|}^{k'} \to G$ by $f(x) = \sum_{i=1}^{k'} x_i a_i$. Each term $x_i a_i$ is computed using $O(\log x_i) = O(\log|G|) = \mathrm{poly}(n)$ sums (by a double-and-add algorithm), so $f$ has a polynomial complexity. $\mathbb{Z}_{|G|}^{k'}$ is abelian and we know its decomposition so we can apply the HSP algorithm to find a set of generators for the hidden subgroup $H = \mathrm{Ker}\, f$. Alternatively, these generators can be obtained using the algorithm of [BucNei1996]. It can be shown that from this set of generators we can find (classically) in $\mathrm{poly}(n)$ a set of generators $\overline{u_1}, \ldots, \overline{u_k}$ of $\mathbb{Z}_{|G|}^{k'}/H$ that satisfies:

$$\mathbb{Z}_{|G|}^{k'}\Big/H = \langle\overline{u_1}\rangle \oplus \ldots \oplus \langle\overline{u_k}\rangle \quad (4.6)$$

Using the isomorphism $G \cong \mathbb{Z}_{|G|}^{k'}/H$, we get $G = \langle v_1\rangle \oplus \ldots \oplus \langle v_k\rangle$ where $v_i = f(u_i)$. We define an abelian HSP by $f_i : \mathbb{Z}_{|G|} \to G$ and $f_i(x) = xv_i$ (again, $f$ has a $\mathrm{poly}(n)$ complexity). As in Shor's algorithm, the hidden subgroup is $t_i \mathbb{Z}_{|G|}$ where $t_i$ is the order of $v_i$. Finally, we get $G \cong \mathbb{Z}_{t_1} \times \mathbb{Z}_{t_2} \times \mathbb{Z}_{t_3} \times \ldots \times \mathbb{Z}_{t_k}$.

## 4.2. Fourier Transform over Finite Groups

In order to generalize the previous algorithm to any finite group, we need a way to define Fourier Transform. First, we give a brief overview of representation theory of groups, based on [Ser1971].

**Definition 4.1 (representation)**

A *representation* of a group $G$ is a group homomorphism $\rho : G \to \mathrm{GL}(V)$ where $\mathrm{GL}(V)$ is the group of automorphisms over a finite dimensional vector space $V$. The *dimension* $d_\rho$ of the representation $\rho$ is the dimension of $V$. A subspace $W \subseteq V$ is invariant iff $\forall g \in G$, $\rho(g)(W) \subseteq W$. If the only invariant subspaces are $\{0\}$ and $V$, we say that $\rho$ is *irreducible*. Two representations $\rho : G \to \mathrm{GL}(V_\rho)$ and $\tau : G \to \mathrm{GL}(V_\tau)$ of $G$ are *equivalent* or *isomorphic* iff there is an isomorphism $\psi : V_\rho \to V_\tau$ such that $\tau(g) = \psi \circ \rho(g) \circ \psi^{-1}$ for any $g \in G$. Given a representation $\rho$, we define the *character* $\chi_\rho$ by $\chi_\rho(g) = \text{tr}(\rho(g))$. ◇

In [Ser1971], it is shown that any finite group $G$ has only a finite number of irreducible pairwise non-isomorphic representations. More precisely, if we denote $\hat{G}$ a complete set of such representations and $V_\rho$ and $d_\rho$ the associated vector space and dimension (for any $\rho \in \hat{G}$), we have the following equality from *2.4 Décomposition de la représentation régulière - Corollaire 2*:

$$\frac{1}{|G|} \sum_{\rho \in \hat{G}} d_\rho \chi_\rho(g) = \delta_{g,1} \quad (4.7)$$

In particular for $g = 1$, because $\chi_\rho(1) = \text{tr}\left(I_{d_\rho}\right) = d_\rho$, we have:

$$|G| = \sum_{\rho \in \hat{G}} d_\rho^{\,2} \quad (4.8)$$

For each $\rho$, we can choose a basis of $V_\rho$ and then $\rho(g)$ is represented by a matrix of size $d_\rho$. Again, in view of formula 4.8, we can group all the coefficients $\left(\rho_{ij}(g)\right)_{1 \leq i,\, j \leq d_\rho,\, g \in G}$ in a square matrix of dimension $|G|$. To make this matrix unitary, we add the requirement that each $\rho(g)$ is unitary. We will see that it is always possible to choose the basis such that this requirement is satisfied. Once such a basis is chosen and when no confusion is possible, we will also denote $\rho(g)$ the matrix representation of $\rho(g)$ in this basis.

**Definition 4.2 (general Fourier Transform)**

Let $G$ be a group and $\hat{G}$ a complete set of pairwise non-isomorphic representations $\rho : G \to \text{GL}(V_\rho)$ of size $d_\rho$. For any $\rho$, we fix a basis of $V_\rho$ such that for each $g \in G$ we have a unitary matrix representation $\left(\rho_{ij}(g)\right)_{1 \leq i,\, j \leq d_\rho}$ of $\rho(g)$. Then the *Fourier Transform* over $G$ in the given basis is defined as a linear operation over a Hilbert space of dimension $|G|$. More precisely, if we denote the canonical basis either by group elements $(|g\rangle)_{g \in G}$ or by representations/coordinates $(|\rho ij\rangle)_{\rho \in \hat{G}, 1 \leq i,\, j \leq d_\rho}$, we have:

$$\forall\, g \in G,\ F_G |g\rangle = \frac{1}{\sqrt{|G|}} \sum_{\rho \in \hat{G}} \sqrt{d_\rho} \sum_{i,j=1}^{d_\rho} \rho_{ij}(g) |\rho ij\rangle \quad (4.9)$$

or, if $\hat{G} = \{\rho^1, \cdots, \rho^M\}$ and $d_k = d_{\rho_k}$, by the images of basis vectors:

$$F_G|g\rangle = \frac{1}{\sqrt{|G|}} \begin{pmatrix} \sqrt{d_1}\left(\rho^1_{11}(g)\right) \\ \sqrt{d_1}\left(\rho^1_{12}(g)\right) \\ \vdots \\ \sqrt{d_1}\left(\rho^1_{1d_1}(g)\right) \\ \sqrt{d_1}\left(\rho^1_{21}(g)\right) \\ \vdots \\ \sqrt{d_1}\left(\rho^1_{2d_1}(g)\right) \\ \vdots \\ \sqrt{d_1}\left(\rho^1_{d_1 d_1}(g)\right) \\ \sqrt{d_2}\left(\rho^2_{11}(g)\right) \\ \vdots \\ \sqrt{d_M}\left(\rho^M_{d_M d_M}(g)\right) \end{pmatrix} \quad (4.10)$$

◇

**Proposition 4.3 (Fourier Transform is unitary)**

$F_G$ is a unitary transform i.e. the columns $(F_G|g\rangle)_{g \in G}$ of $F_G$ form an orthonormal family.

proof: We have $\chi_\rho\left((g')^{-1}g\right) = \text{tr}\left(\rho(g')^{-1}\rho(g)\right) = \text{tr}\left(\rho(g')^\dagger \rho(g)\right) = \langle \rho(g'), \rho(g)\rangle_F = \sum_{i,j=1}^{d_\rho} \left(\rho_{ij}(g')\right)^* \left(\rho_{ij}(g)\right)$.

So the hermitian product of two such columns $F_G|g'\rangle$ and $F_G|g\rangle$ is exactly:

$$(F_G|g'\rangle)^\dagger (F_G|g\rangle) = \frac{1}{|G|} \sum_{\rho \in \hat{G}} d_\rho \sum_{i,j=1}^{d_\rho} \left(\rho_{ij}(g')\right)^* \left(\rho_{ij}(g)\right) = \frac{1}{|G|} \sum_{\rho \in \hat{G}} d_\rho \chi_\rho\left((g')^{-1}g\right) \quad (4.11)$$

and we conclude with formula 4.7. □

**Proposition 4.4 (representations are unitarizable)**

For each $\rho \in \hat{G}$, there is a basis $\left(f_i^\rho\right)_{1 \leq i \leq d_\rho}$ of $V_\rho$ such that every $\rho(g)$ is unitary.

proof: Let $\left(e_i^\rho\right)_{1 \leq i \leq d_\rho}$ be an arbitrary basis of $V_\rho$. We have the canonical hermitian product given by $\left\langle \sum_{i=1}^{d_\rho} a_i e_i^\rho, \sum_{i=1}^{d_\rho} b_i e_i^\rho \right\rangle = \sum_{i=1}^{d_\rho} a_i^* b_i$. We define $\langle a, b\rangle_G = \sum_{g \in G} \langle \rho(g)(a), \rho(g)(b)\rangle$ which is still a inner product because $\rho(g)$ is an automorphism. Using Gram-Schmidt process, we construct a new basis $f_1^\rho, \ldots, f_{d_\rho}^\rho$ orthonormal for $\langle .,.\rangle_G$ in which we express $\rho(g)$. On the one hand, we have by construction $\langle a, b\rangle_G =$ $\langle \rho(g)(a), \rho(g)(b) \rangle_G$ and in particular $\langle \rho(g)(f_i^\rho), \rho(g)(f_j^\rho) \rangle_G = \langle (f_i^\rho), (f_j^\rho) \rangle_G$. On the other hand, $\rho(g)(f_j^\rho) = \sum_{k=1}^{d_\rho} \rho_{kj}(g) f_k^\rho$ and hence $\langle \rho(g)(f_i^\rho), \rho(g)(f_j^\rho) \rangle_G = \sum_{k=1}^{d_\rho} \rho_{k,i}^*(g) \rho_{k,j}(g)$. Finally we have:

$$\begin{aligned}\rho(g)^\dagger \rho(g) &= \left( \sum_{k=0}^{d_\rho} \rho_{k,i}^*(g) \rho_{k,j}(g) \right)_{1 \leq i, j \leq d_\rho} \\ &= \left( \langle \rho(g)(f_i^\rho), \rho(g)(f_j^\rho) \rangle_G \right)_{1 \leq i, j \leq d_\rho} \\ &= \left( \langle (f_i^\rho), (f_j^\rho) \rangle_G \right)_{1 \leq i, j \leq d_\rho} \\ &= (\delta_{i,j})_{1 \leq i, j \leq d_\rho} \\ &= I_{d_\rho}\end{aligned} \qquad (4.12)$$

□

Let's see how the previous definition generalizes the abelian case:

**Example 4.5 (Abelian Fourier Transform)**

Suppose $G$ is abelian and define for all $g \in G$ the 1-dimensional (i.e. $d_{\rho_g} = 1$) representation $\rho_g : G \to \mathrm{GL}(\mathbb{C}^1) \cong \mathbb{C}^*$ by $\rho_g(h) = \chi_g(h)$ where $\chi_g$ is the character defined in the previous chapter. The representations are clearly unitary. Note that $\chi_{\rho_g}(h) = \chi_g(h)$ so the definition of the character is consistent with what we have seen earlier. Suppose that two representations $\rho(g)$ and $\rho(g')$ are isomorphic. This means that there is a $\lambda \neq 0$ such that $\rho_g(h) = \lambda \rho_{g'}(h) \lambda^{-1} = \rho_{g'}(h)$ for all $h \in G$. With the notation of previous chapter, we have

$$\frac{\rho_g}{\rho_{g'}}(h) = 1 = \left( \prod_{i=1}^{k} e^{\frac{2i\pi(g_i - g_i')h_i}{t_i}} \right) \qquad (4.13)$$

For $j$ going from 1 to $k$, we evaluate the equality at $h = h^j = (\delta_{i,j})_{1 \leq i \leq k}$ we get $g_j - g_j' = 0$ and finally $g = g'$. So $\hat{G} = \{\rho(g) | g \in G\}$ is a set of pairwise non-isomorphic 1-dimensional representations. It is complete since it satisfies the equality of formula 4.8. Using the identification $|g\rangle \cong |\rho_g 11\rangle$, and $(\rho_g)_{11}(h) = \chi_g(h)$ we recover the definition of the abelian Fourier Transform. ⌑

**Remark 4.6 (irreducible representations of non-abelian groups)**

In *3.1 Sous-groupes commutatifs - Théorème 9* of [Ser1971], it is shown that $G$ is abelian iff all its irreducible representations are 1-dimensional. So in the nonabelian case, we will always have to deal with at least one representation of dimension greater than 1. △

The proposition 4.4 above shows that there is at least one basis of $V_\rho$. Actually, we obtain all of them by

unitary change of basis:

**Proposition 4.7**

Let $\tau$ be an irreducible representation of $G$ and let $\tau(g)$ denote an unitary matrix representation. Then the possible unitary matrix representations of all other isomorphic representations $\rho$ are given by $\forall\, g \in G$, $\rho(g) = U\tau(g)U^{\dagger}$ for any unitary matrix $U$.

proof: First, it is clear that given a unitary matrix, then $\rho(g) = U\tau(g)U^{\dagger}$ is still unitary and is the matrix representation of $\rho = \tau$ obtained using $U$ as a change-of-basis matrix. Conversely, suppose $\rho$ is isomorphic to $\tau$ and let $\rho(g)$ denote an unitary matrix representation. Then there is an invertible matrix $M$ (the matrix representation of an isomorphism between $\tau$, $\rho$ in their respective basis) such that $\forall\, g \in G\, \rho(g) = M\tau(g)M^{-1}$. This gives:

$$I = \rho(g)^{\dagger}\rho(g) = \left(M\tau(g)M^{-1}\right)^{\dagger}\left(M\tau(g)M^{-1}\right) = \left(M^{\dagger}\right)^{-1}\tau(g)^{-1}M^{\dagger}M\tau(g)M^{-1} \quad (4.14)$$

and finally $\tau(g)M^{\dagger}M = M^{\dagger}M\tau(g)$ for any $g \in G$. Since $\tau$ is irreducible, the Schur's lemma (see *Le lemme de Schur ; premières applications - Proposition 4* of [Ser1971]) implies that $M^{\dagger}M = \lambda I$ for some $\lambda \in \mathbb{C}$. More precisely, $\lambda = \frac{1}{d_\tau}\mathrm{tr}\left(M^{\dagger}M\right) = \frac{1}{d_\tau}\|M\|_F^2 \in \mathbb{R}_+^*$. We conclude the proof by taking $U = \frac{1}{\sqrt{\lambda}}M$. $\square$

In this section, we have defined an operation $F_G$ for any given group. We know that it is unitary so a valid quantum. However, if we want to to use it in a HSP algorithm, we need to implement it efficiently i.e. using $\mathrm{poly}(n)$ elementary gates. As mentioned above, there is an approximate efficient implementation of $F_N$ for cyclic group. By taking the tensor product of such operations, we get an approximate efficient implementation for any abelian group $G$. Other efficient implementations are known for various groups, see [Lom2004] for an overview. The exact description of the quantum gate $F_G$ depends on the design used for the implementation. Nevertheless, we will suppose in what follows that the gate implementing $F_G$ works on $n$ qubits. The input $|g\rangle$ and output $|\rho i j\rangle$ are encoded by the integers $0, \ldots, |G| - 1$. The possible remaining basis states are not touched by the gate.

### 4.3. Weak and Strong Fourier Sampling

For the generalized version of the Fourier Sampling circuit, we replace the first Fourier Transform by a gate $V$ that computes a superposition $\frac{1}{\sqrt{|G|}}\sum_{g \in G}|g\rangle$ from the $|0^n\rangle$ state. There is no canonical implementation for $V$ but if the elements of $G$ are encoded by the integer $0, \ldots, |G| - 1$ (as we have assumed in previous section) we can use the following circuit:

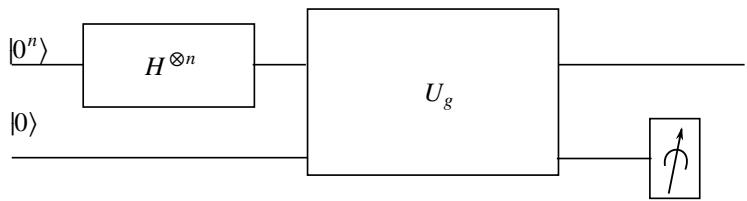

Figure 4.2

We use a hadamard gate to create a superposition over $0, \ldots, 2^n - 1$ and a function $g$ to select the values less than $|G|$: $g(i) = 1$ if $0 \leq i < |G|$ and $g(i) = 0$ otherwise (It can be implemented efficiently, just by comparing the binary decomposition of $i$ and $|G|$). We succeed to create the superposition if the result of the measurement is 1. We have $2^{n-1} < |G| \leq 2^n$, so this happens with probability $\frac{|G|}{2^n} > \frac{1}{2}$.

The remaining of the circuit is similar to what we have previously seen, with the use of a general Fourier Transform gate $F_G$:

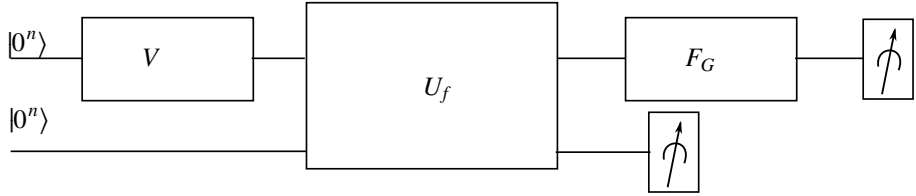

Figure 4.3

After applying the gate $U_f$ we have the superposition

$$\frac{1}{\sqrt{|G|}} \sum_{g \in G} |g\rangle |f(g)\rangle \quad (4.15)$$

We then measure the second register and get $y = f(g^0)$. We obtain a superposition on a coset $g^0 H$:

$$\frac{1}{\sqrt{|H|}} \sum_{h \in H} |g^0 h\rangle \quad (4.16)$$

Creating this superposition is the starting point for all HSP algorithms currently known. Hence we define:

**Definition 4.8 (coset sampling, the "standard method", coset state)**

*Coset sampling* also known as the "*standard method*" for HSP consists in creating many states given by formula 4.16, where $g^0 \in G$ is chosen uniformly at random. These states are called *coset states*. ◇

For Fourier Sampling, the next step is to apply the Fourier Transform $F_G$. This gives the state

$$\sum_{\rho \in \hat{G}} \sum_{i,j=1}^{d_\rho} \sqrt{\frac{d_\rho}{|G|}} \left( \frac{1}{\sqrt{|H|}} \sum_{h \in H} \rho(g^0 h) \right)_{ij} |\rho i j\rangle = \sum_{\rho \in \hat{G}} \sum_{i,j=1}^{d_\rho} \sqrt{\frac{d_\rho}{|G|}} \left( \rho(g^0 H) \right)_{ij} |\rho i j\rangle \quad (4.17)$$

where we have introduced $\rho(g^0 H) = \frac{1}{\sqrt{|H|}} \sum_{h \in H} \rho(g^0 h)$. Finally, we perform a measurement on the first register.

**Definition 4.9 (Fourier Sampling Algorithm)**

The *Fourier Sampling* algorithm for the HSP consists in running the circuit of figure 4.3 several times and trying to deduce information on the hidden subgroup $H$. We distinguish two forms of Fourier Sampling:

- In the *Weak Fourier Sampling* for the HSP, we observe a representation $\rho \in \hat{G}$ with probability $P(\rho) = \sum_{i,j=0}^{d_\rho} P(\rho, i, j)$
- In the *Strong Fourier Sampling*, we observe a representation $\rho \in \hat{G}$ as well as coordinates $i, j$ with probability $P(\rho, i, j)$.

What is the distribution probability? First the conditional probability given $g^0$ is $P\left(\rho,i,j / g^0\right) = \frac{d_\rho}{|G|} \left|\rho(g^0 H)\right|_{ij}^2$. Hence for the Weak Fourier Sampling, we have $P\left(\rho / g^0\right) = \sum_{i,j=1}^{d_\rho} P\left(\rho,i,j / g^0\right) = \frac{d_\rho}{|G|} \left\|\rho(g^0 H)\right\|_F^2$ and using $\rho(g^0 H) = \rho(g^0) \rho(H)$ and the fact that $\rho(g^0)$ is unitary:

$$P(\rho) = \frac{d_\rho}{|G|} \|\rho(H)\|_F^2 \quad (4.18)$$

It was proven in [GriSchVaz2000] that $\rho(H)$ is $\sqrt{|H|}$ times a projection matrix $P$. Moreover, $\rho(H)$ is hermitian:

$$\rho(H)^\dagger = \frac{1}{\sqrt{|H|}} \sum_{h \in H} \rho(h)^\dagger = \frac{1}{\sqrt{|H|}} \sum_{h \in H} \rho(h^{-1}) = \frac{1}{\sqrt{|H|}} \sum_{h' \in H} \rho(h') = \rho(H) \quad (4.19)$$

Hence $\frac{1}{|H|} \|\rho(H)\|_F^2 = \text{tr}\left(P^\dagger P\right) = \text{tr}\left(P^2\right) = \text{tr}(P) = \frac{1}{\sqrt{|H|}} \text{tr}(\rho(H))$ and formula 4.18 can be rewritten:

$$P(\rho) = \frac{d_\rho}{|G|} \sqrt{|H|} \text{tr}(\rho(H)) \quad (4.20)$$

In particular, this probability does not depend on the choice of the basis of $V_\rho$!

For the Strong Fourier Sampling, we first note that the $|G|$-dimensional vectors $\left(\rho(g^0)_{ik}\right)_{g^0 \in G}$ for $1 \leq k \leq d_\rho$ are orthogonal of norm $\sqrt{\frac{|G|}{d_\rho}}$:

$$\begin{aligned}
\left\langle \left(\rho(g^0)_{ik}\right)_{g^0 \in G}, \left(\rho(g^0)_{ik'}\right)_{g^0 \in G} \right\rangle &= \sum_{g^0 \in G} \rho(g^0)_{ik}^* \rho(g^0)_{ik'} \\
&= \sum_{g^0 \in G} \rho\left((g^0)^{-1}\right)_{ki} \rho(g^0)_{ik'} \quad (4.21) \\
&= \frac{|G|}{d_\rho} \delta_{kk'} \delta_{ii}
\end{aligned}$$

where the last line is *2.2 Le lemme de Schur ; premières applications - Corollaire 3* of [Ser1971]. We use this fact to simplify the expression of $P(\rho, i, j)$ as described in [GriSchVaz2000]. Because $g^0$ has been chosen uniformly it is the mean of $P\left(\rho,i,j / g^0\right)$ over the elements of the group $G$:

$$P(\rho, i, j) = \frac{d_\rho}{|G|^2} \sum_{g^0 \in G} \left|\rho(g^0 H)_{ij}\right|^2$$

$$= \frac{d_\rho}{|G|^2} \left\|\left(\rho(g^0 H)_{ij}\right)_{g^0 \in G}\right\|^2$$

$$= \frac{d_\rho}{|G|^2} \left\|\left(\left(\rho(g^0)\rho(H)\right)_{ij}\right)_{g^0 \in G}\right\|^2$$

$$= \frac{d_\rho}{|G|^2} \left\|\left(\sum_{k=0}^{d_\rho} \rho(g^0)_{ik}\rho(H)_{kj}\right)_{g^0 \in G}\right\|^2 \quad (4.22)$$

$$= \frac{d_\rho}{|G|^2} \left\|\sum_{k=0}^{d_\rho} \rho(H)_{kj}\left(\rho(g^0)_{ik}\right)_{g^0 \in G}\right\|^2$$

$$= \frac{d_\rho}{|G|^2} \sum_{k=0}^{d_\rho} \left(\left|\rho(H)_{kj}\right|^2 \frac{|G|}{d_\rho}\right)$$

$$= \frac{1}{|G|} \left\|\rho(H)_j\right\|^2$$

and we finally obtain an expression which is totally independent of $i$:

$$P(\rho, i, j) = \frac{1}{|G|} \left\|\rho(H)_j\right\|^2$$
$$P(\rho, ., j) = \frac{d_\rho}{|G|} \left\|\rho(H)_j\right\|^2 \quad (4.23)$$

This means that measuring the row provides no information and the Strong Fourier Sampling can be reduced to observing $\rho, j$. Note that summing over all the $j$, we recover formula 4.18.

**Remark 4.10 (measurement of $f\left(g^0\right)$)**

As in the abelian case, the measurement of $y = f\left(g^0\right)$ is only used to produce a superposition over one coset. As noted in [GriSchVaz2000], we may discard information by not using this value later. For example, just counting the number of distinct values after many samplings gives an indication on $\frac{|G|}{|H|}$ so on the size of the hidden subgroup $H$. In particular, it is easy to determine whether $H$ is a proper subgroup of $G$ (i.e. the ratio above is at least 2) by repeating several measurements until we find two distinct values or get $k$ times the same value. In the former case we know with certainty that $H$ is a proper subgroup of $G$ and in the latter case we know that $H = G$ with probability at least $1 - 2^{-k}$. For completeness, we give the distribution probability. If $x_1, \ldots, x_{|G|/|H|}$ is a complete set of coset representatives, $P(\rho, i, j, f(x_k)) = \sum_{g^0 \in G} P(x_k) P(g^0/f(x^k)) P(\rho, i, j/g^0) = \sum_{h \in H} \frac{|H|}{|G|} \frac{1}{|H|} \frac{d_\rho}{|G|} \left|(\rho(x_k h H))_{ij}\right|^2$ and finally

$$P(\rho, i, j, f(x_k)) = d_\rho \frac{|H|}{|G|^2} |(\rho(x_k H))_{ij}|^2 \quad (4.24)$$

It seems difficult to construct an algorithm from these values since we do not know the $x_k$ corresponding to the measured value $f(x_k)$. △

As said in remark 4.6, all the irreducible representations are 1-dimensional in the abelian case. Consequently, the Strong Fourier Sampling is the same as the Weak Fourier Sampling: we only take into account the irreducible representations $\rho_{g_j}$ measured. More precisely, we measure random elements $g^1$, ..., $g^{O(n)}$ in $H^\perp = \{g \in G | \chi_g(H) = 1\} = \{g \in G | H \subseteq \text{Ker } \rho_g\}$. Then we solve the systems of equations $\chi_{g^j}(h) = \chi_h(g^j) = 1$ to get elements $h$ in $H$. Said otherwise, we are looking to random elements in $H = \bigcap_{j=1}^{O(n)} \text{Ker } \rho_{g_j}$. In the next section, we will see that more generally, considering the intersection of irreducible representations measured gives information on $H$ and allows to solve the HSP in some particular cases.

### 4.4. The Dedekindian Hidden Subgroup Problem and its extensions

In this section, we see how the Weak Fourier Sampling can be used to extend the abelian HSP. The idea is to measure $\rho_1$, ..., $\rho_{O(n)}$ irreducible representations and to consider the intersection $\bigcap_{j=1}^{O(n)} \text{Ker } \rho_j$. Let's consider first what is happening in an example:

**Example 4.11 (Fourier Sampling on the dihedral group $D_4$)**

The dihedral group $D_4$ is the group of isometries of the plane generated by the reflection $s$ about the x-axis and the rotation $r$ of 90°. It is composed of 4 rotations $r^k$ and 4 reflections $r^k s$ (for $0 \leq k \leq 3$). It satisfies $r^4 = s^2 = srsr = \text{Id}$. A set of complete irreducible representations is given by the table of figure 4.4 (adapted from *5.3 Le groupe diédral $D_n$* of [Ser1971]):

| $\rho$ | $\rho(r^k)$ | $\rho(r^k s)$ | Ker $\rho$ |
|---|---|---|---|
| $a$ | 1 | 1 | $D_4$ |
| $b$ | 1 | -1 | $\{\text{Id}, r, r^2, r^3\}$ |
| $c$ | $(-1)^k$ | $(-1)^k$ | $\{\text{Id}, r^2, s, sr^2\}$ |
| $d$ | $(-1)^k$ | $(-1)^{k+1}$ | $\{\text{Id}, r^2, sr, sr^3\}$ |
| $e$ | $\begin{pmatrix} (-\mathrm{i})^k & 0 \\ 0 & \mathrm{i}^k \end{pmatrix}$ | $\begin{pmatrix} 0 & (-\mathrm{i})^k \\ \mathrm{i}^k & 0 \end{pmatrix}$ | $\{\text{Id}\}$ |

Figure 4.4

We are considering two hidden subgroups $H_1 = \{1, rs\}$ and $H_2 = \{1, r^2\}$. Using formula 4.20 and formula 4.23, it is easy to get:

|   | $H = H_1$ | | $H = H_2$ | |
|---|---|---|---|---|
| $\rho$ | $\rho(H)$ | $P(\rho)$ | $\rho(H)$ | $P(\rho)$ |
| $a$ | $\sqrt{2}$ | $\frac{1}{4}$ | $\sqrt{2}$ | $\frac{1}{4}$ |
| $b$ | $0$ | $0$ | $\sqrt{2}$ | $\frac{1}{4}$ |
| $c$ | $0$ | $0$ | $\sqrt{2}$ | $\frac{1}{4}$ |
| $d$ | $\sqrt{2}$ | $\frac{1}{4}$ | $\sqrt{2}$ | $\frac{1}{4}$ |
| $e$ | $\frac{1}{\sqrt{2}}\begin{pmatrix} 1 & -i \\ i & 1 \end{pmatrix}$ | $\frac{1}{2}$ | $\begin{pmatrix} 0 & 0 \\ 0 & 0 \end{pmatrix}$ | $0$ |

Figure 4.5

and the probability distribution observed by a Strong Fourier Sampling is:

|   | $P(a, ., 1)$ | $P(b, ., 1)$ | $P(c, ., 1)$ | $P(d, ., 1)$ | $P(e, ., 1)$ | $P(e, ., 2)$ |
|---|---|---|---|---|---|---|
| $H_1$ | $\frac{1}{4}$ | $0$ | $0$ | $\frac{1}{4}$ | $\frac{1}{4}$ | $\frac{1}{4}$ |
| $H_2$ | $\frac{1}{4}$ | $\frac{1}{4}$ | $\frac{1}{4}$ | $\frac{1}{4}$ | $0$ | $0$ |

Figure 4.6

Suppose we repeat enough time the Weak Fourier Sampling. For $H = H_1$, we will measure the representations $a, d, e$. The intersection of their kernels is $\{\text{Id}\} \neq H$. For $H = H_2$, we will measure the representations $a, b, c, d$. The intersection of their kernels is $\{\text{Id}, r^2\} = H$. ▫

Consequently, the direct application of the abelian method does not always work. Note that $H_2$ is normal (it is the of center $D_4$) but $H_1$ is not (for example $s(rs)s^{-1} = sr = r^3s \notin H_1$). In general, $\text{tr}(\rho(g^{-1}Hg)) = \text{tr}(\rho(g)^{-1}\rho(H)\rho(g)) = \text{tr}(\rho(H))$ so by formula 4.20, the Weak Fourier Sampling can not distinguish $H$ with one of its conjugate $g^{-1}Hg$. However, we can still hope that it will work for a normal subgroup $H$. This is what Hallgren, Russel and Ta-Shma proved in [HalRusTa-2000]:

**Theorem 4.12 (Normal Hidden Subgroup)**

Let $\rho_1, ..., \rho_s$ be $s = c \log(|G|)$ samples of Weak Fourier Sampling for a normal hidden subgroup $H$. We have

$$P\left(H = \bigcap_i \operatorname{Ker} \rho_i\right) \geq 1 - e^{-\frac{(c-2)^2}{2c} \log(|G|)} \quad (4.25)$$

□

As a consequence, we can use this to solve the Dedekindian Hidden Subgroup Problem i.e. over groups that have only normal subgroups. Of course, this includes the Abelian HSP that we have previously studied. The non-abelian Dedekind groups are called Hamiltonian and they are of the form $G = \mathbb{Z}_2^k \times A \times \mathbb{Q}_8$ where $A$ is an abelian group with all elements of odd order. The appendix B.2 of [HalRusTa-2000] gives a precise algorithm for the Hamiltonian HSP: an efficient implementation of the Weak Fourier Sampling over $G$ and a way to pick $O(n)$ random solutions to the system $\rho_i(h) = I_{d_{\rho_i}}$ in order to get a set of generators for $H$.

By calling several times the Weak Fourier Sampling (possibly over subgroups of $G$), the Dedekindian HSP can be generalized. However the possibility to compute the Fourier Transform, to solve systems $\rho_i(h) = I_{d_{\rho_i}}$ and to do the same for any of its subgroup involved in the algorithm, depends on the underlying group $G$. First, as noted in [GriSchVaz2000], running this algorithm for any $H$ (not necessarily normal) gives with high probability the highest subgroup of $H$ which is normal in $G$. We can also solve the case where "almost all" subgroups of $G$ are normal i.e. the intersection of normalizers $M_G$ is big enough or, equivalently, that $\frac{|G|}{|M_G|}$ is small. For example $M_G = G$ for an abelian group and $\frac{|G|}{|M_G|} = 1$. In [GriSchVaz2000], Grigni et al. presented an efficient HSP algorithm when $\frac{|G|}{|M_G|} \in \exp(O(\sqrt{\log(\log|G|)}))$. They give an application to the semi-direct product $\mathbb{Z}_3 \rtimes \mathbb{Z}_{2^n}$. Gavinsky [Gav2004] strengthened their result to $\frac{|G|}{|M_G|} \in \operatorname{poly}(\log|G|)$. Actually, he proved that we can even just assume $\frac{|G|}{|HM_G|} \in \operatorname{poly}(\log|G|)$ so that the algorithm still works when $H$ is large.

Note that if the group is given as a black-box group without necessarily unique encoding, there is an alternative way to find the normal subgroup $H$ as described in [IvaMagSan2001]. The algorithm does not require the assumptions on Fourier Transform and systems above. Its complexity is given by the input size + a number $v(^G/_H)$ that we will define later. We will come back to this method in the last part of this report.

# 5. The Dihedral and Symmetric Hidden Subgroup Problems

## 5.1. The Dihedral Hidden Subgroup Problem

For any $N \geq 1$, the dihedral group $D_N$ is the group of isometries of the plane generated by the reflection $s$ about the $x$-axis and the rotation $r$ of angle $\frac{2\pi}{N}$. It is composed of $2N$ elements: $N$ rotations $r^k$ and $N$ reflections $r^k s$ ($0 \leq k \leq N-1$). The elements satisfy the relation $r^N = s^2 = srsr = \text{Id}$. We have already seen the case $N = 4$ in an example of Fourier Sampling. Alternatively, we can describe $D_N$ as a semi-direct product $\mathbb{Z}_N \rtimes \mathbb{Z}_2$, where $(a, b)$ represents the isometry $r^a s^b$. We have $r^{a_1} s^0 r^{a_2} s^{b_2} = r^{a_1+a_2} s^{0+b_2}$ and $r^{a_1} s^1 r^{a_2} s^{b_2} = r^{a_1-a_2}(r^{a_2} s r^{a_2}) s^{b_2} = r^{a_1-a_2} s^{1+b_2}$, so the law of this semi-direct product is given by

$$(a_1, b_1)(a_2, b_2) = \left(a_1 + (-1)^{b_1} a_2, b_1 + b_2\right) \quad (5.1)$$

and consequently the inverse operation is

$$(a, b)^{-1} = \left((-1)^{b+1} a, b\right) \quad (5.2)$$

**Definition 5.1 (DHSP)**

The *dihedral HSP* (DHSP) is the hidden subgroup problem for the dihedral group $G = D_N \cong \mathbb{Z}_N \rtimes \mathbb{Z}_2$. An efficient algorithm for the dihedral HSP has a complexity $\text{poly}(\log(N))$ (this is equivalent to our general definition because $\log(|G|) = \log(2N) = 1 + \log(N)$). $\diamond$

DHSP is by itself a natural candidate for finding efficient quantum algorithm based on a nonabelian HSP: on the one hand it is a nonabelian group with a simple structure (so we can hope it is not too difficult to solve) and on the other hand $\langle (d, 1) \rangle_{0 \leq d \leq N-1}$ is an exponential number of subgroups (so the brute-force guessing is infeasible). We will see another motivation in the next section.

What are the possible hidden subgroups $H$? Consider the cyclic subgroup $G' = \mathbb{Z}_N \times \{0\}$ of $G$. Then $H' = G' \cap H$ is a subgroup of $G'$ so there is a divisor $r$ of $N$ such that $H' = (r\mathbb{Z}_N) \times \{0\}$. If $H' \neq H$ then there exists $(d, 1) \in H$. If $(a, 1)$ is another element of $H$, then $(d, 1)(a, 1) = (d - a, 0) \in H'$. We have $(a, 1) = (d, 1)(d - a, 0) \in (d, 1)H'$. As a consequence, $H = H' \cup (d, 1)H'$. Note that if $d = rq + d'$ is the euclidean division of $d$ by $r$, $(d', 1) = (-rq, 0)(d, 1) \in H$. Replacing $d$ by $d'$, we can assume $0 \leq d < r$. Finally, we have:

**Proposition 5.2 (subgroups of $D_N$)**

The subgroups of $D_N$ are:

- $H_r = (r\mathbb{Z}_N) \times \{0\} = \{(rl, 0) \mid 0 \leq l < \frac{N}{r}\}$ for a divisor $r$ of $N$
- $H_{r,d} = H_r \cup (d, 1) H_r = \left\{(rl, 0), (d + rl, 1) \mid 0 \leq l < \frac{N}{r}\right\}$ for a divisor $r$ of $N$ and some $0 \leq d < r$.

Moreover, the dihedral HSP reduces to efficiently find $d$ when $H = H_{r,d}$ and $r$ is known.

proof: The first part has been discussed above. The value $r$ can be found with high probability in $O(\text{poly}(\log N))$ by the cyclic HSP algorithm using the oracle $f_{|G'}$. Hence we may assume that $r$ is known. Suppose we have an algorithm that finds $d$ with high probability when $H = H_{r,d}$ and $r$ is known. We can suppose that this algorithm always returns a value $d_0$. If $f(d_0, 1) = f(0, 0)$ then we return $H = H_{r,d_0}$ otherwise we return $H = H_r$. Because the sub-routines succeed with high probability, so does the whole algorithm. □

Can we use the general algorithms based on the Weak Fourier Sampling, that we have previously seen? We note that $(a, b)(rl, 0)(a, b)^{-1} = \big((-1)^b rl, 0\big)$ so $H_r$ is normal. We also have $(a, b)(d + rl, 1)(a, b)^{-1} = \big(2a + (-1)^b(d + rl), 1\big)$ so $H_{r,d}$ is normal iff for all $a, b \in \mathbb{Z}_N$ we have $2a + (-1)^b d \equiv d \mod (r)$ (which is obviously true for $r$ being 1 or 2). The case $a = 1$ and $b = 0$ shows that $r$ can not be more than 2 if we want this equality to hold. As a consequence, the normal subgroups are $H = H_r$ (for any divisor $r$ of $N$), $H = H_{1,0} = G$ and (if $N$ is even) $H = H_{2,0}$ or $H = H_{2,1}$. It is easy to check that they correspond to the intersection of the kernels of representations measured in theorem 9.9 of Appendix B. We notice that $M_G \subseteq N(H_{N,0}) = \{(a, b) \mid 2a \equiv 0 \mod (N)\}$ so $1 \leq |M_G| \leq 4$. As a consequence, $\frac{|G|}{|HM_G|} = \frac{|G||M_G \cap H|}{|H||M_G|} = \frac{2N}{N/r} \frac{|M_G \cap H|}{|M_G|} = \Theta(r)$. So Gavinski's HSP algorithm is successful iff $r = \text{poly}(\log N)$.

Ettinger and Høyer proved in [EttHøy1998] that the dihedral HSP reduces to the case where $H = H_{N,d} = \langle (d, 1) \rangle$. To see that, suppose we start with the reduced case of the theorem i.e. $H = H_{r,d}$ and $r$ is known. The elements of $G/H_r$ are $(i, 0)H_r$ and $(i, 1)H_r$ for $0 \leq i < r$ while the elements of $H/H_r$ are given by $H_r, (d, 1)H_r$. Hence, moving to the quotient group we have the hidden subgroup problem for $G/H_r \cong D_r$ and hidden subgroup $\langle (d, 1) \rangle$. We see that we have a solution to the hidden subgroup problem with complexity $\text{poly}(\log(N), r)$ ($\text{poly}(N)$ to find $r$ and $\text{poly}(r)$ to find $d$ working in $D_r$). In particular the algorithm is already efficient in the case $r = \text{poly}(\log N)$ solved by Gavinski's HSP algorithm. We have proved the important reduction:

**Proposition 5.3 (reduction to finding a slope)**

The dihedral HSP reduces to efficiently find the slope $d$ when $H = \langle (d, 1) \rangle$. □

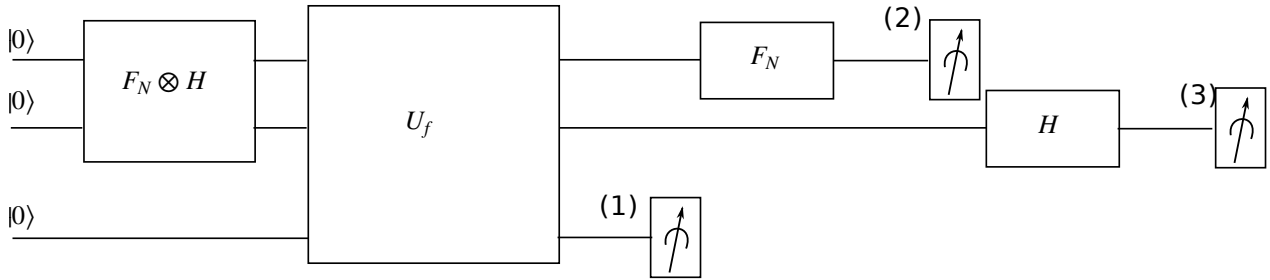

Figure 5.1

Ettinger and Høyer used the same circuit as in HSP over $\mathbb{Z}_N \times \mathbb{Z}_2$ and proved that this yields an algorithm for the dihedral HSP. In this quantum circuit, we distinguish two registers for encoding the elements of $D_N$ and perform three separate measurements.

As usual, the first part of the circuit produces a superposition

$$\sum_{a=0}^{N-1} \sum_{b=0}^{1} |a\rangle|b\rangle|f(a,b)\rangle \quad (5.3)$$

The first measurement yields a $f(a_0, b_0)$ and makes the two first registers collapse to

$$\frac{1}{\sqrt{2}}\left(|a_0\rangle|b_0\rangle + \left|a_0 + (-1)^{b_0}d\right\rangle|b_0 + 1\rangle\right) = \begin{cases} \frac{1}{\sqrt{2}}(|a_0\rangle|0\rangle + |a_0 + d\rangle|1\rangle) & \text{if } b_0 = 0 \\ \frac{1}{\sqrt{2}}(|(a_0 - d)\rangle|0\rangle + |(a_0 - d) + d\rangle|1\rangle) & \text{if } b_0 = 1 \end{cases} \quad (5.4)$$

since $a_0$ and $b_0$ are chosen uniformly at random, this is equivalent to have a state

$$\frac{1}{\sqrt{2}}(|x\rangle|0\rangle + |x + d\rangle|1\rangle) \quad (5.5)$$

for some $x \in \mathbb{Z}_N$ chosen uniformly at random. Applying the Fourier transform $F_N$ to the first register gives the state

$$\frac{1}{\sqrt{2N}} \sum_{k=0}^{N-1} \left( e^{\frac{2i\pi kx}{N}} |k\rangle|0\rangle + e^{\frac{2i\pi k(x+d)}{N}} |k\rangle|1\rangle \right) \quad (5.6)$$

The second measurement returns a $k \in \mathbb{Z}_N$ chosen uniformly at random and make the second register collapse to the state

$$\frac{1}{\sqrt{2}} e^{\frac{2i\pi kx}{N}} \left( |0\rangle + e^{\frac{2i\pi kd}{N}} |1\rangle \right) \quad (5.7)$$

Finally, we apply an hadamard gate on the second register to get

$$\frac{1}{2} e^{\frac{2i\pi kx}{N}} \left( (|0\rangle + |1\rangle) + e^{\frac{2i\pi kd}{N}} (|0\rangle - |1\rangle) \right) = \frac{1}{2} e^{\frac{2i\pi kx}{N}} e^{\frac{i\pi kd}{N}} \left( \left( e^{\frac{i\pi kd}{N}} + e^{-\frac{i\pi kd}{N}} \right) |0\rangle + \left( -e^{\frac{i\pi kd}{N}} + e^{-\frac{i\pi kd}{N}} \right) |1\rangle \right)$$

$$= e^{\frac{2i\pi kx}{N}} \left( \cos\left(\frac{k\pi d}{N}\right) |0\rangle - i\sin\left(\frac{k\pi d}{N}\right) |1\rangle \right)$$

(5.8)

The measurement given by (2) and (3) is then:

$$P(k, j) = \begin{cases} \frac{1}{N} \cos^2\left(\frac{k\pi d}{N}\right) & \text{if } j = 0 \\ \frac{1}{N} \sin^2\left(\frac{k\pi d}{N}\right) & \text{if } j = 1 \end{cases} \quad (5.9)$$

The algorithm of Ettinger and Høyer starts by checking the values $d = 0$ or $d = \frac{N}{2}$ (just by comparing $f(0, 0)$ with $f(0, 1)$ and $f\left(\frac{N}{2}, 1\right)$). Otherwise, they show that there is some $m = O(\log N)$ such that from $k_1, \ldots, k_m$ samples of the previous algorithm we can find $d$, just by searching the minimum of a function $x \mapsto g(x, k_1, \ldots, k_m)$ over the domain $\{1, 2, \ldots, \lfloor \frac{N}{2} \rfloor\}$. The whole algorithm allows to solve the dihedral HSP using $O(\log N)$ calls to $f$. However, to find the minimum of the function $g$ we test $O(N)$ elements.

The state at the output of coset sampling is important since it is the base of all the other DHSP algorithms. Thus we define the following problem, to which DHSP reduces:

**Definition 5.4 (Dihedral Coset Problem)**

Let $N \in \mathbb{N}$ and $d \in \mathbb{Z}_N$. The *dihedral coset problem* (DCP) is to find the value of $d$ given a black box that outputs a state given in formula 5.5 for a random $x \in \mathbb{Z}_N$. ◇

Note that the circuit of figure 5.1 is not the Fourier Sampling based on group representations. Appendix B studies in detail all the possible distribution obtained by this latter method. One of the main results is that the distribution probability of formula 5.9 can be obtained from the output of a Strong Fourier Sampling in a particular basis and so Ettinger and Høyer 's algorithm applies. The second important result is that Strong Fourier Sampling is less general than DCP, in the sense that a black-box for DCP can simulate a Strong Fourier Sampling. This includes the case where we apply Strong Fourier Sampling with any hidden subgroup $H$ and not only $H_{N,d}$. Actually, it is quite remarkable that applying Strong Fourier Sampling on the initial group is essentially the same as applying Strong Fourier Sampling on the quotient group. Some attempts to solve DHSP via Strong Fourier Sampling are given in Appendix C and suggest that this method is not likely to work. Indeed, Fourier Sampling uses only measurement on one coset state at once, while we will see later that we require a polynomial number of them. In appendix G, we propose an entirely new approach, based on uniform superpositions over large subsets of $f(D_N)$. In particular, we can solve the case $N = 2^n$ if we have an efficient process to create for $b = 0, 1$ the states $\frac{1}{\sqrt{N'}} \sum_{i=0}^{N'-1} |f(2i, b)\rangle$.

In the next section, we will see in more details the different results obtained for DCP.

## 5.2. The Dihedral Coset Problem

The next successful step after Ettinger and Høyer was the discovery by Kuperberg [Kup2003] of a subexponential-time algorithm to find the slope $d$ and thus the hidden subgroup. First note that ignoring the phase factor, formula 5.7 allows to generate states:

$$|\Psi_k\rangle = \frac{1}{\sqrt{2}} \left( |0\rangle + e^{\frac{2i\pi kd}{N}} |1\rangle \right) \quad (5.10)$$

We now give a rough description of Kuperberg's algorithm in the particular case $N = 2^n$. In that case, if we are able to create $|\Psi_{2^{n-1}}\rangle = \frac{1}{\sqrt{2}} \left( |0\rangle + (-1)^d |1\rangle \right)$ then a measure in the Hadamard basis gives the parity of $d$. Kuperberg noted that the hidden subgroup is necessarily included in the subgroup $H_{2,0}$ (if $d$ is even) or $H_{2,1}$ (if $d$ is odd). Both are isomorphic to $D_{2^{n-1}}$ so we can restrict to the appropriate subgroup and use an iterative algorithm for finding the other digits of $d$.

Suppose we have two states $|\Psi_k\rangle$ and $|\Psi_l\rangle$. Taking their tensor product, applying a CNOT gate and ignoring the phase factor gives a state $|\Psi_{k\pm l}\rangle$ on the first register where $\pm$ is randomly determined by the measurement of the second register. If both $k$, $l$ are multiple of $2^i$ (i.e. have 0's as their $i$ least significant bits) then one of $k \pm l$ is multiple of $2^{i'}$ for some $i' > i$.

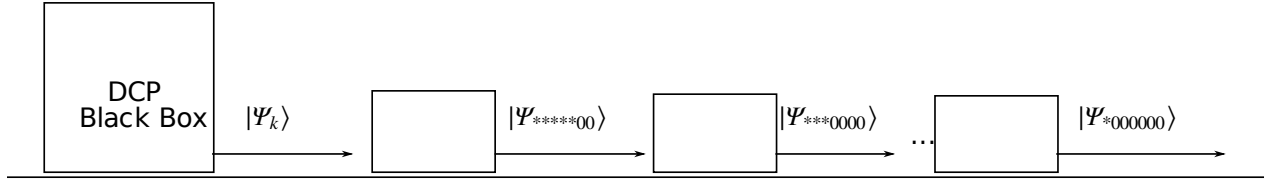

Figure 5.2

The idea of Kuperberg is to use a pipeline as represented in figure 5.2. At the input, many $|\Psi_k\rangle$'s are created by the DCP black box. At each stage $i$, we have states $|\Psi_k\rangle$ such that the $n_i$ least significant bits of $k$ are zeros. We make many combinations as described above and we get at stage $i+1$ a certain amount of states $|\Psi_{k'}\rangle$ such that the $n_{i+1} > n_i$ least significant bits of $k$ are zeros. At the output of the pipeline, we get the state $|\Psi_{2^{n-1}}\rangle$ with high probability.

The running time of Kupenberg's algorithm is $2^{O(\sqrt{\log N})}$ which is subexponential but unfortunately, the algorithm also requires $2^{O(\sqrt{\log N})}$ quantum space. Shortly after that Kupenberg designed its algorithm, Regev [Reg2004] gave an improved version that requires only polynomial space at the price of increasing the running time to $2^{O(\sqrt{\log(\log N)\log N})}$. Note that in both algorithms the query complexity is actually the same as the running time.

In [BacChiDam2005] quantum-information theory is used to see how much information we can get from a DCP black box. If we let $|\varphi_{x,d}\rangle = \frac{1}{\sqrt{2}}(|x\rangle|0\rangle + |x+d\rangle|1\rangle)$ then the density matrix output by the DCP black box is $\rho_d = \frac{1}{N}\sum_{x \in \mathbb{Z}_N} |\varphi_{x,d}\rangle\langle\varphi_{x,d}|$. So if we use $k = \nu \log(N)$ outputs from this black box, our purpose is to identify the density matrix $\rho_d^{\otimes k}$ among all the *a priori* equiprobable values of $d$. Bacon, Childs and van Dam found the optimal measurement for that purpose (the so-called "pretty good measurement") and showed that we can determine $d$ with exponentially small probability if $\nu < 1$ and a constant probability if $\nu > 1$. Since the algorithm of Ettinger and Høyer gives an algorithm using linear (in $\log(N)$) oracle calls, this means that the query complexity of DCP is exactly $\Theta(\log(N))$. Note that Bacon, Childs and van Dam have a similar result in the case where we try to determine only the parity of $d$.

Finally, let's mention a relationship between DCP and the subset sum problem. Recall that this problem is to find a subset of a given set of $k$ integers that sums to another given integer $r$. It is well-known that the decision version of the problem is NP-complete. However, in the discussion below, we fix a density $k > \log(N)$. We can still hope that it is possible to solve some instances at that density and so that it will help us to get better algorithms for DHSP.

**Definition 5.5 (subset sum problem)**

Let $N > 0$ be an integer. The *subset sum problem* at a fixed density $k > 0$ is, given a $x \in \mathbb{Z}_N^k$ and a target value $t \in \mathbb{Z}_N$, to find an element of $S_t^x = \{b \in \mathbb{Z}_2^k \text{ such that } b.x = t \mod (N)\}$. A *legal* instance of the problem is such that $S_t^x \neq \emptyset$. ◇

In [Reg2003], Regev proved that if it is possible to solve a fraction $\frac{1}{\text{poly}(\log N)}$ of the legal instances of the subset sum problem if $k > 4 + \log(N)$, then we have a solution to DCP. Some algorithms have been proposed after Regev's paper [FlaPrz2004] [Lyu2004], giving an alternative solution for DHSP with subexponential complexity.

Bacon, Childs and van Dam gave a similar result in [BacChiDam2005]. We define the uniform superposition of solutions to a legal instance of the subset sum problem with input $(t, x)$:

$$|S_t^x\rangle = \frac{1}{\sqrt{|S_t^x|}} \sum_{b \in S_t^x} |b\rangle \quad (5.11)$$

if we have a quantum circuit that operates on input states $|t, x\rangle$ and sends the legal instances to $|S_t^x, x\rangle$ then we can implement the optimal measurement and thus solve DCP. Producing these states $|S_t^x\rangle$ is stronger than the Regev's condition: not only we need to be able to solve all the legal instances but also for a given input we need to gather all the solutions in one big entangled state. Conversely, Bacon, Childs and van Dam showed that an implementation of the optimal measurement according to their particular schema yields a solution to the subset sum problem.

Note that these DCP solutions based on the subset sum problem are solving DHSP by coset sampling over the whole group. In contrast, Kuperberg's algorithm for DHSP works recursively: we need a DCP black-box which is able to sample over subgroups of $D_N$ containing $H$. However, there is an alternative non recursive version where we use the phase estimation algorithm to approximate $d$.

### 5.3. Dihedral HSP, Lattice Problems and Cryptosystems

In the two previous sections, we have given an overview of the DHSP results known so far. While we can try to find efficient HSP algorithms for their own sake, Regev showed that the case of the dihedral group is particularly interesting: it is related to some lattice problems whose difficulty is assumed in some cryptographic systems. This section, based on [Reg2003] and [MicReg2008], explains more precisely this relationship.

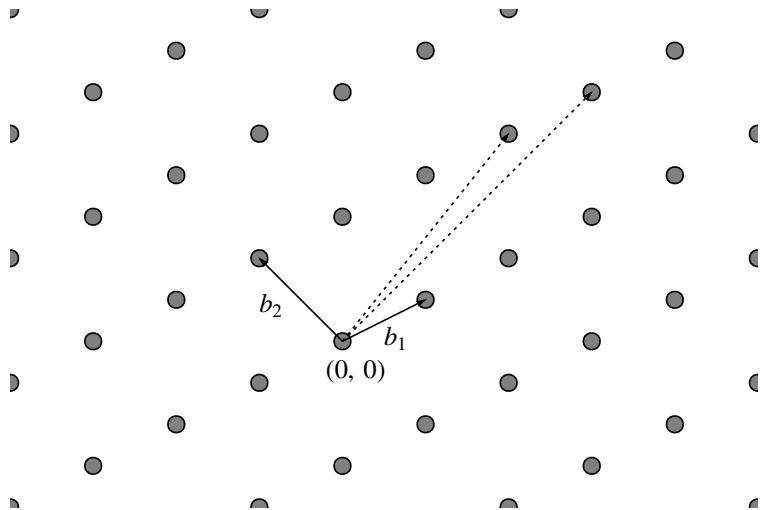

Figure 5.3

**Definition 5.6 (lattice)**

Let $n > 0$ be an integer and $b_1, b_2, ..., b_n$ a basis of $\mathbb{R}^n$ (or equivalently a matrix $[b_1 \mid b_2 \mid ... \mid b_n]$ $= B \in \mathrm{GL}_n(\mathbb{R})$). We define the *lattice* generated by the basis as the set of all linear combinations of the basis vectors with integer-valued coefficients:

$$\mathscr{L}(B) = \{ Bx = \sum_{i=1}^{n} x_i b_i \mid x \in \mathbb{Z}^n \} \quad (5.12)$$

◇

An example of a 2-dimensional lattice i.e. generated by a basis $b_1, b_2$ of $\mathbb{R}^2$ is given in figure 5.3 (notice that an alternative basis is also given). Lattices are involved in many problems assumed to be hard. The one that we will particularly be interested in is the following:

**Definition 5.7 (f(n)-uniqueSVP)**

The Shortest vector problem (SVP) is to find in the lattice a non zero vector of minimum length. In the $f(n)$-uniqueSVP (uniqueSVP) we have the promise that such a vector is necessarily shorter by a factor $f(n)$ than all other non-parallel vectors. ◇

As we have seen at the beginning, Shor's algorithms allow to break cryptosystems based on the hardness of factoring or discrete logarithm's computation. One kind of alternative cryptosystems proposed is based on lattices. In Appendix E, we summarize in one table the overview of [MicReg2008].

In order to establish a connection between one of the $f(n)$-uniqueSVP and DCP, we need to modify the black-box in a way that allows errorous output. So we introduce:

**Definition 5.8 (DCP with failure parameter p)**

Let $N > 0$ be an integer and $p > 0$. The DCP over $D_N$ with failure parameter $p$ is the problem of finding a hidden slope $d \in \mathbb{Z}_N$ given a black box that behaves almost always like a DCP black box. More precisely, this black box produces a DCP-state $\frac{1}{\sqrt{2}}(|x\rangle|0\rangle + |x+d\rangle|1\rangle)$ for some uniformly random $x \in \mathbb{Z}_N$ with probability at least $1 - \frac{1}{(\log N)^p}$ and an arbitrary $|a\rangle|b\rangle \in D_N$ otherwise. ◇

We can now state the main theorem, due to Regev [Reg2003]:

**Theorem 5.9**

Let $n > 0$ an integer and define $N = 2^{(4n+1)n}$. If there is a solution to the DCP over $D_N$ with failure parameter $p$ then there is solution to the $\Theta\left(n^{\frac{1}{2}+2p}\right)$-uniqueSVP. □

Regev said that if we have a solution to the exact DCP, then by taking $p$ large enough the black-box outputs only DCP-states. He concluded that a solution to DCP gives a solution to the $\text{poly}(n)$-uniqueSVP. Let's state a slightly stronger version of its corollary:

**Corollary 5.10**

If there is a solution to DCP over $D_{2^{(4n+1)n}}$ whose query complexity can be expressed as a polynomial of degree $D$, then there is a solution to $\Theta\left(n^{\frac{1}{2}+2D}\right)$-uniqueSVP.

proof: the probability that the DCP with failure parameter $p$ produces only DCP-states after $\sum_{i=0}^{D} a_i (\log N)^i$ oracle calls is at least

$$\left(1 - \frac{1}{(\log N)^p}\right)^{\sum_{i=0}^{D} a_i (\log N)^i} = \prod_{i=0}^{D}\left[\left(1 - \frac{1}{(\log N)^p}\right)^{(\log N)^i}\right]^{a_i} \quad (5.13)$$

When $N \to +\infty$, the limit of the bracket term is 0 if $i > p$, $e^{-1}$ if $i = p$ and 1 otherwise. So the above value is bounded below by a constant $c > 0$ iff $D \leq p$. In that case, we can repeat the procedure about $\frac{1}{c}$ times to be sure that we solve DCP over $D_{2^{(4n+1)n}}$ with failure parameter $p = D$ and so the $\Theta\left(n^{\frac{1}{2}+2D}\right)$-uniqueSVP. □

The corollary gives the optimal degree of the approximate polynomial for uniqueSVP we can get from Regev's algorithm, assuming that the DCP algorithm does not accept failure at all. By the result of [BacChiDam2005], a DCP algorithm can not work with sublinear query complexity, so the best we can do is $\Theta\left(n^{5/2}\right)$-uniqueSVP. If we look to figure 12.1, this would be a rather minor result since it affects only Ajtai-Dwork cryptosystem improved by Goldreich/Goldwasser/Halev, which is not likely to be used in practice. Moreover it is important to note that an algorithm for uniqueSVP only invalidates the proof of security but, contrary for example to Shor's algorithm, does not necessarily give a way to break them.

However, solving DCP would give a HSP-related algorithm that invalidates the security proof of a cryptosystem and hence gives hope that more algorithms of this kind could be found. According to figure 12.1, an interesting direction would be to find algorithms for other problems such that SVP, SIVP or LWE. Considering specific version of lattices used in cryptosystems such that "ideal lattices" would maybe help finding efficient algorithms.

It is worth looking to the ideas of Regev's algorithm. First Regev introduces the "two points problem" which is essentially DCP where the scalars $x$, $x+d$ are replaced by two vectors in $\mathbb{Z}_M^n$ with constant difference. The two point problem reduces to DCP over $D_{(2M)^n}$ by considering the mapping:

$$(a_1, a_2, \ldots, a_n) \mapsto a_1 + a_2 2M + \ldots + a_n (2M)^{n-1} \quad (5.14)$$

This is clearly one-to-one and so (modulo some additional wires) can be implemented as a unitary transform permuting the basis states. In general, it is not clear that it can be done efficiently but in Regev's case, $M = 2^{4n}$ is the power of 2, so this is just adding wires and permuting them. As announced above, we are considering DCP over $D_{2^{(4n+1)n}}$. In particular, we only need a solution to DCP over a power of 2, which may be easier than the general $N$. We also note another way to solve this step: we can directly find the difference $v \in \mathbb{Z}_M^n$ if we have an algorithm for a DCP-like problem over $\text{Dih}\left(\mathbb{Z}_{2^{4n}}^n\right) = \mathbb{Z}_{2^{4n}}^n \rtimes \mathbb{Z}_2$ in place of $D_N \cong \mathbb{Z}_N \rtimes \mathbb{Z}_2$. This corresponds to a generalized DHSP over $\mathbb{Z}_M^n$ with hidden subgroup $\langle (v, 1) \rangle$.

Then, Regev creates a polynomial number of instances for the two points problem. Among all these instances, there is at least one giving an output from which we can extract the shortest vector. So we only run the procedure many times and keep the shortest vector inside the lattice. The routine uses a funtion $F = g \circ f$ where $f$ maps elements of $\mathbb{Z}_M^n$ to lattice vectors and is very similar to coset sampling: we start by a superposition over $\mathbb{Z}_M^n$, apply $U_F$ and measure the result image of $F$. Regev gives two versions with different $g$ (one based on cube partition of the space and a more efficient based on sphere partition) but in any case it is chosen such that with high probability, we are going to have a superposition over two very close lattice points encoded in a state $\frac{1}{\sqrt{2}}(|x\rangle|0\rangle + |x+v\rangle|1\rangle)$.

This similarity between the coset sampling procedure and Regev's algorithm is really important. While all the studies that have been made so far on DHSP were based on the DCP problem i.e. a standard coset sampling on $H = \langle (d, 1) \rangle$, it is well possible that modifying our coset sampling procedure will give better results to solve DHSP. Hopefully, the modifications can be applied in a straightforward manner to Regev's procedure, and so our relation with uniqueSVP will still hold. The example 5.11 gives possible modifications we have experimented, but they have not lead to improvements.

**Example 5.11 (modified coset samplings)**

- Modifying the oracle. Fix $k > 0$ and define the function $f_1(a, b) = f(a, b \mod (2))$ where $a \in \mathbb{Z}_N, b \in \mathbb{Z}_{2k}$ ($f$ is the DCP oracle). Then the coset sampling using $f_1$ as oracle produces a state $\frac{1}{\sqrt{2k}} \sum_{i=1}^{2k} (|x_i\rangle|2i\rangle + |x_i + d\rangle|2i+1\rangle)$ for some uniformly random elements $x_i \in \mathbb{Z}_N$.
- Modifying the input state on the oracle register: fix some $k > 0$. Rather than sending the state $|0\rangle$ on the

oracle register (i.e. the last register of $U_f$) send a superposition of $k$ basis states $\frac{1}{\sqrt{k}}\sum_{i=1}^{k}|y_k\rangle$ for some uniformly random distinct elements $y_k$. Then the coset sampling produces a state $\frac{1}{\sqrt{2k}}\left(\left(\sum_{i=1}^{k}|x_i\rangle\right)|0\rangle + \left(\sum_{i=1}^{k}|x_i+d\rangle\right)|1\rangle\right)$ for some uniformly random distinct elements $x_i \in \mathbb{Z}_N$.

- Making the superposition non-uniform: input the states $\sum_{a=0}^{N-1}\sum_{b=0}^{1}p_{a,b}|a\rangle|b\rangle$ for a distribution probability $p_{a,b}$. The coset sampling produces a state $\frac{1}{\sqrt{2}}\left(p_{x,0}|x\rangle|0\rangle + p_{x+d,1}|x+d\rangle|1\rangle\right)$.

One important variant is DCP over another hidden slope problem for the dihedral group, derived from the initial $D_N$. For example in Kuperberg's algorithm, we use a coset sampling over $D_N$ to determine the parity $b$ of $d$, over $D_{N/2} \cong H_{2,b} \subseteq D_N$ and so forth... Similarly for $\text{Dih}\left(\mathbb{Z}_{2^{4n}}^n\right)$, we can get information on $v$ and move to a smaller subgroup. For example if we determine the parity of $b$ of the $i$th coordinate of $v$ then we can work in $H_{(1,1,\ldots,1,2,1,\ldots,1),b}$ (with a 2 at the $i$th position) instead, which is isomorphic to another generalized dihedral group $\text{Dih}\left(\mathbb{Z}_N^{i-1}\times\mathbb{Z}_{N/2}\times\mathbb{Z}_N^{N-i-1}\right)$ and by induction, we continue to move to dihedral groups of the form $\text{Dih}(A)$ with smaller and smaller complexity. Of course, if we are only working on a specific class of group we have to ensure that we remain in this class at each induction step, contrary to the previous example.

In the case of $A = \mathbb{Z}_{2^{4n}}^n$, this reduction to a smaller group can be translated in a straightforward way to Regev's algorithm: we create the superposition over the new subgroup of $A$ rather than on the whole $A$. For the dihedral group, we have to look carefully to what is happening with the mapping of formula 5.14. Determining the digit of $d$ in the increasing order of their significance is doing the same for each $a_i$ for an increasing $i$ (again, this works well because $M$ is a power of 2). Hence we can again create the superposition over subgroups of the corresponding $A$. The mapping sends them to the decreasing sequence of groups described above for Kuperberg's algorithm. Finally, we have the following corollary:

**Corollary 5.12 (class-preserving recursive DCP algorithm is allowed)**

The previous connections between DCP and uniqueSVP hold for class-preserving recursive DCP algorithms over some generalized dihedral groups. More precisely, we mean an algorithm based on coset sampling over a class of generalized dihedral group $\text{Dih}(A)$ which includes $\text{Dih}\left(\mathbb{Z}_{2^{4n}}^n\right)$ and such that the recursive step moves to a subgroup of the same class. A recursive DCP algorithm is also allowed for $A = \mathbb{Z}_{2^n}$ group if the inductive step consists in determining the parity $b$ of $d$ and moving to $H_{2,b} \cong D_{N/2}$. $\square$

As a final remark, Regev's procedure generates a superposition of two lattice points with constant difference. Maybe we could increase the number of points in the superposition and use another group with larger coset states (for example semi-direct product $A \rtimes \mathbb{Z}_p$ with $A$ abelian). Allowing more points in one partition could weaken the uniqueSVP promise.

### 5.4. The Symmetric Hidden Subgroup Problem

For any $n \geq 1$, let $S_n$ denote the symmetric group i.e. the group of permutations on a set of $n$ elements. This group is of order $n!$, so $\log|S_n| = \sum_{i=2}^{n}\log(i)$ and since $1 \leq \log(i) \leq i$ for $i \geq 2$, we have $n-1 \leq \log|S_n| \leq \frac{n(n+1)}{2} - 1$.

This means that $n$ can be used as a measure of efficiency since $\log|S_n| = \text{poly}(n)$ (more precisely, the Stirling formula gives $|S_n| = \sqrt{2\pi n}\left(\frac{n}{e}\right)^n(1 + o(1))$ so $\log|S_n| \underset{n \to \infty}{\sim} n \log n$).

**Definition 5.13 (SHSP)**

The *symmetric HSP* (SHSP) is the hidden subgroup problem for the symmetric group $S_n$. An efficient algorithm for the symmetric HSP has a complexity $\text{poly}(n)$. ◇

By Cayley's theorem, every group $G$ is isomorphic to a subgroup of $S_{|G|}$. Hence to find a subgroup $H$ of $G$, one can apply a SHSP to the corresponding subgroup of $S_{|G|}$. However, an efficient algorithm for $S_{|G|}$ would be $\text{poly}(|G|)$ while an efficient algorithm for $G$ needs to be $\text{poly}(\log(|G|))$.

A more exciting application is that an efficient algorithm for SHSP would give a solution to the graph isomorphism problem, which has remained unsolved for many decades. We recall:

**Definition 5.14 (Graph isomorphism and automorphism problems)**

Let $(V_1, E_1)$ and $(V_2, E_2)$ be two (undirected) graphs. They are said to be *isomorphic* iff there is a bijection $\varphi : V_1 \to V_2$ such that for any two vertices $a, b \in V_1$ we have $\{a, b\} \in E_1$ iff $\{\varphi(a), \varphi(b)\} \in E_2$. The *graph isomorphism problem* is to determine whether two such graphs are isomorphic, in time polynomial in the size of the input parameters (the number of vertices for example). The graph automorphism problem is to determine whether a graph has a non-trivial automorphism (i.e. there exists a $\varphi \neq \text{Id}$ which is a permutation of the vertices of the graph). ◇

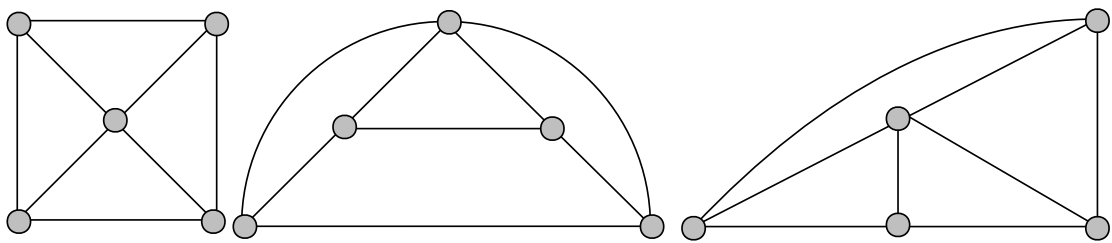

Figure 5.4

The figure 5.4 shows an example of three isomorphic graphs. Clearly, the automorphisms are given by the bijections fixing the vertex of degree 4.

Notice that the definition above is the decisional version of the graph isomorphism problem. It turns out that finding an explicit isomorphism or counting the number of isomorphism has the same complexity, so we can restrict ourselves to the decisional version. This version is clearly in the class NP, but it is known neither to be NP-complete nor to belong to the class P. Many problems are computationally equivalent to the graph isomorphism problem. For instance, it is the case of the graph automorphism problem, to the weaker version with connected graphs or to the stronger version with directed graphs. See, [Köb1993] for details.

Let's consider first the reduction of the graph automorphism problem to SHSP. Let $(V, E)$ be a graph and define $H$ to be the set of its automorphisms. Obviously, the identity is an automorphism, the inverse of an automorphism is an automorphism and the composition of two automorphisms is again an automorphism. Hence $H$ is a subgroup of the permutations of $V$ that can itself be viewed as $S_n$ for $n = |V|$. We define the oracle $f$ to be the function that send $\varphi \in S_n$ to $f(\varphi) = (\varphi(V), \varphi(E))$. As indicated in [Lom2004], $f$ can be computed efficiently. Suppose for instance that we have fixed an enumeration $\{0, \cdots, n-1\}$ of the vertices $V$ and we represent the graph $(V, E)$ by an ordered list of pairs $(a, b)$ where $a \leq b \in V$ and $\{a, b\} \in E$. Given $(V, E)$, we can compute $(\varphi(V), \varphi(E))$ efficiently by replacing each $(a, b)$ by $(\varphi(a), \varphi(b))$ and then applying classical sorting algorithms on the whole structure.

Finally, we have $H$ a subgroup of $S_n$ and clearly $\forall \varphi_1, \varphi_2 \in S_n, f(\varphi_1) = f(\varphi_2) \Leftrightarrow \varphi_1 H = \varphi_2 H$. So an efficient algorithm for DHSP gives a $\text{poly}(n)$ algorithm for determining $H$. Actually, we only need a weaker version of DHSP where we want to determine whether $H$ is non-trivial i.e. the graph has a non-trivial automorphism.

When the two graphs $(V_1, E_1)$ and $(V_2, E_2)$ are *rigid* i.e. do not have a non-trivial automorphism, there exists a reduction from the graph isomorphism problem to a somewhat simpler version of HSP. First, if the graphs do not have the same number of vertices, then they are clearly not isomorphic and we are done. So let $n$ be their common number of vertices. It is easy to check whether each graph is connected in time $O(n^2)$. Again, if one is connected and the other is not then we are done. If both are not connected, then their complement graphs are connected and using these complement graphs instead changes neither the answer to the graph automorphism nor the rigidity of the graphs. Hence we suppose that the two graphs are connected and rigid.

We now build a 2-component graph of $2n$ vertices by taking the disjoint union of $(V_1, E_1)$ and $(V_2, E_2)$. We suppose $V_1 = \{1, \cdots, n\}$ and $V_2 = \{n+1, \cdots, 2n\}$. The permutations of $V_1 \cup V_2$ can be viewed as the group $S_{2n}$. An isomorphism $\varphi : V_1 \to V_2$ between the two initial graphs can be seen as an element of $S_n$. If we introduce the the involution $s_n$ permuting $i$ and $n + i$ for all $1 \leq i \leq n$, $\varphi$ can actually be seen as the automorphism $(\varphi, \varphi^{-1}) s_n \in S_{2n}$ of the disjoint union of the initial graphs.

As above the set $H$ of all automorphisms of the union graph is a subgroup of $S_{2n}$. Since we know that the union graph is made of two connected rigid components, we have either $H = \{\text{Id}\}$ if the two initial graphs are not isomorphic or otherwise $H = \{\text{Id}, \sigma\}$ where $\sigma = (\varphi, \varphi^{-1}) s$ is an involution of $S_{2n}$ permuting the two components. Hence we have obtained a variant of DHSP where the graph is either trivial or a conjugate of $H_0 = \{\text{Id}, s\}$ and whose solution gives an efficient algorithm to determine whether two rigid graphs are isomorphic.

We can actually simplify the problem any further: $H$ is always included in the group $G$ generated by $s_n$ and the permutations of $S_{2n}$ fixing each component (i.e. elements of $S_n \times S_n \subseteq S_{2n}$). We have $(a, b) s_n = s_n(b, a)$ and $s_n$ is of order $2$ so any element of $G$ can be written $(a, b) s_n^c$ for some $(a, b) \in S_n \times S_n$ and $c \in \mathbb{Z}_2$. If $\tau$ is the transposition of $S_2$ then the law of $G$ is defined by:

$$(a_1, b_1, c_1)(a_2, b_2, c_2) = ((a_1, b_1) \circ \tau^{c_2}(a_2, b_2), c_1 + c_2) \quad (5.15)$$

This is called the wreath product of $S_n$ and $\mathbb{Z}_2$ and denoted by $G \cong S_n \wr \mathbb{Z}_2 = (S_n \times S_n) \rtimes \mathbb{Z}_2$. Finally, we now have an instance of HSP over the group $S_n \wr \mathbb{Z}_2$ and we only have to distinguish the cases $H$ trivial (the two graphs are not isomorphic) or $H = \{\text{Id}, (a, a^{-1}, 1)\}$ ( $a \in S_n$ is the isomorphism between the two graphs).

We can notice that we have a reduction from the rigid graph isomorphism problem to the simple HSP group $A_{2n}$. If $\varepsilon$ denotes the signature of permutation, we have $\varepsilon(\sigma) = \varepsilon(\varphi)^2 \varepsilon(s_n) = 1(-1)^n$. If $n$ is even, then we can directly work in $A_{2n}$. Otherwise, let $m = n+1$, add to each initial rigid graph another connected component composed of only one single node, and combine them into a big graph of $2m$ nodes. If we define the transposition $\tau = (m, 2m)$, then the automorphism group $H$ of the big graph is a subgroup of $S_{2m}$ and is similar to the previous case: either $H = \langle \tau \rangle$ or $H = \langle \tau, \sigma \rangle$ where $\sigma = (\varphi, m \mapsto m, \varphi^{-1}, 2m \mapsto 2m) s_m$. The problem is now to determine whether the subgroup $H \cap A_{2m}$ of $A_{2m}$ is trivial or $\langle \sigma \rangle$.

Attempts to solve SHSP, even the restricted versions have not reached any success so far. The first question is whether the classical Fourier Sampling method is helpful. Some negative results were proved for the weak Fourier Sampling [GriSchVaz2000] [HalRusTa-2000], extended to strong Fourier Sampling [MooRusSch2005] and finally to an arbitrary POVM over many entangled coset states [HalRoeSen2005]. More precisely, the last result shows that we need to perform measurement on at least $\Omega(\log(|S_n|)) = \Omega(n \log n)$ entangled states at once.

One model for using such an entanglement is Kuperberg's algorithm for DHSP. It can be interpreted as a repetition of combine-and-measure operations on representation state (i.e. given representation states $\rho_i$ and $\rho_j$, apply measurement on their tensor product $\rho_i \otimes \rho_j$ to get a new representation state) until we reach a representation from which one can extract useful information on the hidden subgroup. [MooRusSni2006], proved however that such an algorithm can not succeed for the symmetric group unless it takes $e^{\Omega(\sqrt{n})}$ time. This does not give any meaningful improvement over the best known classical algorithms: $e^{O(\sqrt{n \log n})}$ for general graphs [BabLuk1983] or $e^{O(n^{1/3} \log^2 n)}$ for strongly regular graph [Spi1996].

Despite intensive study, no quantum algorithm for SHSP are known. It is worth noting that the structure of the symmetric group is very complex ($S_n$ contains all the groups of order $n$ as subgroups) and is likely to be much more difficult to solve. It is possible that a straightforward method (i.e. performing coset sampling over the $S_n$ or $S_n \wr \mathbb{Z}_2$) is not appropriate and we should rather distinguish many particular or reduced cases as in the classical algorithms mentioned above, which rely on classification of finite simple groups. Solving HSP for other groups will provide us more techniques and probably help us to find a general method. For example, O'Nan–Scott theorem gives the form of the maximal subgroups of $S_n$ and we could use the "maximal subgroup reduction" method given in the next section.

# 6. The General Hidden Subgroup Problem

## 6.1. General approach to the Hidden Subgroup Problem

In this section, we recall the two fundamental theorems of group theory describing all finite groups. We show how they can be used to solve the general HSP under certain conditions. We study the special cases of dedekindian, dihedral and symmetric groups in that framework.

A group $G$ is said to be *simple*, if it contains no other normal subgroups than $\{1\}$ and $G$. Finite simple groups are the "building blocks" for all finite groups and their classification are given by a theorem of group theory. More precisely, the theorem states that the possibilities for a finite simple group are:

- Cyclic groups $\mathbb{Z}_p$ for $p$ prime.
- Alternating groups $A_n$ for $n \geq 5$.
- Simple groups of Lie type.
- 26 sporadic simple groups.

For an arbitrary finite group $G$, a *composition series* is a sequence of subgroups of $G$

$$\{1\} = H_0 \triangleleft H_1 \triangleleft H_2 \triangleleft \ldots \triangleleft H_n = G \quad (6.1)$$

such that each $H_i$ is a maximal normal subgroup of $H_{i+1}$. The integer $n$ is called the *composition length*, and is clearly $O(\log|G|)$. Each $H_{i+1}/H_i$ is called a *composition factor* and is by definition a simple group. Any finite group has a composition series: informally, we can start with $\{1\} \triangleleft G$ and insert normal subgroup $H_i$ until we get a composition series. Moreover, the Jordan–Hölder theorem says that all the composition series of a group are equivalent i.e. they all have the same length and are made of the same composition factors (up to isomorphism and permutation).

The consequence of these two theorems is that we know how any finite group $G$ is built. Let's see how they can be used to solve the general HSP. Suppose $H$ is a subgroup hidden by $f$ and $N \triangleleft G$ a normal subgroup. Then $\overline{H} = \{hN, h \in H\}$ is a subgroup of $G/N$ and $H \cap N$ a subgroup of $N$. We can consider the HSPs over $G/N$ and $H \cap N$ respectively and get two sets of generator $\overline{H} = \langle h_i N \rangle$ and $H \cap N = \langle x_i \rangle$. Finally we obtain a set of generators $\langle h_i, x_j \rangle = \overline{H} H \cap N = H$ for the hidden subgroup. Of course, this quotient reduction is only relevant if $\{1\} \subsetneq N \subsetneq G$ i.e. is only possible if the group is not simple. Moreover, if $n_1$, $n_2$ are the length of two composition series for $G/H$ and $H \cap N$ we have $n = n_1 + n_2$. Hence if we repeat the previous quotient reduction to these subgroups we will reach simple groups in $O(\log|G|)$ steps. Note that the Jordan–Hölder theorem essentially says that there are no privileged choice for the normal subgroup $N$ at each step. Finally, we have the following theorem:

**Theorem 6.1 (solution to the general HSP)**

We have a solution to the general HSP if

- We have an efficient algorithm for HSP over simple groups.

- For any hidden subgroup problem $G$, $H$, $f$ over a non-simple group and a normal subgroup $\{1\} \not\subseteq N \subsetneq G$, we have an efficient way to build polynomial-time oracles $\overline{f}$ and $f_{|H \cap N}$ over $G/N$ and $N$ hiding the subgroups $\overline{H}$ and $H \cap N$.

Thus, one research direction is to try to solve HSP over simple groups. We can also restrict in a first time to *solvable* groups (i.e. such that the composition factors are abelian, or more precisely cyclic groups of prime order) and use the HSP algorithm that we already have for the composition factors.

The second research direction is to find reduction methods. We have already seen many times the following subgroup and quotient reduction:

**Definition 6.2 (subgroup reduction, quotient reduction)**

Let $G$, $H$, $f$ be a HSP problem. We have the following reductions:

- *Subgroup reduction*: Suppose we can find in polynomial time a subgroup $G_1 \subseteq G$ that contains $H$ and an isomorphism $G_2 \stackrel{g}{\cong} G_1$ such that $g$ has polynomial complexity. The oracle $f \circ g$ over $G_2$ hides the group $g^{-1}(H)$. A solution to HSP over $G_2$ gives a set of generators $u_1, \ldots, u_n$ for $g^{-1}(H)$ and so a set of generators $g(u_1), \ldots, g(u_n)$ for $H$.

  Moreover, if $G_1 \subsetneq G$, then $|G_1| \leq \frac{1}{2}|G|$.

- *Quotient reduction*: Suppose we can find in polynomial time a normal subgroup $H_1 \subseteq H \subseteq G$, a set of generators $u_1, \ldots, u_n$ for $H_1$ and an isomorphism $G_2 \stackrel{g}{\cong} G/H_1$ such that $g$ has polynomial complexity. The oracles $f \circ g$ hides the group $g^{-1}(H/H_1)$. A solution to HSP over $G_2$ gives a set of generators $v_1, \ldots, v_n$ for $g^{-1}(H/H_1)$ and so a set of generators $g(v_1), \ldots, g(v_n)$ for $H/H_1$. Then the union of $u_i$, $g(v_j)$ generates $H$.

  Moreover, if $H_1 \neq \{1\}$ then $|G/H_1| \leq \frac{1}{2}|G|$.

Note that the theorem 6.1 uses a combination of generalized versions of these reduction methods with $N = G_1 = H_1$ and without the constraints $H \subseteq G_1 = N$ and $N = H_1 \subseteq H$. For the subgroup reduction part, the oracle does not need to be changed and will hide the expected group $H \cap N$ instead. However, it is less clear how to build the oracle for the quotient reduction part because two elements $g_1, g_2 \in G$ that belong to the same class in $G/N$ may not belong to the same preimage of $f$. Note that if $H \neq G$, then it is always included in a maximal subgroup $G_1$ and thus subgroup reduction is always possible. This is particularly interesting in the case of the simple group, where quotient reduction is not possible. However, the complete description of maximal groups is only known for some groups.

**Theorem 6.3 (iterative maximal subgroup reduction)**

We have a solution to HSP over a group $G$ if we have an algorithm to find a maximal subgroup $G_1$ containing $H$, apply subgroup reduction to $G_1$ and repeat the operation inductively. We stop when we no longer can find a maximal subgroup i.e. $H = G$. $\square$

Let's consider the groups we have previously studied. First, in the Dedekindian HSP we work on a group whose subgroups are all normal. An algorithm is given in appendix F for the particular case of abelian group, but should also work for hamiltonian ones. The idea is to use Weak Fourier Sampling to get a sequence of subgroups of $G$ containing $H$ that become smaller and smaller. Because all groups are normal, we have a subsequence of a composition series:

$$\{1\} \triangleleft H \triangleleft G_m \triangleleft \ldots \triangleleft G_0 = G \quad (6.2)$$

In that case, $H/G_k$ is the trivial group and $H \cap G_k = G_k$, so the algorithm of theorem 6.1 is really only using subgroup reduction. The abelian algorithm of appendix F also shows the importance of "nice" description of groups, such that the one provided by black box groups. In our definition 3.3 of the hidden subgroup problem, we say that we want to determine the hidden subgroup $H$ and definition 3.5 asks for a set of generating elements. However in the abelian case, to be able to iterate the reductions we use at each step a direct decomposition $G_k = \oplus_{i=1}^{d_{k+1}} \langle a_i^{k+1} \rangle$ so that we can have the polynomial-time isomorphism mentioned in definition 6.2 and thus efficient oracles on the new groups.

Another interesting case is the reduction we have made so far for the dihedral group. The possible composition series are given in figure 6.1, where the dots correspond to composition series of the isomorphic groups indicated.

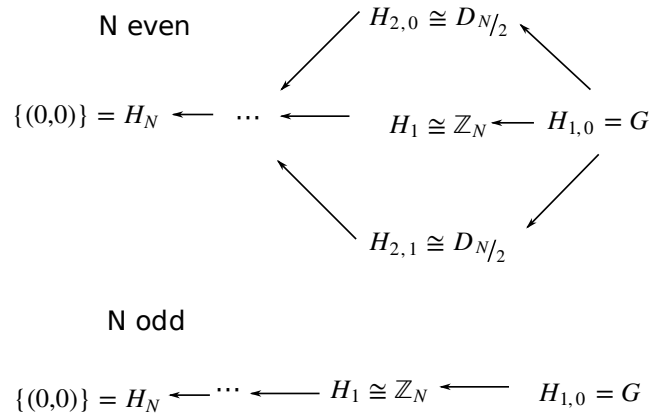

Figure 6.1

As usual, we consider the case $H = H_{r,d}$ which contains the normal subgroup $H_r$. Because $H_r = H_1 \cap H$, we can apply a subgroup reduction to determine $r$: we move in $H_1$ which is isomorphic to $\mathbb{Z}_N$ and apply a cyclic HSP algorithm. Since we now have a normal subgroup included in $H$, we can use a quotient reduction. We have $G/H_r \cong D_N/\mathbb{Z}_{N/r} \cong D_r$ so we still have a similar composition series diagram and we now assume $r = N$. The hidden subgroup becomes $H_{N,d}$ and we want to determine $d$. It easy to see that none of the candidates for quotient reduction (i.e. normal subgroups of $G$ included in the hidden subgroup) provide any further simplification. As a consequence, we are turning our attention to subgroup reduction. The possibilities are $H_{a,b}$ where $a$ is a divisor of $N$ and $d \equiv b \bmod(a)$. In that case, we have $H_{a,b} \cong D_{N/a}$ and the new hidden subgroup is generated by a slope $(q, 1)$ where $q$ is the quotient of the division of $d$ by $a$. We repeat subgroup reduction until we have determined entirely $d$ and again we need only $O(\log|G|)$ reduction steps. Note that variations of this kind are proposed in Kuperberg's paper [Kup2003]. The key step for solving DHSP is thus to determine the remainder of $d$ modulo a divisor $a$ of $N$. For $N = 2^n$ the procedure will be to determine the bits of $d$, from the least to the most significant ones. Finally, we recover the techniques that have been used so far. Note that the dihedral is solvable, so we already have a solution for the simple groups involved. Hence a solution to DHSP will either be an algorithm to find the parity of $d$ (or any other remainder of $d$) or by new reduction techniques.

There is an alternative way to see the first quotient reduction by $H_r$. Using theorem 9.9 one can see that, except in the degenerated cases $H = H_{2,d}$, the intersection of the kernels of the representations measured by a Weak Fourier Sampling over the dihedral group is exactly $H_r$. Actually, we know that in general the Weak Fourier Sampling gives the largest normal subgroup of $G$ which is included in $H$. Hence a general method is to apply Weak Fourier Sampling to find that subgroup and then use a quotient reduction. In the quotient group, we can still repeat this procedure. After $O(\log|G|)$ reductions of this kind, we will not be able to reduce the underlying group any further: the largest hidden subgroup of $H$ normal in $G$ is the trivial group $\{1\}$. If this group is also $H$, then we can determine the original hidden subgroup. Otherwise, $H$ is not necessarily simple and we need to use subgroup reduction or other methods as in the case of the dihedral group. We have:

**Proposition 6.4 (iterative quotient reduction)**

A repetition of quotient reduction using the largest normal subgroup included in the hidden subgroup leads to a hidden subgroup problem where the reduced hidden subgroup $H$ does not contain any non-trivial normal subgroup. If the reduced hidden subgroup is trivial then we can solve HSP using this method. □

Finally, let's consider the symmetric group $S_n$. We may discard small values of $n$ and assume $n \geq 5$. In that case, $\{\text{Id}\} \triangleleft A_n \triangleleft S_n$ is a composition series and in particular $S_n$ is not solvable. Moreover, consider the case of our reduction of rigid graph isomorphism problem in $S_{2n}$, where $H$ is trivial or $H = \langle \sigma \rangle$ for some involution $\sigma = (\varphi, \varphi^{-1})s_n$ with signature $\varepsilon(\sigma) = (-1)^n$. If we split the HSP problem between $S_{2n}/A_{2n} \cong \mathbb{Z}_2$ and $A_{2n}$, we have two cases:

- $n$ is even, so $H \cap A_{2n} = H \subseteq A_{2n}$ and all the information is in $A_{2n}$. However, this group is simple so we can not reduce it any further using quotient groups.
- $n$ is odd, so $H \cap A_{2n} = \{\text{Id}\}$ and all the information is in $S_{2n}/A_{2n} \cong \mathbb{Z}_2$. The HSP is trivial on this group and all the hardness consists in the construction of an oracle.

Actually, we have even seen in the previous section that we can reduce the rigid graph isomorphism problem to the simple group $A_{2n}$.

**Possible algorithm for the general HSP**

Input: The group $G$ the oracle $f$. Output: The hidden subgroup $H$.

1. Use at most $k$ coset measurements (remark 4.10) to determine whether $H = G$ with probability of success at least $1 - 2^k$ or $H$ is a proper subgroup of $G$ with certainty. In the former case, return $G$.
2. If we have an efficient HSP algorithm over $G$, execute it and return the subgroup $H$ found.
3. If $G$ is simple, determine a maximal subgroup $G'$ containing $H$, and use it to apply subgroup reduction. Call this algorithm recursively and return $H$.

4. Apply Weak Fourier Sampling to determine a maximal normal subgroup $H'$ of $G$ included in $H$. If $H'$ is non-trivial, use it to apply quotient reduction. Call this algorithm recursively and return $H$.
5. Otherwise, find a way to reduce the problem by using a normal subgroup $\{1\} \nsubseteq N \subsetneq G$, subgroup reduction or any other methods... Call this algorithm recursively and return $H$.

Figure 6.2

As a conclusion of this section, we have seen a general approach to the HSP based on solution to the simple HSP and various reduction methods. A possible algorithm is proposed in figure 6.2. For the problems related to the dihedral and symmetric groups, it seems however that all the known reductions have already been used.

## 6.2. Known solutions to the Hidden Subgroup Problem

In this section, we review miscellaneous solutions and techniques obtained for non-abelian groups. The HSP table of appendix A completes this section with a good overview of the state of the art.

The groups for which an efficient algorithm is known are mainly semi-direct products of abelian groups, so we start by recalling the definition. Given two groups $G_1$, $G_2$ and an automorphism $\varphi : G_2 \to G_1$, the *semi-direct product* $G_1 \rtimes_\varphi G_2$ is the group with the operation

$$(a_1, b_1)(a_2, b_2) = (a_1(\varphi(b_2)(a_2)), b_1 b_2) \quad (6.3)$$

For example, we have seen the dihedral group $D_N \cong \mathbb{Z}_N \rtimes_\varphi \mathbb{Z}_2$ where $\varphi(b)(a) = (-1)^b a$ and a subgroup of the symmetric group $S_n \wr \mathbb{Z}_2 = (S_n \times S_n) \rtimes_\varphi \mathbb{Z}_2$ where $\varphi(c)(a, b) = \tau^c(a, b)$ and $\tau$ permutes the two coordinates. The latter is an example of a *wreath product* $G_1 \wr G_2$. Each element $b \in G_2$ defines a permutation of the elements of $G_2$ by $x \mapsto bx$. Hence we can identify it with a permutation $\sigma_b$ of $S_{|G_2|}$. We then define $\varphi(b)(a_1, ..., a_{|G_2|}) = (a_{\sigma_b(1)}, ..., a_{\sigma_b(|G_2|)})$ and $G_1 \wr G_2 = G_1^{|G_2|} \rtimes_\varphi G_2$.

The first non-abelian HSP considered was DHSP. Although still unsolved, many methods were introduced, so let's recall them:

- enumerating all the subgroups of $G$: they can all be described by two parameters $d < r$.
- using the abelian Fourier Sampling on abelian subgroups: working on $\mathbb{Z}_N \times \{0\}$ allows to find $r$.
- using quotient reduction: using the normal subgroup $H_r$ yields the standard reduction for DHSP.
- applying the abelian Fourier Sampling on the whole group, viewed as a cartesian product: this produces coset states and reduces the problem to DCP.
- using group reduction: as we have seen in previous section, determining the remainder of $d$ modulo a divisor $a$ of $N$ is like moving to a subgroup isomorphic to $D_{N/a}$.
- pipeline of coset states: this allows to determine $d$ in subexponential time.
- pretty good measurement: we consider the tensor product of $k$ coset states $\rho_d^{\otimes k}$ and use the optimal measurement to identify it among an equiprobable repartition of $d \in \mathbb{Z}_N$. If it is efficiently implementable for some $k > \log(N)$ then we can find $d$.
- superposition of many oracle values: a new approach to determine $d$ proposed in appendix G.

We have presented in details Fourier Sampling based on Fourier Transform over finite groups. It is a natural and elegant extension of the abelian case and was presented as a promising method in the review paper of Lomont [Lom2004]. The non-abelian Fourier Transform has been used as a key element for the Dedekindian HSP [HalRusTa-2000], $\mathbb{Z}_2^n \wr \mathbb{Z}_2$ [RötBet1998], $\mathbb{Z}_3 \rtimes_\varphi \mathbb{Z}_{2^n}$ [GriSchVaz2000], the affine group $\mathrm{Aff}(1, p)$ as well as some particular cases of $q$-hedral groups [MooEtAl2005]. More recently it has been used to solve the Weyl-Heisenberg group [KroRöt2008] and to find 1-point stabilizer of some finite Lie groups [DenMooRus2008]. However, most efficient non-abelian HSP algorithms currently known do not require it and we know that it would not work for all groups. Again, we recall the methods:

- Weak Fourier Sampling to find normal subgroups or more generally the largest normal subgroup of $G$ contained in $H$. As previously said, it is $H_r$ for the dihedral group (except for $H = H_{2,d}$) and so it is an alternative way to determine $r$.
- Strong Fourier Sampling. It was proved in [MooEtAl2005] that it helps to solve some HSPs on which the weak version fails.
- Weak Fourier Sampling over subgroups. They are used in the extensions proposed in [GriSchVaz2000] and [Gav2004].

The other fundamental methods are given by black box results and we refer to [Lom2004] for a review. Recall that the degree of a group $G$ is defined to be the smallest integer $k > 0$ such that there is a one-to-one homomorphism $\varphi : G \to S_k$. For example, we have $k \leq |G|$ by Cayley's theorem, $k = p$ for a cyclic group $G = \mathbb{Z}_p$ of prime order and more generally $k = \sum_{i=1}^m |A_i|$ for an abelian group with decomposition $A = A_1 \times A_2 \times \ldots \times A_m$ where each $A_i$ is cyclic of prime order. We have seen in previous section that every group has a composition series and we define:

**Definition 6.5 (nu of G)**

For any group $G$, $\nu(G) > 0$ is the smallest positive integer bounding the degrees of the nonabelian composition factors. ◇

In particular, if $G$ is solvable all the composition factors are abelian (and even cyclic) so we have $\nu(G) = 1$. In general, we will be interested in groups for which $\nu(G) = \mathrm{poly}(\log|G|)$. [IvaEtAl2001] contains various interesting algorithms, including:

- For a blackbox group with unique encoding, we can find generators of a hidden normal subgroup $N$ in time $\nu(G/N)$ + input size.
- For a blackbox group with unique encoding, we can solve HSP in time the input size + $[G, G]$.

Actually, an extension of the second point is given in [MooEtAl2005].

- Let $G_1 \triangleleft G$, $G/G_1 \cong G_2$. Assume that $G_1$ is of size polynomial in $\log|G_2|$ and that we have an efficient HSP algorithm for $G_2$, then we also have an efficient HSP algorithm for $G$.

Unfortunately, this construction can not be iterated more than a constant number of times whereas we would like to repeat this a polynomial number of time for the algorithm of figure 6.2. This extension was used in [DenMooRus2008] to solve some particular cases of $\mathrm{Aff}(1, p^m)$. Note that the particular case of the commutator is $G_1 = G'$, $G_2$ is the abelianization of $G$, to which we can apply the abelian HSP and we have $|G_1| = \mathrm{poly}(\log|G|)$

= poly(log|$G_1$|, log|$G_2$|) = poly(log|$G_2$|).

Inui and Le Gall studied the case $\mathbb{Z}_{p^n} \rtimes_\varphi \mathbb{Z}_q$ for $p$, $q$ prime [InuLeG2004] and gave a complete classification. One of the class contains groups isomorphic to $\mathbb{Z}_{p^n} \rtimes_\varphi \mathbb{Z}_p$ ($p^n \neq 2^2$ and $\varphi(b)(a) = a(p^{n-1}+1)^b$) which they solved using enumeration of subgroups and blackbox techniques. Chi, Kim and Lee [ChiKimLee2006] noted that the algorithm also works for some groups $\mathbb{Z}_{2p^n} \rtimes_\varphi \mathbb{Z}_p$ and extended the algorithm to the class $\mathbb{Z}_N \rtimes_\varphi \mathbb{Z}_p$ under some conditions. The idea is to use a factorization of $N$ to factor out the group and then apply Inui and Le Gall's algorithm. Using a similar approach, Cosme and Portugal [CosPor2007] solved some HSPs over $\mathbb{Z}_N \rtimes_\varphi \mathbb{Z}_{p^2}$. Inui and Le Gall also proposed a different approach for some cases of $\mathbb{Z}_{p^m}^n \rtimes_\varphi \mathbb{Z}_p$, based on blackbox techniques [InuLeG2004].

A general proposal for the semi-direct products $A \rtimes_\varphi \mathbb{Z}_p$ for some abelian group $A$ and $p$ prime is made in [BacChiDam2005bis]. The first step is similar to the reduction of the dihedral HSP: you use the abelian HSP algorithm over $A$ to find $H_1 = H \cap (A \times \{0\})$ and use a "quotient reduction" ($H_1$ may not be normal, but they find a way to workaround that issue). This reduces the problem to a semi-direct product $A_2 \rtimes_\varphi \mathbb{Z}_N$ with $A_2$ a subgroup of $A$. The hidden subgroup is either trivial or the subgroup of size $N$ generated by $(d, 1)$ for some $d \in A_2$. Next, we use the pretty good measurement with a product of $m$ coset states and reduce the problem to finding equations of a system with $m$ unknowns and some parameters uniformly chosen at random. The authors called it the matrix sum problem and solved it for various groups $A$. In particular, they obtained efficient HSP for $G = \mathbb{Z}_p^k \rtimes_\varphi \mathbb{Z}_p$ ($p$ prime, $k$ constant) and $\mathbb{Z}_N \rtimes_\varphi \mathbb{Z}_q$ ($\frac{N}{q} = \text{poly}(\log N)$ and $q$ prime).

More blackbox techniques have also been developed. Using the hidden shift problem, the HSP over some semi-direct products $\mathbb{Z}_{p^k}^n \rtimes_\varphi \mathbb{Z}_2$ was solved in [FriEtAl2002]. They also introduced orbit coset and orbit superposition problems and use it to solve HSP over some solvable groups. More precisely, they defined:

**Definition 6.6 (smoothly solvable)**

An abelian group is *smoothly solvable* if it can be decomposed into the direct product of a subgroup of bounded exponent and a subgroup of polylogarithmic size. A solvable group is *smoothly solvable* if its derived series is of bounded length and has smoothly abelian factor groups. ◇

They obtained:

- HSP can be solved in solvable groups having smoothly solvable commutator subgroups

Ivanyos, Sanselme, Santha solved HSP over extraspecial groups [IvaSanSan2007] and later extended their result to nil-2 groups [IvaSanSan2007bis] i.e. such that $[[G, G], G] = \{1\}$. They first used the concept of hiding procedure defined in [FriEtAl2002] which generalizes the hiding oracle $f$:

- hiding procedure for a subgroup $H$ of $G$: for every $g_1, \ldots, g_n \in G$, given the input $|g_1\rangle|g_2\rangle\ldots|g_N\rangle|0\rangle$ it outputs $|g_1\rangle|g_2\rangle\ldots|g_N\rangle|\Psi_{g_1}^1\rangle|\Psi_{g_2}^2\rangle\ldots|\Psi_{g_N}^N\rangle$, where $\{|\Psi_g^i\rangle, g \in G\}$ form a hiding set of unit states i.e. $|\Psi_g^i\rangle$, $|\Psi_{g'}^i\rangle$ are orthogonal if $g$, $g'$ are in distinct left cosets of $H$ and equal otherwise.

The abelian HSP algorithm still works when the oracle is replaced by such a procedure. Then, they reduced the HSP over nil-2 groups to those having exponent $p$. They showed that in that case we can find a hidden subgroup $H$ if we have a hiding procedure for $H[G, G]$ and succeeded in constructing such a hiding procedure.

Finally, note that all the positive results previously mentioned in this section work for groups which are in some sense almost abelian. This is the case for the extensions of the Dedekindian HSP which work for groups with a large amount of normal subgroups, of the semi-directs product of abelian groups $G = A \rtimes B$ which can be broken down into $A \cong A \times \{0\} \triangleleft G$ and $G/_{A \times \{0\}} = \{0\} \times B \cong B$. More generally, the results obtained from the black box methods break down the groups into abelian pieces through a normal series. However, it does not seem possible to apply this method for groups which contain a non-cyclic simple group in their composition series such that $A_n$ or $S_n$. One attempt in that direction is [DenMooRus2008], with algorithms to find some hidden subgroups over the simple group $\text{PGL}(2, p^m)$ and other related finite groups of Lie type. More precisely, we have a group $G$ acting on a set $S$ and have the promise that $H$ is a one point stabilizer $G_s, s \in S$. Their idea is to restrict to one of this stabilizer $G_{s_0}$, apply a Strong Fourier Sampling on this group to determine $G_{s_0} \cap G_s$ and deduce $s$. However to make this work, they need some additional hypotheses that do not seem to be satisfied by many groups. Moreover, it is still open whether we can find an efficient HSP taking into account the other subgroups. This may be hard, for example some maximal subgroups of $\text{PGL}(2, p^m)$ are dihedral groups for which we do not have solutions yet.

## 6.3. The Hidden Subgroup Problem and Efficient Algorithms

In this section we discuss how HSP is related to some efficient algorithms with concrete applications. We have already mentioned in the first part of this report how particular abelian HSP have been used by Shor to make efficient algorithms for factoring numbers and computing discrete logarithms. In appendix D, we give a reduction of Monotone 1-in-3 3SAT to GapCVP$^\infty$ where one step uses the abelian HSP algorithm to find the kernel of a group homomorphism. Although the quantum part is not actually needed here, this provides another example where a HSP can be used. Hallgren solved particular cases of HSP over the infinite abelian groups $\mathbb{R}$ and $\mathbb{R}^r$ [Hal2002], [Hal2005]. This allowed him to design polynomial-time algorithms for finding solutions $(x, y) \in \mathbb{Z}^2$ to Pell's equation $x^2 - dy^2 = 0$ (where $d$ is a positive non-square integer) to which the factoring problem reduces. He also solved many important number fields problems.

The two non-abelian cases always mentioned are dihedral and symmetric groups. The HSP over these groups (or their variants: generalized dihedral HSP, alternating HSP etc) is expected to solve a lattice problem used in cryptography and the graph isomorphism problem. However, the first application requires the dihedral HSP to be solved by coset sampling while the second looks like out of the scope of all the techniques currently known. Hence, there is another point of view which consists in building cryptographic tools assuming the hardness of the HSP. For example, a classical one-way function is proposed in [MooRusVaz2007] and is at least as hard to invert as solving the Graph Isomorphism Problem. Also, inverting the function reduces to HSP over $\text{GL}(n, q)$ which is believed to be hard, given the negative results on its subgroup $S_n$.

[Dam1988] contains a proposal for a hard problem with application in cryptography: Given $l + 1 = O(\log p)$ successive evaluations of the Legendre symbol $\left(\frac{s}{p}\right), \left(\frac{s+1}{p}\right), ..., \left(\frac{s+l}{p}\right)$ predict the next value $\left(\frac{s+l+1}{p}\right)$. The authors of [DamHallp2002], proposed the related *shifted Legendre symbol problem*: given access to an oracle $f(x) = \left(\frac{x+s}{p}\right)$ determine the hidden shift $s$. Obviously, an algorithm to solve the *shifted Legendre symbol problem* using only oracle calls for values $0 \leq x \leq l$ provides a solution to the problem of [Dam1988]. The

authors of [DamHallp2002] solved the *shifted Legendre symbol problem* as well as other related hidden shift problems. They indicate that they can break certain cryptosystems by a reduction to the shifted Legendre symbol problem. They say that these cryptosystems can also be broken by Shor's algorithm, but the two attacks on the cryptosystems appear to use completely different ideas. Hence this *shifted Legendre symbol problem* shows another relation between cryptosystems and HSP: it can be reduced to the HSP over $G = D_p$ (under certain assumptions) and as we will see in the next section it can also be reduced to a HSP over $G = \mathbb{Z}_p \wr \mathbb{Z}_2$.

Recall that in Grover's algorithm we have $N$ elements indexed from 0 to $N-1$, an oracle $\varphi : \{0, \ldots, N-1\} \to \{0, 1\}$ taking the value 1 for $M \leq \frac{N}{2}$ elements and we want to find one of these elements. Here, we take $M = 1$ and denote $j_0$ the element that satisfies $\varphi(j_0) = 1$. Lomonaco and Kauffman proposed a comparaison between this case and Shor's algorithm [LomKau2006]. They described Shor's algorithm as an infinite HSP over $\mathbb{Z}$ with a subgroup $p\mathbb{Z}$ hidden by an oracle $f$. They defined a "push" method, used to reduce the problem to some HSP over $\mathbb{Z}_q$ (for some integer $q$) and a new oracle $\tilde{f}$. Next, they considered the HSP over $S_N$ with hidden subgroup $H = \text{Stab}_{j_0}$ of permutations stabilizing $j_0$. They used the subgroup $\text{Stab}_0$ and their "push" method to get a new problem where the oracle $\tilde{f} = \varphi$ is just Grover's oracle. However, from the definition of a "push" given of their paper, it is not obvious that their algorithm is applyable since we are loosing the group structure. They say that the definitions of the "push" and Fourier Sampling can be generalized but still the algorithm fails to provide a solution [Lom2010]. Anyway, Grover's algorithm can be interpreted as an *efficient* HSP algorithm for otherwise we would have an exponential speedup instead of the known optimal quadratic speedup [Zal1997]. Here, we are working over $G = S_N$ and an efficient algorithm is expected to be polynomial in $N \log N$. With the promise $H = \text{Stab}_{j_0}$, we only need to find the $j$ satisfying $f(\sigma_j) = f(\text{Id})$ where we use the cycle $\sigma_j = (0, 1, \ldots, j-1, j+1, \ldots, N-1)$ moving all elements but $j$. This can be done in $O(N)$ time classically (brute-force search) or even in $O(\sqrt{N})$ time quantumly (Grover's algorithm) and thus the HSP in that case is efficiently solvable. This was actually noted in [DenMooRus2008], where efficient algorithms are provided for 1-point stabilizer subgroups of $\text{SL}(2, q)$, $\text{PGL}(2, q)$, $\text{PSL}(2, q)$. Interestingly, the big picture of their algorithm is similar to the "push" method above: rather than working over the whole group, they reduce the sampling to a particular 1-point stabilizer for which they can apply the non-abelian Fourier Sampling. Note that the size (or number of 1-point stabilizers) of these three groups is polynomial in $q$ and the efficient HSP algorithm polynomial in $\log q$. Thus, the following open question: can we interpret these algorithms as concrete search problems with an exponentially fast solution?

### 6.4. Variants and Extensions

In this section, we present various problems related to the Hidden Subgroup Problem. Indeed, there have been several proposals to interpret the efficient algorithms of the first part of this report, other than the abelian HSP. New problems have been defined and solved for some particular cases. These problems are often related to HSP and their study has sometimes lead to solutions to instances of HSP. More generally, it is expected that these problems will also provide new quantum algorithms exponentially faster than the classical ones.

One extension is to allow $G$ to be an *infinite* group: it was actually already the case for the original Shor's algorithm which reduces the HSP over the infinite group $G = \mathbb{Z}$ to the HSP over some cyclic group $\mathbb{Z}_q$, for some $q$ large enough. In general, the abelian HSP algorithm works for any finitely generated abelian group such that $G = \mathbb{Z}^n$. In order to solve various problem, Hallgren had to introduce HSP over $\mathbb{R}$ and $\mathbb{R}^r$ [Hal2002], [Hal2005]. The HSP he considered is finding a hidden subgroup $H = \langle \tau \rangle$ of $\mathbb{R}$ generated by an irrational $\tau$ (or said otherwise, determining an irrational period $\tau$ of a function $f$) as well as finding a hidden lattice $H$ of $\mathbb{R}^r$. In both cases, the rough idea is to construct a computable discretized version of the oracle $f$ and show that we can approximate generators for $H$.

Other extensions to HSP have been proposed in three different papers, but they do not seem to have subsequently been studied. One proposal is the *Hidden Subhypergroup Problem* [AmiKalRoo2006] where the group $G$ is generalized to an hypergroup i.e. the product of two elements is a *subset* of $G$ rather than a single element. A second proposal is the *Hidden Symmetries Problem* [SchUnr2003] where we generalize the condition that $f$ is constant on each left coset to a more general property $\forall\, x \in G$, $V(f(x), f(U(x)))$. For example, saying that $f$ hides $H$ is simply taking $V$ to be the equality and $U(x) = xH$. The third proposal is the *Hidden Covering Space Problem* [OsbSev2004] which generalizes the version of HSP given as a coset sampling black-box. This proposal involves topological spaces and is a bit more complicated to define. We will just say that we consider Cayley graph of groups endowed with their natural topology.

A variation that has been extensively studied is the *Hidden Shift Problem*. In this problem, we have two injective functions $f_0$, $f_1$ on $G$ satisfying $\forall\, x \in G$, $f_1(x) = f_0(x + s)$ for some $s \in G$. The purpose is to determine the hidden shift $s$. The problem was solved for many cases, including smoothly solvable groups [FriEtAl2002], bent functions [Röt2008], functions close to a quadratic [Röt2009], Legendre symbol [DamHallp2002] as well as more general multiplicative characters on finite fields and on ideals $\mathbb{Z}/n\mathbb{Z}$ [DamHallp2002]. When $G$ is abelian, the problem is equivalent to HSP over the generalized dihedral group $G \rtimes \mathbb{Z}_N$. In particular, for $G = \mathbb{Z}_N$ the problem is equivalent to DHSP. A decision version is considered in [ChiWoc2005]: either a shift $s$ exists or the images of $f_0$, $f_1$ are disjoint. We are asked to determine in which case we are. A solution to this decision version for the symmetric group $G = S_n$ would provide an efficient algorithm for the rigid graph isomorphism problem. A *Generalized Hidden Shift Problem* is considered in [ChiDam2005] and solve for the cyclic group $G = \mathbb{Z}_N$ under certain conditions. The generalization consists in having $M$ injective functions $f_0, \ldots, f_{M-1}$ satisfying $\forall\, x \in G$, $f_i(x) = f_0(x + is)$. Note that the generalized hidden shift problem reduces to HSP over the wreath product $G \wr \mathbb{Z}_M$ [FenZha2006]. Hence, the two solutions to the hidden shift problem from Rötteler mentioned above are particular case of $G = \mathbb{Z}_2^n$ which is solved from the efficient algorithm he obtained with Beth [RötBet1998]. When the functions $f_0$, $f_1$ are not injective, the set of elements $s \in G$ satisfying the shift equality can be proved to be a coset of some subgroup $H$ [DamHallp2002]. Finding this set is then called the *Hidden Coset Problem*. It turns out that this problem is actually polynomial-time equivalent to HSP [FenZha2006]. If $H$ is normal, we can work in the quotient group $G/H$ and reduce the problem to the hidden shift problem [Ip2002], [DamHallp2002].

To solve some HSP instances, Friedl et al. [FriEtAl2002] used several HSP-related problems. In addition to the Hidden Shift Problem discussed above (which is called hidden translation in their paper), they introduced the *Hidden Stabilizer Problem, Orbit Coset Problem*, and *Orbit Superposition Problem*. For the first one, we have a group $G$ acting on a set of pairwise orthogonal quantum states. Given such a state we are asked to find its stabilizer subgroup. The second one is a generalization: given two states $|\Psi_0\rangle$ and $|\Psi_1\rangle$ we have to answer whether the intersection of their orbits is empty or find an element $g \in G$ sending $|\Psi_0\rangle$ to $|\Psi_1\rangle$ as well as the stabilizer subgroup of $|\Psi_1\rangle$. One interesting feature of Orbit Coset Problem on a group $G$ is the following self-reductibility property: if $N$ is a solvable normal subgroup of $G$, then the problem reduces to the Orbit Coset Problem on $G/N$ and on subgroups of $N$. Notice that such a property would be really appreciated for HSP to design an algorithm similar to the one of figure 6.2. Finally, the Orbit Superposition Problem asks to create the uniform superposition over the orbit of a given state $|\Psi\rangle$. For solvable groups, it reduces to Orbit Coset Problem. Again, in [FenZha2006] it is proved that Orbit Coset is polynomial-time equivalent to the Hidden Subgroup Problem when we allow the oracle $f$ to be a quantum function.

As many other problems, decision and search versions of HSP have also been imagined. [FenZha2006] defines $\text{HSP}_D$ to be the problem of deciding whether $H$ is trivial and $\text{HSP}_S$ to be the problem of findind a nontrivial element of $H$, if there is one. As we have previously seen, this is of particular importance for the dihedral and symmetric HSP. In the paper, they give a reduction of HSP to $\text{HSP}_D$ for the dihedral group (under certain conditions) and a polynomial-time equivalence between $\text{HSP}_S$ and $\text{HSP}_D$ for permutation groups. The latter equivalence suggests that the problem is hard, since we have this property for problems considered to be difficult such that NP-complete or graph isomorphism. Another proposal for a decision version of HSP is the *Hidden Subgroup Test Function* where we are given $G$, $H$, as well as a function $f$ on $G$ and want to decide whether the function hides the subgroup $H$. In [AblVas2009], the function $f$ is encoded as a particular "input sequence". Similarly in [FriEtAl2002bis], the *Large Period Problem* is studied: we are given $G$, a subgroup $K$ and a function $f$. We want to decide whether $K$ is a proper superset of some subgroup $H \subseteq G$ hidden by $f$. The problem is solved for abelian subgroups. The paper also defines the *Common Coset Range Problem*: given $H \subseteq G$ and two functions $f_0$, $f_1$ on $G$ do we have the equality $\forall\, x \in G,\ f_1(xH) = f_0(xH)$? Again, the problem is solved for some particular cases.

Let's consider now the case of the abelian HSP dealing with the vector space $G = F_q^m$ over the finite field $F_q$, which is studied in [ChiSchVaz2007]. Any subgroup $H$ is a linear subspace of $G$ and cosets $g + H$ are parallel affine subspaces. Thus in the abelian HSP over $F_q^m$ we have an oracle constant on each affine subspace, taking distinct values on different ones and want to find a subspace $H$. If $H$ is a hyperplane, there are coefficients $q_i \in F_q$ such that the linear polynomial $P(x) = \sum_{i=0}^{m} q_i x_i$ satisfies $H = \text{Ker}P$. Moreover, $P$ is constant on each coset and takes distinct values on different ones. Thus the oracle can be written $f(x) = \pi(P(x))$ for some fixed unknown permutation $\pi$ of $F_q$. The *Hidden Polynomial Problem* is to determine the polynomial $P$ whose degree is still bounded, but we now allow this degree to be greater than 1. Similarly to the case of the hyperplane $H$, we assume that the constant term is $P(0) = 0$, for otherwise that term can not be determined, since $\pi$ is arbitrary. The problem can also be viewed as finding the hidden hypersurface defined by the equation $P(x) = 0$ and the cosets are replaced by level hypersurfaces (defined by equations $P(x) = k$). One special case is when the polynomial is of the form $P(x) = x_m - Q(x_1, ..., x_{m-1})$ i.e. the hidden hypersurface is defined by the parametric equation $x_m = Q(x_1, ..., x_{m-1})$. In that case, the level hypersurfaces are just obtained by translating along the $x_m$-axis. This problem, called *Hidden Polynomial Function Graph Problem* is considered in [DecWoc2007], [DecDraWoc2007] and solved for quadratic and cubic polynomial $Q$.

The authors of [ChiSchVaz2007] also consider other hidden structure problems on $G = F_q^m$. They define the *Hidden Radius Problem*: determine an unknown radius $r$ given a blackbox that outputs superposition over the points of a sphere of radius $r$ whose center is chosen uniformly at random. Similarly they define the *Hidden Flat of Center*: determine a flat given a blackbox that outputs superposition over unit sphere whose center is chosen uniformly at random in the hidden flat. They give a partial solution to the former and a complete solution to the latter. Other algorithms of this kind are given in [Mon2008] for the group $G = \{0, 1\}^n$. All these problems are included in the more general category of *Hidden Shifted Subset Problems*. We have a group $G$, a subset of elements $S$ and a subset of shifts $T$ the black box outputs uniform superposition of element over the set $S + t$ for a $t \in T$ chosen uniformly at random. The purpose is to determine some property of $S$ or $T$. To compare their power with the one of a classical computer, we would like an approach similar to HSP where we just have an oracle $f$ whose preimages are the possible superpositions output by the black box. This does not seem possible in the general case, since the superpositions considered may intersect but some workarounds are used for the particular problems mentioned above.

# 7. Conclusion

In this report we have studied the Hidden Subgroup Problem, whose solution for various groups is expected to provide new efficient quantum algorithms. In a first part, we have reviewed the classical efficient Quantum algorithms of Simon and Shor and identified their main features. We have introduced the Hidden Subgroup Problem and showed how the previous algorithms can be defined in that framework.

In a second part, we have been interested in the standard method based on coset sampling. We have presented the well-known solution to the abelian case which includes the algorithms seen in the first part. We have described precisely a natural extension to the non-abelian case, by defining a Quantum Fourier Transform over finite groups as well as Weak/Strong Fourier Sampling. We have shown that Weak Fourier Sampling is able to find normal hidden subgroups and thus provides a solution to HSP over dedekindian groups. We have mentioned some extensions when the group is, in some sense, not too far from a dedekindian group. However, the Weak Fourier Sampling can not distinguish conjugate groups and thus it is not applyable to an arbitrary group.

In a third part, we have turned our attention to the dihedral and symmetric hidden subgroup problems. We have given a complete presentation of the structure of the dihedral group and recalled the reduction of the dihedral HSP to the search of a hidden slope $d$. We have presented Kupenberg's subexponential algorithm to find this slope and mentioned the results about the optimal measurement for dihedral coset sampling. We have studied Regev's algorithm for the uniqueSVP and noticed that it can be modified to work for some generalized dihedral groups and "recursive" coset sampling. We have also shown that the approximation factor obtained for the uniqueSVP is strongly related to the degree of the polynomial bounding the complexity of the coset sampling algorithm used. Finally, we have presented the classical reduction from graph isomorphism to symmetric hidden subgroup problems but have also recalled the negative results on the symmetric group: we need at least a linear number of entangled coset states at once but using an algorithm similar to Kupenberg's one does not help.

In the fourth and last part, we have considered the general HSP. We have suggested a general approach to the hidden subgroup problem based on the mathematical description of finite groups. We have shown that we could reduce the general case to a solution over simple groups and construction of some efficient oracles. These two points are precisely where we are stuck on for dihedral and symmetric groups. However, several efficient algorithms have been discovered for other non-abelian groups. We have given an overview of these solutions which are essentially with respect to "almost abelian" groups in some sense or the other. We have described some efficient algorithms with concrete application, including applications in cryptography, which are related to the hidden subgroup problem. Finally, we have mentioned variants and extensions to the HSP which have been solved for some particular cases and are expected to yield new efficient quantum algorithms.

As a conclusion, we have provided a good overview of the hidden subgroup problem going from its origin to the state of the art as of 2010. We have also brought some contributions to the topic. While the general problem still remains difficult, we hope that our work will help people to have a better understanding of the subject and guide them toward some research directions.

# Appendix

## Appendix A: Table of Hidden Subgroup Problems

In the description, we give the underlying group $G$ as well as the expected complexity $\text{poly}(\log|G|)$ of an efficient algorithm. For convenience, the letters $p$, $p_i$, $q$ denote *prime numbers*. Otherwise, letters are chosen to give an indication of their magnitudes with respect to the size of the group: capital letters $N$, $M$, $N_i$ are $O(|G|)$, minuscule letters $n$, $m$ are $O(\log|G|)$ and $k$ is a constant $O(1)$. When some homomorphisms $\varphi$ used for a semi-direct product are explicitly given, it means that they do not cover all the possibilities unless otherwise specified.

Note: some notions and techniques are defined elsewhere in this report. For blackbox groups, this includes unique encoding and input size (definition 3.7), $v(G)$ (definition 6.5) and smoothly solvable (definition 6.6).

| Name | Description | Examples and applications | Methods and remarks | References |
|---|---|---|---|---|
| Cyclic HSP | $G \cong \mathbb{Z}_N$<br><br>Polynomial in $\log N$ | • Shor's factoring algorithm<br>• Order of elements in a group<br>• Decomposition of abelian group | Enumeration of Subgroups, Fourier Sampling over cyclic group | For review, see for instance this report, [Dam2001], [Lom2004] or [NieChu2007] |
| Abelian HSP | Abelian group i.e.<br><br>$G \cong \mathbb{Z}_{N_1} \times \mathbb{Z}_{N_2} \times \mathbb{Z}_{N_3} \times \ldots \times \mathbb{Z}_{N_n}$<br><br>Polynomial in $\log N_1$, ..., $\log N_n$ | • Cyclic HSP<br>• Simon's problem<br>• Discrete logarithm | Fourier Sampling over abelian group<br><br>We assume we know the decomposition of the group. We can determine it if the group is given as a black-box with unique encoding | See also [Sim1994] and [Sho1995], [Kit1995]<br><br>To determine the decomposition see [CheMos2001] |
| Hamiltonian HSP | Hamiltonian groups i.e. $G = \mathbb{Z}_2^n \times A \times \mathbb{Q}_8$ where $A$ is an abelian group with all elements of odd order<br><br>Polynomial in $n$, $\log|A|$ | | Weak Fourier Sampling | Solved in [HalRusTa-2000] |
| Dedekindian HSP and extensions. | $G$ has only normal subgroups i.e. abelian or hamiltonian<br><br>Extensions:<br>• Finding a normal subgroup $H$<br>• A group $G$ such that $\frac{|G|}{|M_G|} \in \text{poly}(\log|G|)$ | • Abelian HSP<br>• Hamiltonian HSP | Weak Fourier Sampling<br><br>Some extensions use Weak Fourier Sampling over subgroups of $G$<br><br>Extensions are given in an "abstract" way. We assume an efficient implementation of Weak Fourier Sampling as well | Solved in [HalRusTa-2000]<br><br>Extensions in [GriSchVaz2000] and [Gav2004]<br><br>See also blackbox groups in [IvaMagSan2001] |

| Name | Description | Examples and applications | Methods and remarks | References |
|---|---|---|---|---|
| | • Finding a subgroup $H$ such that $\frac{|G|}{|HM_G|} \in \text{poly}(\log|G|)$ | | as a way to solve the systems $\rho_i(h) = I_{d_{\rho_i}}$ for the representations measured $\rho_i$<br><br>In the context of a blackbox group, a normal subgroup $H$ can always be found wih complexity the input size + $v(G/H)$ | |
| Dihedral HSP | $G = D_N \cong \mathbb{Z}_N \rtimes_\varphi \mathbb{Z}_2$ where $\varphi(b)(a) = (-1)^b a$ for all $(a,b) \in G$<br><br>Polynomial in $\log N$ | $\Theta\left(n^{\frac{1}{2}+2D}\right)$-uniqueSVP if solved by a "class-preserving recursive algorithm" DCP algorithm of query complexity $O(n^D)$ | Enumeration of Subgroups, Quotient Group Reduction, Subgroup Reduction, Cyclic Fourier Sampling on $\mathbb{Z}_N \times \{0\}$ or Weak Fourier Sampling to find $r$, Abelian Fourier Sampling on the direct product to reduce the problem to DCP, Pipeline of coset states to get subexponential algorithms, Pretty good measurement, Superposition of many oracle values<br><br>To get the uniqueSVP algorithm, it is enough to consider $N = 2^n$ | See [EttHøy1998], [Kup2003] and [Reg2004]<br><br>Relation to uniqueSVP in [Reg2003] and this report<br><br>See this report for the relation with the degree of query complexity as well as what is meant by a "class-preserving recursive algorithm"<br><br>See appendix G for a solution to the case $N = 2^n$ assuming that we have an efficient process to create for a given $b \in \mathbb{Z}_2$ the superpositions of oracles values $|f(2i,b)\rangle$, $i \in \mathbb{Z}_{2^{n-1}}$ |
| Generalized dihedral HSP over $\mathbb{Z}_N^n$ | $G = \text{Dih}(\mathbb{Z}_N^n) = \mathbb{Z}_N^n \rtimes_\varphi \mathbb{Z}_2$ where $\varphi(b)(a) = (-1)^b a$ for all $(a,b) \in G$<br><br>Polynomial in $n, \log N$ | | To get the uniqueSVP algorithm, it is enough to consider $N = 2^{4n}$ | |
| Semidirect HSP $\mathbb{Z}_N \rtimes_\varphi \mathbb{Z}_M$ | $\mathbb{Z}_N \rtimes_\varphi \mathbb{Z}_q$ with $\frac{N}{q} = \text{poly}(\log N)$<br><br>Polynomial in $\log N, \log q$ | • Dihedral HSP<br>• Quasi-dihedral HSP<br>• $q$-hedral groups $\mathbb{Z}_N \rtimes_\varphi \mathbb{Z}_q$<br>• $\text{Aff}(1,p) \cong \mathbb{Z}_p \rtimes_\varphi \mathbb{Z}_{p-1}$ | quotient group reduction, pretty good measurement, Strong Fourier Sampling | Solved in [BacChiDam2005bis]<br><br>See also [MooEtAl2005] for results on $q$-hedral group and $\text{Aff}(1,\mathbb{Z}_p)$<br><br>See [GriSchVaz2000] for $\mathbb{Z}_3 \rtimes_\varphi \mathbb{Z}_{2^n}$ |

| Name | Description | Examples and applications | Methods and remarks | References |
|---|---|---|---|---|
| | $G = \mathbb{Z}_{p^n} \rtimes_\varphi \mathbb{Z}_q$<br><br>Polynomial in $n$, $\log p$, $\log q$<br><br>All these semi-direct products satisfy $\varphi(b)(a) = a(\varphi(1)(1))^b$ for all $(a, b) \in G$<br><br>Moreover, they belong to one of these disjoint classes:<br><br>1. $n \geq 1$ and $q \mid p - 1$<br>2. Dihedral group $D_{2^n} \cong \mathbb{Z}_{2^n} \rtimes_\varphi \mathbb{Z}_2$ and $\varphi(1)(1) = 2^n - 1$<br>3. Quasi-dihedral group $QD_{2^n} \cong \mathbb{Z}_{2^n} \rtimes_\varphi \mathbb{Z}_2$ and $\varphi(1)(1) = 2^{n-1} - 1$<br>4. $P_{p,n} \cong \mathbb{Z}_{p^n} \rtimes_\varphi \mathbb{Z}_p$ where $\varphi(1)(1) = tp^{n-1} + 1$, $p^n \neq 2^2$ and $0 < t < p$<br>5. Direct product $\mathbb{Z}_{p^n} \times \mathbb{Z}_q$ i.e. $\varphi(1)(1) = 1$ | • $\mathbb{Z}_3 \rtimes_\varphi \mathbb{Z}_{2^n}$ where $\varphi(b)(a) = 2^b a$ for $(a, b) \in G$ (the only possible non-trivial semi-direct product) | For class 4: enumeration of subgroups, blackbox methods, pretty good measurement | See [InuLeG2004] for the classification<br><br>Class 1 solved when $\frac{p^n}{q} = \text{poly}(n, \log p)$ in [BacChiDam2005bis]<br><br>Class 4 solved in [InuLeG2004]<br><br>Class 5 solved by abelian HSP |
| | $G = \mathbb{Z}_{2p^n} \rtimes_\varphi \mathbb{Z}_p$ where $\varphi(b)(a) = a(tp^{n-1} + 1)^b$, $p$ odd and $0 < t < p$<br><br>(these are the only possible non-trivial semi-direct product)<br><br>Polynomial in $n$, $\log p$ | | enumeration of subgroups, blackbox methods. | Solved in [InuLeG2004] and [ChiKimLee2006] |
| | $G = \mathbb{Z}_N \rtimes_\varphi \mathbb{Z}_p$ with the factorization $N = \prod_{i=1}^n p_i^{r_i}$ and an odd $p$ which does not divide $p_i - 1$.<br><br>Polynomial in $\log N$, $\log p$ | | enumeration of subgroups, blackbox methods. | Solved in [ChiKimLee2006] |
| | $G = \mathbb{Z}_N \rtimes_\varphi \mathbb{Z}_{p^2}$ with the factorization $N = \prod_{i=1}^n p_i^{r_i}$ such that $r_i > 4$ and an odd $p$ which does not divide $p_i - 1$. | | enumeration of subgroups, blackbox methods. | Solved in [CosPor2007] |

| Name | Description | Examples and applications | Methods and remarks | References |
|---|---|---|---|---|
| | Polynomial in $\log N$, $\log p$ | | | |
| Semidirect HSP $\mathbb{Z}_N^n \rtimes_\varphi \mathbb{Z}_M$ | $G = \mathbb{Z}_p^k \rtimes_\varphi \mathbb{Z}_p$<br><br>Polynomial in $\log p$ | • Semidirect HSP $\mathbb{Z}_N \rtimes_\varphi \mathbb{Z}_M$<br>• Generalized dihedral HSP over $\mathbb{Z}_N^n$<br>• Heisenberg HSP i.e. over the only non-trivial semi-direct product $\mathbb{Z}_p^2 \rtimes \mathbb{Z}_p$ | pretty good measurement | Solved in [BacChiDam2005bis]<br><br>See also [RadRöeSen2005] for Heisenberg group |
| | $G = \mathbb{Z}_{p^k}^n \rtimes_\varphi \mathbb{Z}_2$<br><br>Polynomial in $n$, $\log p$ | | blackbox methods | Solved in [FriEtAl2002] |
| | $G = \mathbb{Z}_{p^m}^n \rtimes_\varphi \mathbb{Z}_p$ where $(\varphi(b)(a))_i$ $= a_i(p^{n-1}+1)^b$<br><br>Polynomial in $n$, $m$, $\log p$ | | blackbox methods | Solved in [InuLeG2004] |
| | $G = \mathbb{Z}_2^n \wr \mathbb{Z}_2 = (\mathbb{Z}_2^n \times \mathbb{Z}_2^n) \rtimes_\varphi \mathbb{Z}_2$ where $\varphi(c)(a,b) = \tau^c(a,b)$ and $\tau$ permutes the two coordinates.<br><br>Polynomial in $n$ | | Fourier Sampling over abelian subgroup, Weak Fourier Sampling | Solved in [RötBet1998] |
| Semidirect HSP $A \rtimes_\varphi \mathbb{Z}_N$ | $G = A \rtimes_\varphi \mathbb{Z}_N$ where $A \cong \mathbb{Z}_{N_1} \times \mathbb{Z}_{N_2} \times \mathbb{Z}_{N_3} \times ... \times \mathbb{Z}_{N_n}$ is abelian<br><br>Polynomial in $\log N_1$, ..., $\log N_n$, $\log N$ | • Generalized dihedral HSP i.e. $N = 2$ and $\varphi(b)(a) = (-1)^b a$ for all $(a,b) \in G$. Also called Hidden Shift Problem over an abelian group $A$<br>• Semidirect HSP $\mathbb{Z}_N^n \rtimes_\varphi \mathbb{Z}_M$ | pretty good measurement | Unsolved<br><br>See [BacChiDam2005bis] when $N$ is a prime |
| $p$-group HSP | $G$ is a $p$-group i.e. all elements are of order a power of $p$ | • $D_{2^n}$ and $\text{Dih}(\mathbb{Z}_{2^{4n}}^n)$, which are the cases relevant for the polynomial uniqueSVP algorithm<br>• Heisenberg HSP<br>• Weyl-Heisenberg HSP | blackbox methods, hiding procedure | Unsolved<br><br>Solved in [IvaSanSan2007bis] for $p$-group of nilpotency class 2<br><br>See also [IvaSanSan2007] for extraspecial HSP and [KroRöt2008] for Weyl-Heisenberg HSP |

| Name | Description | Examples and applications | Methods and remarks | References |
|---|---|---|---|---|
| | | $G = \mathbb{Z}_p^{n+1} \rtimes_\varphi \mathbb{Z}_p^n$ where $\varphi(x)(y) = \begin{pmatrix} I_n & 0 \\ x^t & 1 \end{pmatrix} y$<br>• Extraspecial HSP i.e. the commutator, center and Frattini subgroup of $G$ are equal and of order $p$ | | |
| nilpotent HSP | $G$ is nilpotent i.e. there is $m$ such that $A_{m+1} = \{1\}$ for the sequence $A_1 = G$, $A_{i+1} = [A_i, G]$ | • $p$-group HSP | blackbox methods, hiding procedure<br><br>Can be reduced to the $p$-group HSP if we can determine the decomposition of $G$ as a product of its Sylow subgroups. | Unsolved<br><br>Solved in [IvaSanSan2007bis] for nilpotent group of class $m = 2$ |
| solvable HSP | $G$ is solvable i.e. there is $m$ such that $G^{(m)} = \{1\}$ for the sequence $G^{(0)} = G$, $G^{(i+1)} = \left(G^{(i)}\right)'$ | • nilpotent HSP | In black box with unique encoding, It is possible to create coset states. | Unsolved<br><br>See some algorithms in [Wat2000] and [IvaEtAl2001]<br><br>Solved in [FriEtAl2002] for smoothly solvable groups having a smoothly solvable commutator subgroup. |
| Symmetric HSP. | $G = S_n \wr \mathbb{Z}_2 = (S_n \times S_n) \rtimes_\varphi \mathbb{Z}_2$ where $\varphi(c)(a,b) = \tau^c(a,b)$ and $\tau$ permutes the two coordinates.<br>Polynomial in $n \log n$ | • rigid graph isomorphism problem | For the rigid graph isomorphism, we can restrict to $n$ even and distinguishing the cases $H$ trivial (the graphs are not isomorphic) or $H = \{$ Id, $(a, a^{-1}, 1)\}$ where $a \in S_{n/2}$ is an isomorphism between the two graphs.<br><br>A solution by coset sampling using an arbitrary POVM would require $\Omega(n \log n)$ entangled state at once. In particular, Strong Fourier Sampling fails. | Unsolved<br><br>See [MooRusSch2005] and [HalRoeSen2005] |

| Name | Description | Examples and applications | Methods and remarks | References |
|---|---|---|---|---|
| | $G = S_n$<br>Polynomial in $n \log n$ | • graph isomorphism problem<br>• Alternating HSP | $S_n \wr \mathbb{Z}_2$ and $A_n$ are subgroups of $S_{2n}$ so a solution of HSP over the whole symmetric group implies a solution over $S_n \wr \mathbb{Z}_2$ and $A_n$. | |
| Alternating HSP | $G = A_n$<br>Polynomial in $n \log n$ | • rigid graph isomorphism problem | For the rigid graph isomorphism, we can restrict to $n$ even and distinguishing the cases $H$ trivial (the graphs are not isomorphic) or $H = \{\text{Id}, (\varphi, \varphi^{-1})\sigma\}$ where $a \in A_{n/2}$ is an isomorphism between the two graphs and $\sigma$ is the product of transpositions $\prod_{i=1}^{n/2}\left(i, i+\frac{n}{2}\right)$ | Unsolved<br>See this report for a reduction of the graph isomorphism problem for rigid graphs |
| Wreath HSP<br>$F \wr \mathbb{Z}_m$ | $G = F^m \rtimes_\varphi \mathbb{Z}_m$ where $\varphi(b)$ applies a permutation $\sigma_b$ to the $m$ first coordinates and $\sigma_b$ is an element of $S_{|F|}$ identified with $x \mapsto bx$<br>Polynomial in $m$, $\log(|F|)$ | • Hidden Shift Problem over $F$ for $m = 2$<br>• Generalized Hidden Shift Problem over $F$<br>• rigid graph isomorphism problem for $m = 2$ and $F = S_n$<br>• For $m = 2$ and $G = \mathbb{Z}_p$ solve the shifted Legendre symbol Problem. | Some solutions exist for the related hidden shift problems. | Unsolved.<br>Solved in [RötBet1998] for $F = \mathbb{Z}_2^n$ and $m = 2$. |
| Affine HSP | $G = \text{Aff}(n, p^m) = F_{p^m}^n \rtimes \text{GL}(n, p^m)$<br>Polynomial in $n, m, \log p$ | • $\text{Aff}(1, p) = \mathbb{Z}_p \rtimes_\varphi \mathbb{Z}_{p-1}$ i.e. $n = m = 1$<br>• $\text{Aff}(1, p^m)$ i.e. $n = 1$ | Strong Fourier Sampling | Unsolved<br>Solved for $n = m = 1$ in [MooEtAl2005]<br>See also some particular cases for $n = 1$ in [DenMooRus2008] |
| Lie HSP | $G$ is a finite Lie group. | • General linear groups $\text{GL}(n, p^m)$.<br>• Special linear | Strong Fourier Sampling | Unsolved<br>Solved for 1-point stabilizer subgroups of $\text{SL}(2, p^m)$, |

| Name | Description | Examples and applications | Methods and remarks | References |
|---|---|---|---|---|
| | | groups $\text{SL}(n, p^m)$<br>• Projective general linear group $\text{PGL}(n, p^m)$<br>• Projective special linear group $\text{PSL}(n, p^m)$ | | $\text{PGL}(2, p^m)$, $\text{PSL}(2, p^m)$ in [DenMooRus2008] |
| Simple HSP | Simple group $G$ i.e. without any normal subgroup other than $\{1\}$, $G$:<br>1. Cyclic groups $\mathbb{Z}_p$ for $p$ prime.<br>2. Alternating groups $A_n$ for $n \geq 5$.<br>3. Simple groups of Lie type.<br>4. 26 sporadic simple groups. | • Cyclic HSP: $\mathbb{Z}_n$ for a prime $n$<br>• Alternating HSP: $A_n$<br>• $\text{PGL}(2, p^m)$<br>• Could be a key step for solving the general HSP | • A possible approach could be to use (maximal) subgroup reduction. | Unsolved.<br>Solved for cyclic groups of prime order<br>Solved for 1-point stabilizer subgroups of $\text{PGL}(2, p^m)$ in [DenMooRus2008]<br>See this report for the relation with general HSP and subgroup reduction |
| General HSP | Arbitrary finite group $G$. | • Finding quantum algorithms exponentially faster than their classical counterparts | • We know that a polynomial number of oracle calls is enough to build a state that identifies the hidden subgroup.<br>• In the context of blackbox with unique encoding, solvable with complexity input size + size of the commutator group<br>• A possible approach could be a combination of quotient and subgroup reductions as well as solving the simple HSP. | Unsolved<br>Polynomial query-complexity in [EttHøy1998]<br>See [IvaMagSan2001] for some black box results<br>See this report for a suggestion toward a solution |

## Appendix B: Strong Fourier Sampling and the DHSP

In this section, we study the Fourier Sampling over the Dihedral group. We describe all the possible cases of the Strong Fourier Sampling, show the relation to the dihedral coset problem and in particular how to recover the algorithm of [EttHøy1998]. This similarity between Strong Fourier Sampling and dihedral coset sampling has already been noticed in [Kup2003] and [BacChiDam2005].

We consider a complete set of pairwise non-isomorphic representation, adapted from *5.3 Le groupe diédral $D_n$* of [Ser1971]. If we let $\omega_N = e^{2i\pi/N}$, we define the 2-dimensional irreducible representations $\tau^k(a, 0) = \begin{pmatrix} \omega_N^{-ka} & 0 \\ 0 & \omega_N^{ka} \end{pmatrix}$ and $\tau^k(a, 1) = \begin{pmatrix} 0 & \omega_N^{-ka} \\ \omega_N^{ka} & 0 \end{pmatrix}$ for $0 < k < \frac{N}{2}$. We have two 1-dimensional irreducible representations $\Psi_0(a, b) = 1$ and $\psi_1(a, b) = (-1)^b$. In the case where $N$ is even, we need to add two 1-dimensional representations to get a complete set: $\Psi_2(a, b) = (-1)^a$ and $\Psi_3(a, b) = (-1)^{a+b}$. We start with the case $H = \langle (d, 1) \rangle$.

**Lemma 9.1 (distribution probability of the Weak Fourier Sampling)**

The distribution probability of the Weak Fourier Sampling over $D_N$ for the hidden subgroup $H = \langle (d, 1) \rangle$ is given by

- $P(\psi_0) = \frac{1}{N}$
- $P(\psi_1) = 0$
- $P(\rho) = \frac{2}{N}$ for any 2-dimensional irreducible representation $\rho \in \hat{G}$

and moreover if $N$ is even

- $P(\psi_2) = \begin{cases} 0 & \text{if } d \text{ is odd} \\ \frac{1}{N} & \text{if } d \text{ is even} \end{cases}$

- $P(\psi_3) = \begin{cases} \frac{1}{N} & \text{if } d \text{ is odd} \\ 0 & \text{if } d \text{ is even} \end{cases}$

proof: This is a straightforward application of formula 4.20. For the 1-dimensional representations we have $\Psi_2(H) = \frac{1}{\sqrt{2}}(\Psi_2(0, 0) + \psi_2(d, 1)) = \frac{1}{\sqrt{2}}\left(1 + (-1)^d\right)$ is 0 if $d$ is odd and $\sqrt{2}$ otherwise, so $P(\psi_2) = \frac{1}{2N}\sqrt{2}\Psi_2(H)$ is as claimed. The cases of $\psi_0, \psi_1, \psi_3$ are similar. Any 2-dimensional irreducible representations $\rho \in \hat{G}$ is isomorphic to some $\tau^k$ i.e. satisfies $\rho(g) = M\tau(g)M^{-1}$ for some matrix $M$. We have $\tau^k(H) = \frac{1}{\sqrt{2}}\begin{pmatrix} 1 & \omega_N^{-kd} \\ \omega_N^{kd} & 1 \end{pmatrix}$ so $P(\rho) = \frac{2}{2N}\sqrt{2}\text{tr}(\rho(H)) = \frac{\sqrt{2}}{N}\text{tr}(\tau^k(H)) = \frac{\sqrt{2}}{N}\frac{2}{\sqrt{2}} = \frac{2}{N}$. □

In the probability distribution, measuring a 2-dimensional irreducible representations gives no indication over $H$: they all have equal probability $\frac{2}{N}$ to be observed whatever the value of $d$. Measuring $\psi_0$ yields no information either and so if $N$ is odd, the Weak Fourier Sampling fails. If $N$ is even, we can only get the parity of $d$, if we observe $\psi_2$ or $\psi_3$. Nevertheless, since these representations are measured with a very small probability the Weak Fourier Sampling provides essentially no information:

**Theorem 9.2 (Weak Fourier Sampling fails over $D_N$)**

Suppose we take $M$ samplings of the Weak Fourier Sampling over $D_N$ for a hidden subgroup $H = \langle (d, 1) \rangle$.

- If $N$ is odd, then we get no information on $d$.
- If $N$ is even, the best we can get is the parity of $d$. If $M = \text{poly}(\log(N))$, this information is obtained with a success probability exponentially close to 0 i.e. less than $K \, e^{-\lambda \log(N)}$ for some constants $K, \lambda > 0$.

proof: If $N$ is odd, the distribution probability is entirely independent of $d$ and so it is impossible to get any information from the samplings. If $N$ is even, the only thing we can know is whether $d$ is even (if we measure $\psi_2$) or odd (if we measure $\psi_3$). If we have $M(N)$ samplings this happens with probability $P(N) = 1 - \left(1 - \frac{1}{N}\right)^{M(N)}$. We have:

$$\begin{aligned}
P(N) &= 1 - \left(1 - \frac{1}{N}\right)^{M(N)} \\
&= 1 - \exp\left(\ln\left(1 - \frac{1}{N}\right) M(N)\right) \\
&= 1 - \exp\left(\left(-\frac{1}{N} + o\left(\frac{1}{N}\right)\right) M(N)\right) \\
&= 1 - \exp\left(-\frac{M(N)}{N} + o\left(\frac{M(N)}{N}\right)\right) \quad (9.1) \\
&= 1 - \left(1 - \frac{M(N)}{N} + o\left(\frac{M(N)}{N}\right)\right) \\
&= \frac{M(N)}{N}(1 + o(1)) \\
&= \frac{M(N)}{N}(1 + \varepsilon(N))
\end{aligned}$$

where $\varepsilon(N) \xrightarrow[N \to +\infty]{} 0$. In particular, there is $K' > 0$ such that $(1 + \varepsilon(N)) < K'$. Similarly, if $M = \text{poly}(\log(N))$ then $M(N) 2^{-\frac{\log(N)}{2}} \xrightarrow[N \to +\infty]{} 0$. So there is $K'' > 0$ such that $M(N) 2^{-\frac{\log(N)}{2}} < K''$. Thus $\frac{M(N)}{N} = M(N) 2^{-\frac{\log(N)}{2}} 2^{-\frac{\log(N)}{2}} < K'' 2^{-\frac{1}{2}\log(N)}$. Finally, we get $P(N) < K'K'' 2^{-\frac{1}{2}\log(N)} = K \, e^{-\lambda \log(N)}$ (with $K = K'K''$ and $\lambda = \frac{\ln 2}{2}$). □

What about the Strong Fourier Sampling? To answer this question, we need to describe all possible matrix representations of irreducible 2-dimensional representations. We know at least one complete set of such representations given by $\tau^k$ and we will use proposition 4.7. So we must enumerate all the unitary matrices of dimension 2:

**Proposition 9.3 (unitary matrices of dimension 2)**

The unitary matrices of dimension 2 are the matrices

$$U = e^{i\gamma} \begin{pmatrix} \cos(\theta) & \sin(\theta)e^{i\alpha} \\ \sin(\theta)e^{i\beta} & -\cos(\theta)e^{i(\alpha+\beta)} \end{pmatrix}$$

for $\alpha, \beta, \gamma, \theta \in \mathbb{R}$.

proof: First of all, such a matrix $U$ is clearly hermitian since its columns are orthonormal. Now, suppose $U = \begin{pmatrix} r_{11} e^{i\theta_{11}} & r_{12} e^{i\theta_{12}} \\ r_{21} e^{i\theta_{21}} & r_{22} e^{i\theta_{22}} \end{pmatrix}$ is unitary. Then because the columns/rows are orthonormal we have $r_{11}^2 + r_{12}^2 = r_{11}^2 + r_{21}^2 = r_{22}^2 + r_{12}^2 = r_{22}^2 + r_{21}^2 = 1$ so there exists $\theta \in \mathbb{R}$ such that $r_{11} = r_{22} = \cos(\theta)$ and $r_{12} = r_{21} = \sin(\theta)$. The orthogonality of the column gives $\sin\theta\cos\theta\left(e^{i(\theta_{12}-\theta_{11})} + e^{i(\theta_{22}-\theta_{21})}\right) = 0$. The cases $\sin\theta = 0$ and $\cos\theta = 0$ are included in the general expression of $U$. Otherwise, we have $\theta_{12} - \theta_{11} \equiv \theta_{22} - \theta_{21} + \pi \mod (2\pi)$. If we let $\gamma = \theta_{11}$, $\alpha = \theta_{12} - \gamma$, $\beta = \theta_{21} - \gamma$ we have $e^{i(\theta_{22}-\gamma)} = -e^{i(\alpha+\beta)}$ and we get the expression of the statement. $\square$

For any 2-dimensional irreducible representation $\rho$ isomorphic to some $\tau^k$, the matrix representation of $\rho(0, 0)$ is always the identity. Thus for $H = \langle (d, 1) \rangle$, all we need in order to compute $\rho(H)$ is $\rho(d, 1) = U\tau^k(d, 1) U^\dagger$. It is easy to get an explicit formula, as given in the following lemma.

**Lemma 9.4**

For any representation $\rho$ such that $\rho(d, 1) = U\tau^k(d, 1)U^\dagger$ for some unitary matrix $U = \begin{pmatrix} u_{11} & u_{12} \\ u_{21} & u_{22} \end{pmatrix}$ we have

$$\rho(H) = \tfrac{1}{\sqrt{2}} \begin{pmatrix} 1+x & z \\ z^* & 1-x \end{pmatrix}$$

where $x = 2\mathrm{Re}\left(u_{12}u_{11}^* \omega_N^{kd}\right) \in \mathbb{R}$ and $z = u_{12}u_{21}^*\omega^{kd} + u_{11}u_{22}^*\omega^{-kd}$.

proof: This is just a matrix product:

$$\rho(d,\,1) \;=\; \begin{pmatrix} u_{11} & u_{12} \\ u_{21} & u_{22} \end{pmatrix} \begin{pmatrix} 0 & w_N^{-kd} \\ w_N^{kd} & 0 \end{pmatrix} \begin{pmatrix} u_{11}^* & u_{21}^* \\ u_{12}^* & u_{22}^* \end{pmatrix}$$

$$= \begin{pmatrix} \left(u_{12} w_N^{kd}\right) & \left(u_{11} w_N^{-kd}\right) \\ \left(u_{22} w_N^{kd}\right) & \left(u_{21} w_N^{-kd}\right) \end{pmatrix} \begin{pmatrix} u_{11}^* & u_{21}^* \\ u_{12}^* & u_{22}^* \end{pmatrix}$$

$$= \begin{pmatrix} \left(u_{11}^* u_{12} w_N^{kd} + u_{11} u_{12}^* w_N^{-kd}\right) & \left(u_{12} u_{21}^* w_N^{kd} u_{11} u_{22}^* w_N^{-kd}\right) \\ \left(u_{11}^* u_{22} w_N^{kd} u_{12}^* u_{21} w_N^{-kd}\right) & \left(u_{21}^* u_{22} w_N^{kd} u_{21} u_{22}^* w_N^{-kd}\right) \end{pmatrix} \qquad (9.2)$$

$$= \begin{pmatrix} x & z \\ z^* & -x \end{pmatrix}$$

where we have used the relation of orthogonality $u_{11}^* u_{12} + u_{21}^* u_{22} = 0$ in the last line.

Finally $\rho(H) = \frac{1}{\sqrt{2}}\left(\rho(0,\,0) + \rho(d,\,1)\right) = \frac{1}{\sqrt{2}}\left(I + \begin{pmatrix} x & z \\ z^* & -x \end{pmatrix}\right) = \frac{1}{\sqrt{2}} \begin{pmatrix} 1+x & z \\ z^* & 1-x \end{pmatrix}. \quad \square$

Now, using proposition 9.3, we can simplify the expression of $x$, $z$ above and get the corresponding probability $P(\rho,\,.,\,j)$:

**Proposition 9.5**

For any representation $\rho$ such that $\rho(d,\,1) = U\tau^k(d,\,1) U^\dagger$ for some unitary matrix $U = e^{i\gamma} \begin{pmatrix} \cos(\theta) & \sin(\theta) e^{i\alpha} \\ \sin(\theta) e^{i\beta} & -\cos(\theta) e^{i(\alpha+\beta)} \end{pmatrix}$ we have

$$P(\rho,\,.,\,j) = \frac{1}{N}\left(1 + (-1)^{j+1} \sin(2\theta) \cos\left(\frac{2\pi}{N} kd + \alpha\right)\right) \qquad (9.3)$$

proof: We let $\theta' = \frac{2\pi}{N} kd + \alpha$. With the notation of the previous lemma, we have $x = 2\mathrm{Re}\left(\sin\theta \cos\theta\, e^{i\alpha} \omega_N^{kd}\right) = \sin(2\theta)\cos(\theta')$ and $z = \sin^2\theta\, e^{i(\alpha-\beta)} \omega_N^{kd} - \cos^2\theta\, e^{-i(\alpha+\beta)} \omega_N^{-kd} = e^{-i\beta}\left(\sin^2\theta\, e^{i\alpha} \omega_N^{kd} - \cos^2\theta\, e^{-i\alpha} \omega_N^{-kd}\right)$. We can write:

$$\sin^2\theta\, e^{i\alpha} \omega_N^{kd} - \cos^2\theta\, e^{-i\alpha} \omega_N^{-kd} = \begin{cases} e^{i\theta'} - 2\cos^2\theta \cos(\theta') \\ 2\sin^2\theta \cos(\theta') - e^{-i\theta'} \end{cases} \qquad (9.4)$$

so taking half the sum of the two expressions:

$$\sin^2\theta\, e^{i\alpha}\omega_N^{kd} - \cos^2\theta\, e^{-i\alpha}\omega_N^{-kd} = \frac{1}{2}\left(2\left(\sin^2\theta - \cos^2\theta\right)\cos(\theta') + e^{i\theta'} - e^{-i\theta'}\right) = -\cos(2\theta)\cos(\theta') + i\sin(\theta') \quad (9.5)$$

If $\pm$ is plus if $j = 1$ and minus if $j = 2$ we have:

$$\begin{aligned}\left\|\sqrt{2}\rho(H)_j\right\|^2 &= |z|^2 + (1 \pm x)^2 \\ &= \left(\cos^2(2\theta)\cos^2\theta' + \sin^2\theta'\right) + \left(1 \pm 2\sin(2\theta)\cos\theta' + \sin^2 2\theta\cos^2\theta'\right) \quad (9.6)\\ &= 2(1 \pm \sin 2\theta\cos\theta')\end{aligned}$$

using formula 4.23 we finally conclude:

$$P(\rho, ., j) = \frac{2}{2N}(1 \pm \sin 2\theta\cos\theta') = \frac{1}{N}\left(1 + (-1)^{j+1}\sin(2\theta)\cos\left(\frac{2\pi}{N}kd + \alpha\right)\right). \square$$

Note that parameters $\beta$, $\gamma$ do not appear in the probability expression $P(\rho, ., j)$. Also $\sin(2\theta)$ is really just a parameter between $[-1; +1]$ and we can restrict this value to a smaller interval by playing with $\alpha$. Finally, we can sum up the distribution probabilities of Strong Fourier Sampling for $H = \langle (d, 1) \rangle$, in a very simple manner:

**Theorem 9.6 (distribution probability of the Strong Fourier Sampling)**

The possible distribution probabilities of a Strong Fourier Sampling over $D_N$ for the hidden subgroup $H = \langle (d, 1) \rangle$ are given by:

- $P(\psi_0) = \frac{1}{N}, P(\psi_1) = 0$.
- If $N$ is even $P(\psi_2) = \begin{cases} 0 \text{ if } d \text{ is odd} \\ \frac{1}{N} \text{ if } d \text{ is even} \end{cases}$ and $P(\psi_3) = \begin{cases} \frac{1}{N} \text{ if } d \text{ is odd} \\ 0 \text{ if } d \text{ is even} \end{cases}$.
- For $0 < k < \frac{N}{2}$, $P(\rho^k, ., j) = \frac{1}{N}\left(1 + (-1)^j \lambda_k \cos\left(\frac{2\pi}{N}kd + \mu_k\right)\right)$ with $\lambda_k \in [0,1]$ and $\mu_k \in [0, 2\pi)$

Moreover, such a distribution is obtained by taking $\alpha_k = \mu_k + \pi$, $\theta_k = \frac{\arcsin(\lambda_k)}{2}$ and $U_k = \begin{pmatrix} \cos\theta_k & \sin\theta_k\, e^{i\alpha_k} \\ \sin\theta_k & -\cos\theta_k\, e^{i\alpha_k} \end{pmatrix}$ as change-of-basis matrix in each $V_{\tau^k}: \forall g \in G\, \rho^k(g) = U_k \tau^k(g) U_k^\dagger$.

proof: The case of 1-dimensional representation is as in lemma 9.1. Let $\rho^k$ ($0 < k < \frac{N}{2}$) a complete set of 2-dimensional irreducible representations. We may assume that each $\rho^k$ is isomorphic to $\tau^k$. By proposition 4.7 there are unitary matrices $U_k$ such that $\rho^k(d, 1) = U_k \tau^k(d, 1) U_k^\dagger$. By , the distribution probability is:

$$P(\rho^k, ., j) = \frac{1}{N}\left(1 + (-1)^{j+1}\sin(2\theta_k)\cos\left(\frac{2\pi}{N}kd + \alpha_k\right)\right) \quad (9.7)$$

where $\theta_k$, $\alpha_k$ are the coefficients in $U_k$ as given in proposition 9.3. We can make this expression in the form of the statement by taking $\lambda_k = |\sin(2\theta_k)|$ and (according to the expected sign) $\mu_k = \alpha_k$ or $\mu_k = \alpha_k + \pi$ (modulo $2\pi$).

Conversely, it is easy to check that for the $\rho^k$ given at the end of the statement, we find the probability with the correct $\lambda_k$, $\mu_k$. □

It is worth noting some facts about these distribution probabilities. As in the Weak Fourier Sampling $\psi_0$, $\psi_1$ are useless. $\psi_2$, $\psi_3$ give the parity of $d$, but since they are observed with exponentially small probability we can ignore them when designing an efficient algorithm based on Strong Fourier Sampling. The 2-dimensional irreducible representations have equal probability $\frac{1}{2N}$ to be observed and for each of them we can spread the probability between the two columns. We have only two degrees of freeness to do that: deciding how much the probability of the two columns differ in the extremal cases (amplitude parameter $\lambda_k$) and moving this extremal cases (phase parameter $\alpha_k$). Note that taking $\alpha_k = \frac{2\pi}{N} kd'$ allows to have control over $d$. Since each $\rho^k$ has very small probability to be measured, an efficient algorithm for DHSP needs to consider them as a whole (or at least to consider an exponential number of them). Some attempts are given in Appendix C and where we also give some hints on why the Strong Fourier Sampling approach is likely to fail. Note however, that using a particular basis, we recover the distribution of theorem 9.6 and so the algorithm on which it is based:

**Proposition 9.7**

Take $\lambda = 1$ and $\mu = -\pi$ in the theorem 9.6 (this corresponds to $\rho(g) = H\tau^k(g)H$). Apply a Strong Fourier Sampling and return the following value:

- If $\psi_0$ is measured, return $(0, 0)$
- If $\psi_2$ or $\psi_3$ is measured (so in the case where $N$ is even), return $\left(\frac{N}{2}, 1\right)$
- If $\rho^k$, $j$ is measured, then flip a coin (or measure an EPR state) to decide to return
    ○ $(k, j-1)$
    ○ or $(N-k, j-1)$

Then the distribution probability of the returned value is the same as in formula 5.9.

proof: Let $k'$ denote the parameter in formula 5.9. In the degenerated case $k' = 0$ (respectively $k' = \frac{N}{2}$) $(k', 0)$ (respectively $(k', 1)$) is returned with probability $\frac{1}{N}$ and $(k', 1)$ (respectively $(k', 0)$) with probability $0$. Since $\psi_0$ and $\psi_2 \cup \psi_3$ happens both with probability $\frac{1}{N}$, we get the same result by returning the values of the statement.

Now, note that for any $0 < k' < \frac{N}{2}$ (i.e. $\frac{N}{2} < N - k' < N$), the probability to return $(k', j)$ is the same as the probability to return $(N-k', j)$ in formula 5.9. This is because $\cos^2\left(\frac{\pi}{N}(N-k')d\right) = \left((-1)^d \cos\left(-\frac{\pi}{N}k'\right)\right)^2 = \cos^2\left(\frac{\pi}{N}k'd\right)$ and similarly for sin. By theorem 9.6, the probability distribution of the Strong Fourier Sampling for the 2-dimensional $\rho^k$ is:

$$P\left(\rho^k, ., 1\right) = \frac{1}{N}\left(1 + \cos\left(\frac{2\pi}{N}kd\right)\right) = \frac{2}{N}\cos^2\left(\frac{\pi}{N}kd\right)$$

$$P\left(\rho^k, ., 2\right) = \frac{1}{N}\left(1 - \cos\left(\frac{2\pi}{N}kd\right)\right) = \frac{2}{N}\sin^2\left(\frac{\pi}{N}kd\right)$$

so taking half these values and sharing between $k$ or $N-k$ we recover formula 5.9. □

**Corollary 9.8 (Strong Fourier Sampling solves dihedral HSP)**

It is possible to solve the HSP over $D_N$ using only a $\mathrm{poly}(\log N)$ number of Strong Fourier Samplings (but an exponential post-processing) proof: use the previous proposition to simulate the distribution of formula 5.9 and apply the algorithm of Ettinger and Høyer. □

The previous corollary shows that to extract information, the Strong Fourier Sampling is indeed more powerful than the Weak Fourier Sampling. Another well-known fact (see [GriSchVaz2000]) is that a random choice for the basis of Strong Fourier Sampling is not helpful. Intuitively, this can be seen in the expression of theorem 9.6: with a uniform choice of the phase $\alpha_k \in [0, 2\pi)$ we are totally blurring the parameter $d$. A random approach could still be useful, for instance randomly choosing $\alpha_k = \frac{2\pi}{N} kp$ for an even integer $p$ leaves only the parity of $d$. Again, see Appendix C to get some arguments about why this is likely to fail.

proposition 5.2 indicates that solving DHSP by a Fourier Sampling over $H = \langle (d, 1) \rangle$ yields an algorithm in the most general case. However, it could be possible that the Fourier Sampling can not solve the case $H = \langle (d, 1) \rangle$ while still being able to find other hidden subgroups $H$. We could even hope that DHSP reduces to these subgroups. The next theorem clarify this:

**Theorem 9.9 (distribution probabilities of the Fourier Sampling)**

Let $S_r = \{\frac{N}{r}, 2\frac{N}{r}, \ldots, \lceil \frac{r}{2} - 1 \rceil \frac{N}{r}\}$. The possible distribution probabilities of Fourier Sampling are:

1. For $H = H_r$
   - $P(\psi_0) = P(\psi_1) = \frac{1}{2r}$
   - $P(\psi_2) = P(\psi_3) = \begin{cases} \frac{1}{2r} & \text{if } r \text{ is even} \\ 0 & \text{if } r \text{ is odd} \end{cases}$
   - For $0 < k < \frac{N}{2}, P(\rho^k, ., j) = \begin{cases} \frac{1}{r} & \text{if } k \in S_k \\ 0 & \text{otherwise} \end{cases}$. In particular, $P(\rho^k) = \begin{cases} \frac{2}{r} & \text{if } k \in S_k \\ 0 & \text{otherwise} \end{cases}$

2. For $H = H_{r,d}$
   - $P(\psi_0) = \frac{1}{r}$
   - $P(\Psi_1) = 0$
   - $P(\psi_2) = \begin{cases} \frac{1}{r} & \text{if } r \text{ and } d \text{ are even} \\ 0 & \text{otherwise} \end{cases}$
   - $P(\psi_3) = \begin{cases} \frac{1}{r} & \text{if } r \text{ is even and } d \text{ is odd} \\ 0 & \text{otherwise} \end{cases}$

- For $0 < k < \frac{N}{2}$, $P(\rho^k, ., j) = \begin{cases} \frac{1}{r}\left(1 + (-1)^j \lambda_k \cos\left(\frac{2\pi}{N} kd + \mu_k\right)\right) & \text{if } k \in S_k \\ 0 & \text{otherwise} \end{cases}$ (where $\lambda_k \in [0,1]$ and $\mu_k \in [0, 2\pi]$). In particular, $P(\rho^k) = \begin{cases} \frac{2}{r} & \text{if } k \in S_k \\ 0 & \text{otherwise} \end{cases}$.

The $\lambda_k$, $\mu_k$ can be obtained by the same change-of-basis as in theorem 9.6.

proof: The proof is very similar to the one of theorem 9.6 so we do not give all the detail here. One can check that $\rho^k(H_r) = \frac{1}{\sqrt{|H_r|}} U_k \left[\sum_{l=0}^{\frac{N}{r}-1} \begin{pmatrix} (\omega_N^{-kr})^l & 0 \\ 0 & (\omega_N^{kr})^l \end{pmatrix}\right] U_k^\dagger$. The inner sum is nonzero and of value $\frac{N}{r} I$ iff $k \in S_r$. So $\rho^k(H_r) = \sqrt{\frac{N}{r}} I$ if $k \in S_r$ and zero otherwise. Similarly, $\rho^k(H_{r,d}) = \rho^k(H_{N,d})\rho^k(H_r)$. We find that it is a nonzero matrix iff $k \in S_r$ and in that case $\rho^k(H_{r,d}) = \sqrt{\frac{N}{r}} \rho^k(H_{N,d})$ and we use the computation for $\rho^k(H_{N,d})$ made in the previous section. □

**Corollary 9.10**

The Fourier Sampling over $D_N$ for a hidden subgroup $H \in \{H_r, H_{r,d}\}$ is equivalent to the Fourier Sampling over $D_r$ for the hidden subgroup $H/H_r \in \{\{0\}, \langle (d, 1)\rangle\}$.

proof: The probabilities and matrix representations of observable irreducible representations are the same in both case. □

In the general Strong Fourier Sampling over the dihedral group, the weak part i.e. the measure of the representation labels is essentially the same as doing a cyclic Fourier Sampling over $G' = \mathbb{Z}_N \times \{0\}$. Hence, we can find $r$ and in particular find $H = H_r$. However, the strong version gives the same result as if we were working in $D_r$. Thus we can really restrict ourselves to the case $H = \langle (d, 1)\rangle$ when studying Strong Fourier Sampling.

We conclude this section by showing the converse of proposition 9.7: the states generated by DCP allow to simulate any Strong Fourier Sampling. One consequence is that to solve DHSP, the Dihedral Coset Problem method is more general than Strong Fourier Sampling.

**Proposition 9.11 (DCP simulates Strong Fourier Sampling)**

Any Strong Fourier Sampling over the dihedral group can be simulated using the output of DCP.

proof: The DCP allows to produce states of the form formula 5.7. Applying a unitary gate of the form of proposition 9.3, we obtained $(k, j)$ with probability:

$$\frac{1}{2N}\left(1 + (-1)^j \sin(2\theta)\cos\left(\frac{2\pi}{N} kd + \alpha\right)\right) \quad (9.8)$$

For any $0 < k < \frac{N}{2}$, we choose the same $\theta_k$, $\alpha_k$ for $k$ and $N - k$, in a way that we recover half the probability $P(\rho^k, ., 2 - j)$ of Strong Fourier Transform. If one of $k$, $N - k$ is measured then we return $\rho^k, ., 2 - j$. For $k = 0$, we choose arbitrary $\theta$, $\alpha$ and we return $\psi_0$ if it is measured. Finally if $N$ even and $k = \frac{N}{2}$ we choose $\alpha = 0$ and $\theta = \frac{\pi}{2}$ and return $\psi_{2+j}$. $\square$

## Appendix C: Attempts to solve DHSP by Fourier Sampling

In this section, we describe several attempts to solve DHSP using a Strong Fourier Sampling. We restrict ourselves to the case $H = H_{N,d} = \langle (d, 1) \rangle$ and use only 2-dimensional irreducible representations. As previously mentioned, we need to consider them as a whole since each of them as only a small probability to be observed.

**Approximation of a trigonometric expression**

A first idea is that the sum of the $\lambda_k \cos\left(\frac{2\pi}{N} kd + \mu_k\right)$ can be expressed using trigonometric functions whose argument contains $l = \frac{2\pi}{N} d$. Trying various values of the parameters, we can obtain the values of $l$ modulo $2\pi$ and so the value of $d$. This sum can be approximated by counting the number of observation $(\rho^k, ., 1)$ and so we can hope to get $d$ with a small probability of error. More formally, we define $X$ to be the random variable taking value 1 if a $(\rho^k, ., 1)$ is measured and 0 otherwise. This random variable is bounded in an interval of length 1 so we can apply Hoeffding's inequality: $S = X_1 + X_2 + \ldots + X_m$ then for any $t > 0$, we have

$$P(|S - E(S)| \geq mt) \leq 2\exp\left(-2mt^2\right) \quad (10.1)$$

We start by trying to approximate the trigonometric expression $\tan\left(\frac{l}{2}\right)$:

For $i = 1, 2$, we denote $S^i$ the random variable where we choose for all $k$ $\lambda_k = 1$ and $\mu_k = (1 - i)\frac{\pi}{2} = \mu^{(i)}$. We have $E(S^i) = \frac{m}{N}(M + x_i)$ where $M = \lceil \frac{N}{2} - 1 \rceil$ and $x_i = \sum_{k=1}^{M} \cos\left(\frac{2\pi}{N} d + \mu^{(i)}\right)$ is the sum we want to estimate. We let $\theta = \left(\frac{M}{2} + 1\right)l$. By one oracle call, we can check that $d \neq 0$ and in that case we have:

$$x_i = \frac{\sin\left(\frac{M+1}{2} l\right)}{\sin\left(\frac{l}{2}\right)} \cos\left(\theta - \mu^{(i)}\right) = \left(\frac{\sin\theta}{\tan\frac{l}{2}} - \cos\theta\right)\cos\left(\theta - \mu^{(i)}\right) \quad (10.2)$$

Note that $\sin\left(\frac{M+1}{2} l\right) = 0$ iff $(M+1)d \equiv 0 \mod (N)$. Distinguishing the case where $N$ is even or odd and applying the euclidean algorithm, it is easy to check that $1 \leq \gcd(M+1, N) \leq 5$ and so by at most 5 oracle calls, we can ensure that $\sin\left(\frac{M+1}{2} l\right)$ is nonzero. Similarly, we can ensure that $\cos\left(\theta - \mu^{(2)}\right) = \sin\theta$ and $\cos\left(\theta - \mu^{(1)}\right) = \cos\theta$ are nonzero by discarding solutions to the linear congruence $(M+2)d \equiv (2-i)\frac{N}{2} \mod (N)$. We even get a stronger result: because $M$, $d$ takes integer values, the arguments of these trigonometric functions is far from $0$, $\pi$, $\pm\frac{\pi}{2}$ by at least $\frac{1}{N}$. Hence we can find a constant $a$ such that $\left|\sin\left(\frac{M+1}{2} l\right)\right|$, $|\cos\theta|$, $|\sin\theta|$ are in the interval $[\frac{a}{N}, 1]$.

Now, we have $\tan\theta = \frac{\sin\theta}{\cos\theta} = \frac{x_2}{x_1}$ so $\theta \equiv \pm \operatorname{atan}\frac{x_2}{x_1} \mod (2\pi)$. Reporting in the expression of $x_1$, extracting $\tan\frac{l}{2}$ and after simplification, we get $d = \pm\frac{N}{2\pi} f(x_1, x_2) \mod (N)$ where $f(x_1, x_2) = \operatorname{atan}\frac{x_2}{x_1{}^2 + x_2{}^2 + x_1}$. If we have the exact values of $x_1, x_2$ then we only have to check two values for $d$!

If $m$ is repeated enough time, we measure $S^i = \frac{m}{N}(M + x_i + \varepsilon_i)$ and can assume that the error satisfies $|\varepsilon_i|$ $\leq \varepsilon$ for some $\varepsilon$ sufficiently small. We sketch how to bound the error on $f(x_1, x_2)$, without giving the explicit constant. First using the constant $a$ above, we can find a bound $O\left(\frac{1}{N^2}\right) \leq |x_i| \leq O(N)$, and so again for some $\varepsilon \leq O\left(\frac{1}{N^2}\right)$ we can find a bound $O\left(\frac{1}{N^2}\right) \leq |x_i + \varepsilon| \leq O(N)$. Then bounding the expression of the first partial derivatives of $f$ on the ball $B_\varepsilon(x_1, x_2)$ we finally use the Taylor formula to get $|f(x_1, x_2) - f(x_1 + \varepsilon_1, x_2 + \varepsilon_2)| \leq O(\varepsilon N^7)$. This is a rough bound, so assume we can do better, said the first partial derivative of $f$ are bounded by a constant, and then replace the previous bound by $O(\varepsilon)$. Then the error for $\frac{N}{2\pi} f(x_1, x_2)$ is $O(\varepsilon N)$. We see that we must choose $t$ in Hoeffding's inequality to be $O\left(\frac{\text{poly}(\log N)}{N^2}\right)$ if we want this error $\text{poly}(\log N)$. In that case $d$ is reachable by calling the oracle around two bands of size $\text{poly}(\log N)$ with centers $\pm \frac{1}{2\pi} f(x_1 + \varepsilon_1, x_2 + \varepsilon_2)$. Nonetheless, we are also forced to choose $m = \Omega(N^4)$ i.e an exponential number of sampling calls... Even assuming the size of the band to be $O(N)$ (for example $\frac{N}{2}$ to eliminate half of the possibilities for $d$) gives $m = \Omega(N^2)$.

To summarize we see that, in addition to the technical bound on the first partial derivatives, there are two major difficulties in the previous algorithm: the leading coefficient $\frac{1}{N}$ in the expression of the probability makes the error of the trigonometric function $\Omega(N)$ and the coefficient $\frac{1}{N}$ inside the trigonometric function increases the error again to $\Omega(N^2)$. So we must at least make $m = \Omega(N^2)$ Fourier Sampling if we want this error to be $O(N)$, which is the worse we can tolerate. It seems that trying to extract $d$ with this kind of approach is likely to fail.

Let's mention a way to get $\cos l$ and $\sin l$ with different parameters i.e. two get only one band rather than two. Take the two same values for $\mu_k$ and lambda to be $\lambda_k = \lambda^k$ for $0 < \lambda < 1$. We get

$$x_i = \frac{\lambda^{M+3} \cos((M+1)l) - \lambda^{M+2} \cos((M+2)l) - \lambda^2 + \lambda \cos(l - \mu^{(i)})}{\lambda^2 - 2\lambda \cos l + 1} \quad (10.3)$$

Using several values of $\lambda$ we get system of linear equations with $\cos l, \sin l$ among their unknowns (these unknowns are even related by trigonometric formulas, so we can reduce the number of $\lambda$ needed to make the system invertible). We have not tried to bound the error but anyway the comment above on the two difficulties are still valid.

**Randomized algorithm and parity of d**

Rather than trying to approximate $d$, we try to make the sum of the $\lambda_k \cos\left(\frac{2\pi}{N} kd + \mu_k\right)$ reveal the parity of $d$. For simplicity, we assume $N = 2^n$ for some $n \geq 2$. We take $\lambda_k = 1$ and randomize the algorithm by uniformly choosing at each sampling some value $d'$ and let $\mu_k = \frac{2\pi}{N} k(2d')$. The random variable $X$ now takes the value 1 if $(\rho^k, ., 1)$ is measured for some odd $k$ and 0 otherwise. If $b$ is the parity of $d$, then we now have:

$$E(S) = m \sum_{q=0}^{\frac{N}{2}-1} \frac{1}{N/2} \left( \sum_{p=0}^{\frac{N}{4}-1} \left( \frac{1}{N} \left( 1 + \cos\left(\frac{2\pi}{N}(2p+1)(2q+b)\right) \right) \right) \right) = m \left( \frac{1}{4} + \frac{\delta_{b,0}}{N} \right) \quad (10.4)$$

So to distinguish whether $\frac{\delta_{b,0}}{N}$ is 0 or $\frac{1}{N}$ we need to take $t = \frac{1}{2N}$ and so $m = \Omega(N^2)$. The detail of the calculation shows that the sum over $p$ for the cosinus is nonzero and of value $\frac{N}{4}$ for only two values (namely $2q + d = 0, \frac{N}{2}$). The factor $\frac{1}{N}$ reduces this value to $\frac{1}{4}$. However, if we want to test all the even values, we need to randomize $q$: this introduces a coefficient $\frac{1}{N/2}$ and makes the difference between the cases odd/even exponentially small. So while randomizing allows to test more values, it also decreases the precision by the same order. As previously stated, a stronger randomization of $\mu_k$ is likely to blur the parameter $d$ and so make the information inaccessible.

**Eliminating many values in one call**

Once we have computed $a = f(0, 0)$, then for each $d' \in \mathbb{Z}_N$, we can check $d' = d$ by one call just by checking $f(d', 1) = a$. In this section, we see how many values of $d$ we can eliminate in one call. The idea is to choose $\lambda_k = 1$ and $\mu_k = -\frac{2\pi}{N} l_k$ for some integer $l_k \in \mathbb{Z}_N$. Suppose that $d'$ is such that $\cos\left(\frac{2\pi}{N}(kd' - l_k)\right) = \pm 1$ then the repartition of the probability of $\rho^k$ is $\frac{1}{2N}$ in one column and zero in the other. In that case, measuring $\rho^k$, $j$ eliminates all the $d'$ that satisfy $P(\rho^k, ., j) = 0$. Note that the cosinus takes the value $\pm 1$ iff $kd \equiv l_k \mod (N)$ or $kd \equiv l_k + \frac{N}{2} \mod (N)$. We now suppose $N = 2^n$ for some $n \geq 2$. Each $1 \leq k \leq 2^{n-1} - 1$ can be written in a unique way as $(2x_k + 1)2^{i_k}$ where $0 \leq i_k \leq n - 1$ and $0 \leq x_k \leq 2^{n-i_k-1} - 1$ and conversely any such $(i_k, x_k)$ define a $k$ in the expected domain. In binary decomposition, the number made with first $n - i_l - 1$ bits is $x_k$, the next bit is a 1 and the $i_k$ last bits are 0's. Clearly, we have $\gcd(k, N) = 2^{i_k}$. Because $\frac{N}{2} = 2^{n-1}$ is a multiple of $2^{i_k}$, the previous linear congruence have a solution iff $l_k = y_k 2^{i_k}$ for some $0 \leq y_k \leq 2^{n-i_k} - 1$. Again, in binary decomposition, the number made with first $n - i_k$ bits of $l_k$ is $y_k$ and the $i_k$ last bits are 0's. Let $r_k, s_k$ such that $r_k x_k + s_k 2^{n-i_k} = 1$ then the possible solutions are given by $r_k y_k + \lambda 2^{n-i_k}$ (cos = 1) and $r_k\left(y_k + 2^{n-1-i_k}\right) + \lambda 2^{n-i_k}$ (cos = -1) for $0 \leq \lambda \leq 2^{i_k} - 1$. So measuring a representation $\rho^k$ allows to eliminate with certainty $2^{i_k}$ values. For the 1-dimensional representation, measuring $\psi_2, \psi_3$ eliminates $2^{n-1}$ possibilities and we get no information from $\psi_0, \psi_1$. As a consequence, in mean we can hope to eliminate with certainty in one call of Fourier Sampling:

$$\frac{2^{n-1}}{N} + \frac{2^{n-1}}{N} + \sum_{k=1}^{2^{n-1}-1} \frac{2^{i_k}}{2N} = 1 + \sum_{i_k=0}^{n-1} 2^{n-i_k-1} \frac{2^{i_k}}{2.2^n} = 1 + \frac{\log(N)}{4} \quad (10.5)$$

which is only $\text{poly}(\log(N))$. So even in the ideal case where each call eliminates different values for $d$ (for example we could hope to do so by choosing the $l_k$ in a clever way at each step), repeating this procedure a polynomial number of time will in mean only discard a $\text{poly}(\log(N))$ number of values for $d$!

## Appendix D: Reduction of Monotone 1-in-3 3SAT

In this section, we try to solve a version of the famous boolean satisfiability problem using an abelian HSP. The version considered is the monotone 1-in-3 3-SAT which is known to be NP-complete [Sch1978]. We obtain a reduction of monotone 1-in-3 3-SAT to a lattice problem called GapCVP$^\infty$ using a quantum algorithm. Note however that a (classical) reduction of the subset sum problem to GapCVP$^1$ is already known [MicGol2002]. Moreover, in our algorithm, the HSP solution is only used to find generators for the kernel of a module homomorphism. However, there exists an efficient classical algorithm for this task [BucNei1996]. Nevertheless, it remains an illustration of general techniques based on HSP, where we use the kernel of a group homomorphism as the hidden subgroup.

Monotone one-in-three 3-SAT is a variant of the boolean satisfiability problem where each clause has only three literals and those literals are not negated. We want to find a truth value where only one literal is true in each clause. More formally:

### Definition 11.1 (monotone one-in-three 3-SAT)

Let $x_1, \ldots x_n$ be boolean variables and consider $N > 0$ different clauses defined for all $0 \leq i \leq N - 1$ by $x_{\sigma_i(1)} \vee x_{\sigma_i(2)} \vee x_{\sigma_i(3)}$, where $\sigma_i \in n^{\{1,2,3\}}$ is a combination. Define the predicate:

$$P = \bigwedge_{i=0}^{N-1} \left( x_{\sigma_i(1)} \vee x_{\sigma_i(2)} \vee x_{\sigma_i(3)} \right) \quad (11.1)$$

The *monotone 1-in-3 3-SAT* is the following question: can we find a truth value for the $x_i$'s such that exactly one literal is true in each clause (and hence the whole conjunction $P$ is true). ◇

Note that $N \leq \binom{n}{3} \leq \frac{n^3}{6}$ ($N$ is at most the number of combination of three variables among $n$) and $n \leq 3N$ (each clause contains three variables). As a consequence, a polynomial algorithm for answering the question can be equivalently measured with $N$ or $n$. Without loss of generality, we are now considering the case where the first clause is $x_1 \vee x_2 \vee x_3$ and the variables $x_1$, $x_2$, $x_3$ are not used in other clauses (we will see later why this is needed). We can add such a clause to any given predicate and this changes neither the answer nor the complexity of the initial problem.

Let us define the following function, which is going to be our oracle:

$$f : \begin{cases} \mathbb{Z}_{4N}^n & \to & \mathbb{Z}_{4N} \\ x & \mapsto & \left( \sum_{i=0}^{N-1} \left( \sum_{k=1}^{3} x_{\sigma_i(k)} \right) 4^i \right) \end{cases} \quad (11.2)$$

Note that the expression of $f$ is quite similar to the one of $P$ if the variables are viewed as the coordinates of a vector $x \in \mathbb{Z}_2^n \subseteq \mathbb{Z}_{4N}^n$. This function allows to characterize all the solutions of the monotone 1-in-3 3-SAT via the following lemma:

**Lemma 11.2**

A vector $x \in \mathbb{Z}_2^n$ is a solution to the monotone 1-in-3 3-SAT iff $f(x) = \frac{4^N-1}{3}$.

proof: For such a vector $x \in \mathbb{Z}_2^n$, the $i$-th sum over $k$ is a value between 0 and 3 that indicates the number of literals that are true in the $i$-th clause. So $f(x)$ can be viewed as a representation of a an integer between 0 to $4^N - 1$ in base 4. Hence $x$ is solution to the solution to the monotone 1-in-3 3-SAT iff each digit is 1 i.e. $f(x) = \frac{4^N-1}{3}$. □

The following lemma is almost straightforward from the definition:

**Lemma 11.3**

$f$ is a module homomorphism.

proof: if $x, y \in \mathbb{Z}_{4^N}^n$, then $f(x+y) \equiv \left(\sum_{i=0}^{N-1}\left(\sum_{k=1}^{3}(x+y)_{\sigma_i(k)}\right)4^i\right) \equiv \left(\sum_{i=0}^{N-1}\left(\sum_{k=1}^{3}\left(x_{\sigma_i(k)} + y_{\sigma_i(k)}\right)\right)4^i\right) \equiv f(x) + f(y) \bmod \left(4^N\right)$.

Similarly, We prove that $f(ax) \equiv af(x) \bmod \left(4^N\right)$ for any $a \in \mathbb{Z}$. □

If we define the sub-module $H = \operatorname{Ker} f$ of $\mathbb{Z}_{4^N}^n$ and the point $p = \begin{pmatrix} \frac{4^N-1}{3} \\ 0 \\ \vdots \\ 0 \end{pmatrix}$, we have:

**Theorem 11.4**

The set of solutions of the monotone 1-in-3 3-SAT is exactly $(p + H) \cap \mathbb{Z}_2^n$.

proof: We have $f(g_1) = f(g_2)$ iff $f(g_1 - g_2) = 0$ iff $g_1 - g_2 \in H$. So $f$ distinguishes cosets of $H$. The coset $p + H$ corresponds to all the vectors that take the value $f(p)$. By the restriction on the predicate $P$ we discussed earlier, this value is exactly $\frac{4^N-1}{3}$. By a previous lemma, the solutions are exactly the vectors of $\mathbb{Z}_2^n$ contained in this coset.

**Remark 11.5**

Given two subsets $S_1$ and $S_2$ of $G$ and $g \in G$, the Coset Intersection Problem is to determine whether $\langle S_1 \rangle$ and $g\langle S_2 \rangle$ have a non-empty intersection. However, the problem to decide whether $(p + H) \cap \mathbb{Z}_2^n \neq \emptyset$ does not exactly fit into that framework, because $\mathbb{Z}_2^n$ is not a subgroup of $\mathbb{Z}_{4^N}^n$. △

Using the efficient algorithm for abelian HSP, we get:

**Proposition 11.6**

There exists a quantum algorithm with a polynomial complexity in $n$ that returns a set of generators for $H$ with probability of success exponentially close to 1.

proof: $H = \operatorname{Ker} f$ is a subgroup of $G = \mathbb{Z}_{4^N}^n$. $f$ distinguishes cosets of $H$ and each call to $f$ is an operation on integers of $\lceil \log(4^N) \rceil = O(n)$ bits, so of polynomial complexity. $G$ is an abelian group of order $(4^N)^n$, so we can apply the abelian HSP algorithm and find a set of $t + \log(|G|) = t + 2nN$ vectors that generate $H$ with probability $1 - 2^t$ (we can just take $t = n$). □

**Remark 11.7**

This kind of interpretation as a HSP works for any group homomorphism. Moreover, the kernel is always a normal subgroup, so $G$ needs not be abelian (we already know an efficient HSP algorithm in that case). △

**Remark 11.8**

A classical and non-probabilistic algorithm already exists to find a set of generators for the kernel of a module homomorphism [BucNei1996]. The ring is $R = \mathbb{Z}_{4^N}$ and, similarly to the call to $f$, the elementary "ring operations" are polynomial in $n$. The whole procedure is polynomial. Hence quantum computation is not needed here. △

Rather than searching the solution in $\mathbb{Z}_2^n$ we are now looking into $p + H$ and try to find a vector with only 0 and 1 coordinates. In some sense, we want to find a non-zero vector of smallest coordinates in $p + H$ (i.e short for the infinity norm), although we need some technical modifications to make this statement exact (vectors with -1 coordinates are also short, we work modulo $4^N$...). Before to do that, let's define our lattice problem:

**Definition 11.9 (decisional closest vector problem)**

Let $b^1, ..., b^n \in \mathbb{Q}^n$ (or equivalently $\mathbb{Z}^n$) a basis of $\mathbb{R}^n$. The *lattice* corresponding to this basis is the set of all linear combinations $x = \sum_{i=1}^{n} x_i b^i$ with coefficients $x_i \in \mathbb{Z}$. If $d \in \mathbb{R}$ and $v \in \mathbb{R}^n$, the *decisional closest vector problem with infinity norm* (GapCVP$^\infty$) is to answer whether $\inf_{x \text{ in the lattice}} \|v - x\|_\infty \leq d$. ◇

Note that the previous proposition gives a generating sets $b^i$ of vectors for $H$. According to module theory, since the rank of the image of $f$ is 1, the rank of the module $H$ is $n - 1$. Hence we can extract $n - 1$ vectors $b^1$, ...$b^{n-1} \in \mathbb{Z}^n$ that generate $H$, and this can be done in polynomial time, using classical reduction on the matrix whose columns are the vectors $b^i$. The extracted vectors are independent linear vectors of $\mathbb{R}^n$: suppose there are $a_i$'s such that if $\sum_{i=1}^{n-1} a_i b^i = 0$. We can determine them by a Gaussian elimination they are integer (or rational, but since we get a 1-dimensional space of solutions, we can always make them integer). Hence the equality $\sum_{i=1}^{n-1} a_i b^i = 0$ holds in $\mathbb{Z}_{4^N}$ and get that the $a_i$'s are zeros.

As a consequence, we almost have a GapCVP$^\infty$ version. What we want is to find a vector $x$ in the lattice generated by $b^i$ such that $p + x$ is a vector of $\mathbb{Z}_2^n$ (i.e. with only 0's and 1's entries). Equivalently, we want to determine the minimum distance (for the infinity norm) from a vector of the lattice to the the point $v = -p + \begin{pmatrix} 1/2 \\ 1/2 \\ \vdots \\ 1/2 \end{pmatrix}$. This distance is less or equal to $d = \frac{1}{2}$ iff there is a solution to the monotone 1-in-3 3-SAT. We have two problems to overcome: the $b^i$'s are not a basis of $\mathbb{R}^n$ and we want to allow the result modulo $4^N$. First we complete the $n - 1$ vectors with a vector $b^n$ in order to get a basis. We can assume that $b^n$ is integer-valued and even that all his coefficients are of the form $k_i 4^N$. Next, we move to $\mathbb{R}^{2n}$: we extend $v$, $p$ and each vector $b^i$ with $n$ zeros and for all $1 \leq i \leq n$ we define the vector $b^{n+i}$ that has only two nonzero entries: $4^N$ at position $i$ and a rational $\varepsilon$ at position $n + i$. Clearly, we have a basis of $\mathbb{R}^{2n}$ that we now denote $c_j$ ($p$ and $v$ will denote either the initial vectors or those extended with zero, depending on the context). In matrix form, we can write:

$$C = \begin{pmatrix} b_1^1 & b_1^2 & \cdots & b_1^{n-1} & k_1 4^N & 4^N & 0 & \cdots & 0 \\ b_2^1 & \vdots & & \vdots & k_2 4^N & 0 & \ddots & & \vdots \\ \vdots & \vdots & & \vdots & \vdots & \vdots & & \ddots & 0 \\ b_n^1 & b_n^2 & \cdots & b_n^{n-1} & k_n 4^N & 0 & 0 & \cdots & 4^N \\ 0 & 0 & \cdots & & 0 & \varepsilon & 0 & \cdots & 0 \\ 0 & 0 & & & 0 & 0 & \varepsilon & & \vdots \\ \vdots & \vdots & & & \vdots & \vdots & & \ddots & 0 \\ 0 & 0 & \cdots & & 0 & 0 & \cdots & 0 & \varepsilon \end{pmatrix} \quad (11.3)$$

If we restrict ourselves to the $n$ first coordinates, then the combination of the column of $C$ is exactly the values of $H$ modulo $4^N$. The intuition is that if $\varepsilon$ is small enough, we can almost stay in the subspace of vectors whose $n$ last coordinates are zero, where the vector $v$ lives and can ignore what is happening in the $n$ last coordinates. This is exactly what the following lemma says:

**Lemma 11.10**

Let $R > 0$ be a rational and define $B(R) = \left\{ x = \sum_{j=1}^{2n} u_j b^j \middle| \forall j |u_j| \leq R \right\}$. There exists a rational $\varepsilon > 0$ such that for any $x \in B(R)$ in the lattice, $\|v - x\|_\infty = \max_{1 \leq i \leq n} |x_i - v_i|$.

proof: If $r$ is the shortest distance of the GapCVP$^\infty$ problem, we take $\varepsilon > 0$ any rational less than $\frac{r}{R}$. For any $x \in B(R)$ and $1 \leq i \leq n |x_{n+i} - v_{n+i}| = |u_{n+i} \varepsilon - 0| = |u_{n+i}| \varepsilon \leq R \frac{r}{R} = r \leq \|v - x\|_\infty = \max_{1 \leq i \leq 2n} |x_i - v_i|$. Hence we

have $\|v - x\|_\infty = \max_{1 \leq i \leq n} |x_i - v_i|$. □

**Remark 11.11**

The proof of the previous lemma can be rewritten in a constructive way. For example we will choose an accurate $R$ in the next theorem. We do not know the value of $r$. However, we are working with integer values in the $n$ first coordinates so the minimum values for the $|x_i - v_i|$ is at least $\frac{1}{2}$ and thus $r \geq \frac{1}{2}$. Hence we can take the rational $\varepsilon = \frac{1}{2R}$, which can be computed in polynomial time. △

We can now conclude this section:

**Theorem 11.12**

There is some $\varepsilon > 0$, such that a solution to the initial monotone 1-in-3 3-SAT problem exists iff the answer to the GapCVP$^\infty$ with that parameter is true.

proof: Let $M_b = \left(1 + \sup \left| b_i^j \right|\right)$ (the supremum is taken for $1 \leq i \leq n$ and $1 \leq j \leq n - 1$), $R = \max\{M_b + 2, 4^N - 1\}$ and choose $\varepsilon$ accordingly.

A solution to monotone 1-in-3 3-SAT yields a vector $x = \sum_{j=1}^{2n} u_j b^j$ in the lattice such that $\max_{1 \leq i \leq n} |x_i - v_i| = \frac{1}{2}$. We can assume that the $n - 1$ first $u_j$'s to be between 0 and $4^N - 1$, $u_n = 0$. The coordinates of $\sum_{j=1}^{n} u_j b^j$ are bounded in absolute values by $4^N M_b$. Because we want to come back inside $[-4^N + 1, 4^N - 1]$ where the coordinates $v_i \pm \frac{1}{2}$ are located, the $|u_j|$'s for the $n$ last coordinates need only be less than or equal to $M_b + 2$. Hence we have for all the $u_j$, $|u_j| \leq R$ and thus $x \in B(R)$. By the previous lemma, $\|v - x\|_\infty = \max_{1 \leq i \leq n} |x_i - v_i| = \frac{1}{2} = d$, so the GapCVP$^\infty$ problem is true.

Conversely, suppose the GapCVP$^\infty$ problem is true. Then there is is a closest vector $x = \sum_{j=1}^{2n} u_j c_j$ such that $d = \frac{1}{2} \geq \|v - x\|_\infty \geq \max_{1 \leq i \leq n} |x_i - v_i|$. In the $n$ first coordinates, this means that for all $i$, $|x_i + p_i - \frac{1}{2}| \leq \frac{1}{2}$ and because we are only working with integer values, $x_i + p_i$ is 0 or 1. Hence $p_i + \sum_{j=1}^{n-1} u_j b_i^j \equiv 0$ or 1 modulo $4^N$ i.e. $p + \sum_{j=1}^{n-1} u_j b^j \in (p + H) \cap \mathbb{Z}_2^n$. In that case it is a solution to the monotone 1-in-3 3-SAT. □

**Corollary 11.13**

Monotone 1-in-3 3SAT reduces to GapCVP$^\infty$. □

# Appendix E: Table of Lattice-based Cryptosystems

In this section, we sum up in a table the overview of lattice-based cryptosystems given in [MicReg2008]. The coefficient in uniqueSVP problems are according to [Reg2003bis]. We will list many other problems without defining them: Shortest Vector Problem (SVP), unique Shortest Vector Problem (uniqueSVP), Shortest Independent Vectors Problem (SIVP), Closest vector problem (CVP), Learning With Errors (LWE). Also, more assumptions are often made on these problems: particular structure of the lattice (e.g. uniqueSVP rather than SVP), approximate factors etc.

There is often a trade-off between efficient implementation and security insurance: adding assumptions may make the system more vulnerable but also more realizable in practice. An important feature of lattice-based cryptosystems is that a security proof, when we have one, is based on worst-case hardness rather than average-case hardness.

| Classification in [MicReg2008] | Cryptosystem | Security proof | Practical Implementation |
|---|---|---|---|
| Hash Functions | Ajtai's construction of hash functions, based on $q$-ary lattices. | approximate SIVP, SVP, or CVP over $q$-ary lattices. | • Quite inefficient.<br>• Collision resistant functions. |
| | Ajtai's construction restricted to cyclic lattices. | approximate SIVP over cyclic lattices. | • Efficient<br>• Only one-way functions (not strong enough for practical use). |
| | Ajtai's construction restricted to ideal lattices. | approximate SVP over ideal lattices. | • Efficient (there exists an optimized version, the SWIFFT hash function).<br>• Collision resistant functions. |
| Public Key Encryption Schemes | GGH/HNF | none | Attacks are known. |
| | NTRU | none | Attacks are known. |
| | Ajtai-Dwork cryptosystem, improved by Goldreich/Goldwasser /Halevi. | $O(n^7)$-uniqueSVP | Quite inefficient. |

| Classification in [MicReg2008] | Cryptosystem | Security proof | Practical Implementation |
|---|---|---|---|
| | Ajtai-Dwork cryptosystem improved by Regev. | $O(n^{3/2})$-uniqueSVP | Quite inefficient. |
| | Regev's cryptosystem | LWE | Relatively efficient (can be optimized any further, by restricting to some class of lattices). |
| Digital Signature Schemes | GGH/NTRUSign | none | • Efficient<br>• Attacks are known. For NTRUSign, there are some countermeasures based on perturbation techniques. |
| | Schemes based on preimage sampleable trapdoor functions. | approximate SIVP and more. | Quadratic complexity. |
| | Schemes based on collision resistant hash functions. | approximate SVP in ideal lattices. | Almost linear complexity. |
| Other Cryptographic Primitives | CCA-secure cryptosystems | LWE | / |
| | IBE | LWE | / |
| | OT protocols | LWE | / |
| | Zero-Knowledge proofs and ID schemes | approximate-SVP | / |

Figure 12.1

## Appendix F: Variant for the Abelian HSP algorithm

In this section, we study an alternative algorithm for the abelian hidden subgroup problem. The idea is to use the information obtained after each call to move to a simpler Hidden Subgroup Problem, built on the original one. It does not generalize to arbitrary group, but is a good example of a subgroup reduction method.

**Cyclic Hidden Subgroup Problem**

We have already described an algorithm for the Cyclic HSP in the period-finding submodule of Shor's factoring algorithm. Recall that we do a constant number $k$ of Weak Fourier Samplings on $G = \mathbb{Z}_N$ in order to find a hidden subgroup $H = d\,\mathbb{Z}_N$ for some $1 \leq d \leq N$. For each sampling, we get a number $\lambda \frac{N}{d}$ for some $\lambda \in \{0, \dots d-1\}$. Computing the greatest common divisor of these multiples gives $\frac{N}{d}$ with probability at least $1 - 2^{-k/2}$.

The idea of our algorithm is the following. If we compute the irreducible fraction of $\lambda_k \frac{N}{d}$, the denominator is a divisor $a_1$ of $d$. This gives $H = d\,\mathbb{Z}_N \subseteq a_1 \mathbb{Z}_N \subseteq G = \mathbb{Z}_N$ and we can now work in the group $a_1 \mathbb{Z}_N$ instead. Repeating this several times, we get some $a_k \mathbb{Z}_N$'s which are closer and closer to $H$. Said otherwise, $a_k$ becomes $d$.

More precisely, we define $G_0 = G = \mathbb{Z}_N$, $H_0 = d\mathbb{Z}_N$ and $a_0 = 1$. At each step $k$, we have another cyclic hidden subgroup problem: $G_k = a_k \mathbb{Z}_N \cong \mathbb{Z}_{N_k}$, $H = H_k \cong d_k \mathbb{Z}_{N_k}$ with $N_k = \frac{N}{a_k}$ and $d_k = \frac{d}{a_k}$. Moreover, to build a oracle on $G_k = \mathbb{Z}_{N_k}$ from the initial $f$, we define $f_k(x) = f(a_k x)$.

Now at the step $k$, applying a Weak Fourier Sampling over $G_k$ gives a $\lambda_k \frac{N_k}{d_k}$ for some $\lambda_k \in \{0, \dots, d_k - 1\}$. We compute the irreducible fraction $\frac{p_k}{q_k}$ of $\lambda_k \frac{N_k}{d_k} \frac{a_k}{N} = \frac{\lambda_k}{d_k}$ and get a divisor $q_k$ of $d_k$. We set $a_{k+1} = q_k a_k$ and continue this procedure until $k = m$. Very often, we have $\lambda_k > 0$ (i.e. $q_k > 1$) an so $a_{k+1} \geq 2a_k$. Hence repeating this $O(\log N)$ times, $a_k$ should reach $d$. This is essentially what the following theorem says:

**Theorem 13.1**

If we repeat the procedure above $m = 2c\log N$ times for some $c \geq 2$ then we have $P(a_m = d) \geq 1 - e^{-\frac{(c-1)^2}{c}\log N}$. Moreover, all the operations are done on $\log N$ bits so we have a solution to the cyclic hidden subgroup problem.

proof: First, the probability of failure given that $a_k$ has reached $d$ before the $m$th call is zero. Hence we have:

$$P(a_m \neq d) = P(a_m < d) = P(a_m < d\,/\,\forall\,k < m,\,a_k < d)P((\forall\,k < m,\,a_k < d)) + 0 \leq P(a_m < d\,/\,\forall\,k < m,\,a_k < d)$$
$$(13.1)$$

Now per discussion above, if the number of $\lambda_k \neq 0$ is $\geq \log N$ then

$$a_m = \prod_{k < m} q_k \geq \prod_{k < m,\,\lambda_k \neq 0} 2 \geq 2^{\log N} = N \geq d \quad (13.2)$$

Thus we have

$$P(a_m \neq d) \leq P(|\{k < m, \lambda_k \neq 0\}| < \log N / \forall\, k < m, a_k < d) \quad (13.3)$$

In order to majorate the right hand side probability, we introduce a variable $X_k$ following a Bernoulli distribution with equal failure/success probability $\frac{1}{2}$. Since $a_k < d$, we have $\lambda_k > 0$ with a probability at least $\frac{1}{2}$. Hence if $S = \sum_{k < m} X_k$, we have $P(a_m \neq d) \leq P(S \leq \log N) = P(S - c \log N \leq (1-c)\log N)$ and finally using Hoeffding's inequality:

$$P(a_m \neq d) \leq e^{-\frac{2(c-1)^2 \log^2 N}{m}} = e^{-\frac{(c-1)^2}{c}\log N} \quad (13.4)$$

□

While this algorithm does not really improve the complexity, it is interesting to note how it works: rather than applying Fourier Sampling over $\mathbb{Z}_N$, we gather information on $H$ to "reduce" the group and hidden subgroup at each step. Finally, we reach the case where $H_m = G_m$ is i.e. we know everything about the hidden subgroup.

**Abelian Hidden Subgroup Problem**

We mimic the previous algorithm by building a sequence of subgroups $H = G_m \subseteq G_{m-1} \subseteq \ldots \subseteq G_0 = G$. Since $G_k \subseteq G$ is abelian, we have an isomorphism $G_k \cong \mathbb{Z}_{t_1^k} \times \mathbb{Z}_{t_2^k} \times \ldots \times \mathbb{Z}_{t_{d_k}^k}$. We assume that each factor is nontrivial, so the length $d_k$ of this decomposition is of size $O(\log|G|)$.

We let $G_0 = G$, $H_0 = H$ and $f_0 = f$. For $1 \leq i \leq d_0$, $a_i^0$ is element of $G_0$ with a one at the $i$th position and zeros elsewhere. At step $k$, we apply a Weak Fourier Sampling over $\mathbb{Z}_{t_1^k} \times \mathbb{Z}_{t_2^k} \times \ldots \times \mathbb{Z}_{t_{d_k}^k}$ to find a subgroup $H_k$ hidden by a function $f_k$. This gives an element $g_k \in G_k$ such that $H_k \subseteq \text{Ker } \chi_{g_k} = G_{k+1}$. Note that we have $G_{k+1} \subsetneq G_k$ if $g_k \neq 1$. This happens with probability at least $\frac{1}{2}$ if we have not reached $H$ yet: in that case $H \subsetneq G_k$, so $|G_k| \geq 2|H| \geq 2$ and by formula formula 4.20, $P(1) = \frac{\sqrt{|H|}}{|G_k|} \leq \frac{\sqrt{\frac{|G_k|}{2}}}{|G_k|} = \frac{1}{\sqrt{2|G_k|}} \leq \frac{1}{\sqrt{4}} = \frac{1}{2}$. Because of the isomorphism $G_{k+1} \cong \mathbb{Z}_{t_1^{k+1}} \times \mathbb{Z}_{t_2^{k+1}} \times \ldots \times \mathbb{Z}_{t_{d_{k+1}}^{k+1}}$ there are elements $a_1^{k+1}, \ldots, a_{d_{k+1}}^{k+1}$ of $G_k$ such that we have the direct decomposition $G_{k+1} = \bigoplus_{i=1}^{d_{k+1}} \langle a_i^{k+1} \rangle$. We define $f_{k+1}(x_1, \ldots, x_{d_{k+1}}) = f_k\left(\sum_{i=1}^{d_{k+1}} x_i a_i^{k+1}\right)$.

Note how everything is similar to the cyclic HSP algorithm above. When $G_{k+1} \subsetneq G_k$, we have $|G_{k+1}| \leq \frac{1}{2}|G_k|$ and so we have the equivalent of theorem 13.1 above saying that $O(\log|G|)$ steps are enough. However, we need more care to claim that all the operations are $\text{poly}(\log|G|)$. First, it is easy to show by induction that for $k \geq 1$, an element $\left(x_1^k, \ldots, x_{d_k}^k\right) \in G_k$ corresponds to following element of the initial group $G$:

$$\sum_{i_k=1}^{d_k} x_{i_k}^k \left[ \sum_{j=0}^{k-1} \sum_{i_j=1}^{d_j} \left( \prod_{l=0}^{k-1} a_{i_{l+1}, i_l}^{l+1} a_{i_0}^{0} \right) \right] \quad (13.5)$$

We define the term in brackets to be a vector $u_{i_k}^k \in G$ and these are the vectors we keep in memory. So we start with $u_{i_0}^0 = a_{i_0}^0$ and at each step we compute $u_{i_{k+1}}^{k+1} = \sum_{i_k=1}^{d_k} a_{i_{k+1}, i_k}^{k+1} u_{i_k}^k$ in poly(log|G|). A call to the oracle $f_k$ consists in computing $x = \sum_{i_k=1}^{d_k} x_{i_k}^k u_{i_k}^k$ in poly(log|G|) followed by a call to $f(x)$. Also, with very high probability, we have $H_m = G_m = \mathbb{Z}_{t_1^m} \times \mathbb{Z}_{t_2^m} \times \ldots \times \mathbb{Z}_{t_{d_k}^m}$ and so the elements of $H$ are obtained by $\sum_{i_m=1}^{d_m} x_{i_m}^m u_{i_m}^m$ for $x_{i_m}^m \in \mathbb{Z}_{t_{i_m}^m}$ and we have directly obtained:

$$H = \bigoplus_{i=1}^{d_m} \langle u_i^m \rangle \cong \mathbb{Z}_{t_1^m} \times \mathbb{Z}_{t_2^m} \times \ldots \times \mathbb{Z}_{t_{d_m}^m} \quad (13.6)$$

The more annoying thing is to determine the element $a_i^{k+1}$ in the direct decomposition Ker $\chi_{g_k} = \bigoplus_{i=1}^{d_{k+1}} \langle a_i^{k+1} \rangle$. As we have previously seen, any element $h \in$ Ker $\chi_{g_k}$ is solution to a congruence equation of the form $\sum_{i=1}^{d_k} \alpha_i h_i \equiv 0 \mod (e_k)$ where $e_k = \operatorname*{lcm}_{1 \leq i \leq d_k} (t_i^k)$. We can view $h \in \mathbb{Z}_{e_k}^{d_k}$ and using the kernel algorithm 6.1 of [BucNei1996], we can compute a generating set for Ker $\chi_{g_k}$ in poly(log|G|). Nevertheless, this generating set does not necessarily form a direct sum as wanted. However, we use the technique proposed by Cheung and Mosca in [CheMos2001] to get the decomposition Ker $\chi_{g_k} = \bigoplus_{i=1}^{d_{k+1}} \langle a_i^{k+1} \rangle$ from this generating set in poly(log|G|). Their algorithm uses the cyclic HSP to compute the orders $t_i^{k+1}$ of $a_i^{k+1}$ in poly(log|G|).

To conclude this section, note that the algorithm may also work for Hamiltonian HSP: it is easy to see that the $G_k$'s will remain Dedekindian i.e. isomorphic to either an abelian or a hamiltonian group. The possible irreducible representations are described in [HalRusTa-2000]. However, we have not been able to figure out how to construct the isomorphism from the kernel equations. For more general groups, the $H_m$ obtained at the end will be the largest normal subgroup of $G$ containing $H$, as noted in [GriSchVaz2000]. Thus we may need to continue the procedure, for example by using a quotient reduction as indicated in proposition 6.4.

## Appendix G: Attempts to solve DHSP by superposition of oracle values

In this section, we try to solve the HSP over $D_N = \mathbb{Z}_N \rtimes_\varphi \mathbb{Z}_2$, reduced to the case $H = \langle (d, 1) \rangle$ for some $d \in \mathbb{Z}_N$. Contrary to all the other approaches proposed so far, we will not use coset states of the form $\frac{1}{\sqrt{2}}(|x\rangle|0\rangle + |x+d\rangle|1\rangle)$ but superpositions over many oracle values. In particular, this approach will not *a priori* yield an efficient algorithm for polynomial uniqueSVP.

**Finding the parity of d**

First let's recall the classical inductive method to determine $d$. If $M$ is a divisor of $N$, we let $N' = \frac{N}{M}$ and write the euclidean division $d = Md' + r$. If we are able to determine $r$ in polynomial time, then we can define the oracle $f'(x, y) = f(Mx + r, y)$ for $x \in \mathbb{Z}_{N'}$, $y \in \mathbb{Z}_2$ and obtain a new dihedral HSP with hidden subgroup $\langle (d', 1) \rangle$. Since the prime factorization has only $O(\log N)$ factors, we can repeat this inductive step to get an efficient algorithm. Of course, this is likely to be useful only when the factors are small, for otherwise it may be as much difficult to determine $r$ as to determine the whole $d$. One interesting case is $N = 2^n$ and $M = 2$: at each step, $N$ remains a power of 2 and we only need to determine one digit of $d$.

If we fix $b \in \{0, 1\}$, the elements $(Mi + j, b)$ for $i \in \mathbb{Z}_{N'}$, $j \in \mathbb{Z}_M$ form a complete set of coset representatives, and so the elements $f(Mi + j, b)$ are the $N$ possible values of $f(D_N)$. Moreover, we can consider a partition of $f(D_N)$ by defining for all $j \in \mathbb{Z}_M$ the subsets

$$E_j^b = \{f(Mi + j, b), i \in \mathbb{Z}_{N'}\} \quad (14.1)$$

From the equality $f(g(d, 1)) = f(g)$ we have the relation:

$$\begin{aligned} E_j^1 &= \{f(Mi + j, 1), i \in Z_{N'}\} \\ &= \{f((Mi + j - d, 0)(d, 1)), i \in Z_{N'}\} \\ &= \{f(Mi + j - (Md' + r), 0), i \in Z_{N'}\} \quad (14.2) \\ &= \{f(Mi + (j - r), 0), i \in Z_{N'}\} \\ &= E_{(j-r) \bmod (M)}^0 \end{aligned}$$

In particular for $j = r$, we have the equality $E_0^0 = E_r^1$ and so we can hope to get information on $r$ by comparing the superposition over $E_0^0$ and those over $E_j^1$. We show that it is possible for $M = 2$, but it also works for any polynomial value by an obvious modification. In that case, the uniform superposition over $E_0^0$ and $E_0^1$ are respectively

$$|\psi_0\rangle = \frac{1}{\sqrt{N'}} \sum_{i=0}^{N'-1} |f(2i, 0)\rangle \quad (14.3)$$

and

$$|\phi_0\rangle = \frac{1}{\sqrt{N'}} \sum_{i=0}^{N'-1} |f(2i, 1)\rangle \quad (14.4)$$

If $r = 0$, $|\psi_0\rangle = |\phi_0\rangle$ is the uniform superposition over $E_0^0$. Otherwise, $r = 1$ and $|\phi_0\rangle$ is the uniform superposition over the complementary set $E_0^1$. We let $f_i^b = f(2i + b, 0)$ and measure the product state $|\psi_0\rangle|\phi_0\rangle$ in the basis $|x\rangle|x\rangle_{x \in \mathbb{Z}_N}$, $\frac{1}{\sqrt{2}}(|x\rangle|y\rangle + |x\rangle|y\rangle)_{x < y \in \mathbb{Z}_N}$, $\frac{1}{\sqrt{2}}(|x\rangle|y\rangle - |x\rangle|y\rangle)_{x < y \in \mathbb{Z}_N}$. It turns out that this allows to distinguish the two cases:

- If $r = 0$, the product contains vectors with the same coordinates $|f_i^0\rangle|f_i^0\rangle$ for $i \in \mathbb{Z}_N$. For the other vectors, we have a symmetry between the two coordinates. Hence we can group them by pairs $|f_{i_1}^0\rangle|f_{i_2}^0\rangle + |f_{i_2}^0\rangle|f_{i_1}^0\rangle$ for $i_1 < i_2 \in \mathbb{Z}_N$. Finally, the product state belongs to the space spanned by $|x\rangle|x\rangle_{x \in \mathbb{Z}_N}$ and $|x\rangle|y\rangle + |x\rangle|y\rangle_{x < y \in \mathbb{Z}_N}$.
- If $r = 1$, the product contains vectors $|f_{i_1}^0\rangle|f_{i_2}^1\rangle$ where the two coordinates are not equal. This time, we never have two symmetric vectors i.e. which are the same modulo permutation of the two coordinates. Moreover, each vector $|f_{i_1}^0\rangle|f_{i_2}^1\rangle$ is the expression of a sum/difference (according to the respective order of $f_{i_1}^0$ and $f_{i_2}^1$) of two vectors $|x\rangle|y\rangle + |x\rangle|y\rangle$ and $|x\rangle|y\rangle - |x\rangle|y\rangle$. So the product state belongs half to $|x\rangle|y\rangle + |x\rangle|y\rangle_{x < y \in \mathbb{Z}_N}$ and half to $|x\rangle|y\rangle - |x\rangle|y\rangle_{x < y \in \mathbb{Z}_N}$.

Suppose that we have a procedure to create many states $|\psi_0\rangle$ and $|\phi_0\rangle$. We measure a state in the space $\frac{1}{\sqrt{2}}(|x\rangle|y\rangle - |x\rangle|y\rangle)_{x < y \in \mathbb{Z}_N}$ with probability $\frac{r}{2}$. So repeating the procedure a constant number of times allows to determine $r$ with high probability!

It is not however clear whether we can create the states $|\psi_0\rangle$ and $|\phi_0\rangle$ efficiently. With the usual approach, we can start by creating a superposition over the elements $(2i, b)$ and apply the gate $U_f$. To remove the entanglement between a point and its image by $f$, we apply a Quantum Fourier Transform to the first coordinate and measure it. We get two random numbers $j_1, j_2 \in \mathbb{Z}_{N'}$ as well as the states

$$|\psi_{j_1}\rangle = \frac{1}{\sqrt{N'}} \sum_{i=0}^{N'-1} e^{\frac{2i\pi j_1 i}{N'}} |f(2i, 0)\rangle \quad (14.5)$$

and

$$|\phi_{j_2}\rangle = \frac{e^{\frac{2i\pi j_2 d'}{N'}}}{\sqrt{N'}} \sum_{i=0}^{N'-1} e^{\frac{2i\pi j_2 i}{N'}} |f(2i - r, 0)\rangle \quad (14.6)$$

Note that we can actually ignore the phase factor in the second one. The desired result $j_1 = j_2 = 0$ only happens with exponentially small probability. Nonetheless, we can still apply the previous measurement. After a bit of calculation, we get the probabilities:

| Space | case $r = 0$ | case $r = 1$ |
|---|---|---|
| $\|x\rangle\|x\rangle_{x \in \mathbb{Z}_N}$ | $\frac{1}{N'}$ | 0 |
| $\|x\rangle\|y\rangle + \|x\rangle\|y\rangle_{x < y \in \mathbb{Z}_N}$ | $\frac{2}{N'^2} \sum_{i_2=0}^{N'-1} \sum_{i_1=0}^{i_2-1} \cos^2 \frac{\pi}{N'}(j_2 - j_1)(i_2 - i_1)$ | $\frac{1}{2}$ |
| $\|x\rangle\|y\rangle - \|x\rangle\|y\rangle_{x < y \in \mathbb{Z}_N}$ | $\frac{2}{N'^2} \sum_{i_2=0}^{N'-1} \sum_{i_1=0}^{i_2-1} \sin^2 \frac{\pi}{N'}(j_2 - j_1)(i_2 - i_1)$ | $\frac{1}{2}$ |

If we repeat the procedure several times, we need to take the means over the choice of $j_1, j_2 \in \mathbb{Z}_{N'}$. By classical trigonometric identities, it is possible to write the sums of the first column $\frac{1}{2} + S + o(\frac{1}{N'})$ where $S$ is a sum of cosinus. Unfortunately, evaluating $S$ with a computer suggests that $S = \Theta(\frac{1}{N'})$. Hence, we can not distinguish the two cases with only a polynomial (in $\log N$) number of samplings...

**Finding an approximation of d**

For convenience, we assume $N = 4M$ and define $N' = 2M = \frac{N}{2}$. In this section, we consider for $a \in \mathbb{Z}_N$ and $b \in \mathbb{Z}_2$ the sets of $N'$ successive values

$$F_a^b = \{f(a+i, b), i \in \mathbb{Z}_{N'}\} \quad (14.7)$$

The algorithm is based on the assumption that we can efficiently create uniform superposition over these states. As above, we have $F_a^1 = F_{a-d}^0 = F_s^0$ with $-N < s < N$ and we measure the tensor product of the uniform superpositions over $F_a^1 = F_s^0$ and $F_0^0$ in the basis $\|x\rangle\|x\rangle_{x \in \mathbb{Z}_N}$, $\frac{1}{\sqrt{2}}(\|x\rangle\|y\rangle + \|x\rangle\|y\rangle)_{x < y \in \mathbb{Z}_N}$, $\frac{1}{\sqrt{2}}(\|x\rangle\|y\rangle - \|x\rangle\|y\rangle)_{x < y \in \mathbb{Z}_N}$. We define $l = |F_0^0 \cap F_a^1|$ the number of common elements.

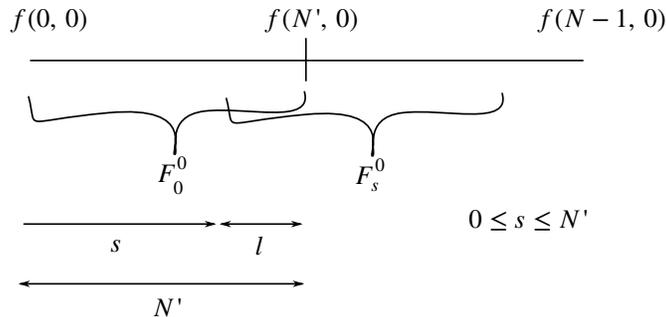

Figure 14.1

The figure 14.1 shows that in the case $0 \le s \le N'$, we have $l = N' - s$. In general, the expression is

$$l = |N' - |s|| = |N' - |d - a|| \quad (14.8)$$

The possible vectors in the product of the uniform superpositions over $F_a^1 = F_s^0$ and $F_0^0$ are:

- $l$ vectors of the form $\frac{1}{N'}(|x\rangle|x\rangle)$ for $x \in F_0^0 \cap F_a^1$ in the intersection.
- $\binom{l}{2} = \frac{l(l-1)}{2}$ vectors of the form $\frac{\sqrt{2}}{N'}\left(\frac{1}{\sqrt{2}}(|x\rangle|y\rangle + |x\rangle|y\rangle)\right)$ for $x, y \in F_0^0 \cap F_a^1$ two distinct elements of the intersection.
- $N'^2 - l^2$ other vectors $|x\rangle|y\rangle$ for $x \neq y$ with one of the vector ouside the intersection $F_0^0 \cap F_a^1$, which can be written as a combination of two vectors of the basis. For example if $x < y$, it is $\frac{\sqrt{2}}{2N'}\left(\frac{1}{\sqrt{2}}(|x\rangle|y\rangle + |x\rangle|y\rangle) + \frac{1}{\sqrt{2}}(|x\rangle|y\rangle - |x\rangle|y\rangle)\right)$

The probability of measuring $|x\rangle|x\rangle$ or $\frac{1}{\sqrt{2}}(|x\rangle|y\rangle + |x\rangle|y\rangle)$ is thus

$$P = \frac{1}{N'^2}\left(l + 2\frac{l(l-1)}{2} + \frac{1}{2}\left(N'^2 - l^2\right)\right)$$
$$= \frac{1}{2}\left[1 + \left(\frac{l}{N'}\right)^2\right] \quad (14.9)$$

and of course the one of measuring $\frac{1}{\sqrt{2}}(|x\rangle|y\rangle - |x\rangle|y\rangle)$ is the complementary:

$$1 - P = \frac{1}{N'^2}\left(\frac{1}{2}\left(N'^2 - l^2\right)\right)$$
$$= \frac{1}{2}\left[1 - \left(\frac{l}{N'}\right)^2\right] \quad (14.10)$$

We can try to repeat $m$ times the measurement in order to approximate $P$ and hence also $l$, $s$ and finally $d$. Of course, we can test several values of $a$ and compare the different approximations of $d$ to refine them. Once again, we use Hoeffding's inequality. We define $X$ to be the random variable taking the value 1 with probability $P$ and 0 otherwise and $S = X_1 + X_2 + \ldots + X_m$. We take $m = (\log N)^{2p+1}$ for some fixed $p$ and $t = \sqrt{\frac{\log N}{m}} = \frac{1}{(\log N)^p}$ i.e. $mt^2 = \log N$. We have:

$$P\left(\frac{1}{m}|S - E(S)| \geq t\right) \leq 2\exp(-2mt^2) = 2\exp(-2\log N) \quad (14.11)$$

So if $l_{\text{exp}}$ is the experimental value of $l$ we have with very high probability, $|l^2 - l_{\text{exp}}^2| < tN'^2 = \frac{N'^2}{(\log N)^p}$. Unfortunately, this does not provide a good precision. Instead, we will turn toward a more modest goal: determine whether $l$ is less than $\frac{N'}{2}$. This is possible if $l_{\text{exp}}$ is sufficiently far from $\frac{N'}{2}$. Note that

$$\left| l_{\exp}^2 - \left(\frac{N'}{2}\right)^2 \right| < tN'^2 \Leftrightarrow l_{\exp}^2 = \left(\frac{N'}{2}\right)^2 + \varepsilon t N'^2 \text{ for some } |\varepsilon|<1$$

$$\Leftrightarrow l_{\exp} = \frac{N'}{2}\sqrt{1+4\varepsilon t} \text{ for some } |\varepsilon|<1 \qquad (14.12)$$

$$\Leftrightarrow \sqrt{1-4t}\,\frac{N'}{2} < l_{\exp} < \sqrt{1+4t}\,\frac{N'}{2}$$

Hence, if $t$ is small enough, we can decide with certainty whether $l$ is less than $\frac{N'}{2}$, except if $l_{\exp}$ is in a small interval around $\frac{N'}{2}$. Unfortunately again, we have $\sqrt{1 \pm 4t} = 1 \pm 2t + o(t)$ so this interval is approximately of size $tN' = \frac{N}{2(\log N)^p}$ which is exponentially large...

The algorithm fails in general but still may give a partial hint. For example, if $\log N \geq 2$ and if we take $p=3$, the answer is correct for at least $l_{\exp}$ out of the intervall $\left]\frac{N'}{3}, \frac{2N'}{3}\right[$. This is in contrast to the approximation proposed in appendix C, for which the error seems too difficult to bound in order to get any information.

As a conclusion, both methods above provide improvement over the standard coset sampling approach. It is not clear how we can create efficiently the uniform superpositions over the sets of oracle values $E_j^b$ or $F_a^b$, but they may provide an alternative research direction to solve the dihedral HSP.

Quantum Information & Computation vol. 9 no. 5&6, pp. 500-512, 2009
http://arxiv.org/abs/0806.3362

**[MooEtAl2005]** Cristopher Moore, Daniel Rockmore, Alexander Russell and Leonard J. Schulman, 2005,
*The Power of Strong Fourier Sampling: Quantum Algorithms for Affine Groups and Hidden Shifts* .
http://arxiv.org/abs/quant-ph/0503095

**[MooRusSch2005]** Cristopher Moore, Alexander Russell, Leonard J. Schulman, 2005,
*The Symmetric Group Defies Strong Fourier Sampling: Part I.*
http://arxiv.org/abs/quant-ph/0501056

**[MooRusSni2006]** Cristopher Moore, Alexander Russell, Piotr Sniady, 2006,
*On the impossibility of a quantum sieve algorithm for graph isomorphism: unconditional results.*
STOC '07: Proceedings of the thirty-ninth annual ACM symposium on Theory of computing, pages 536-545, New York, NY, USA, 2007. ACM Press
http://arxiv.org/abs/quant-ph/0612089

**[MooRusVaz2007]** Cristopher Moore, Alexander Russell, Umesh Vazirani, 2007,
*A classical one-way function to confound quantum adversaries.*
http://arxiv.org/abs/quant-ph/0701115

**[NieChu2007]** Michael A. Nielsen and Isaac L. Chuang 2007,
*Quantum Computation and Quantum Information* (ninth printing).

**[OsbSev2004]** Tobias J. Osborne, Simone Severini, 2004,
*Quantum Algorithms and Covering Spaces.*
http://arxiv.org/abs/quant-ph/0403127

**[RadRötSen2005]** Jaikumar Radhakrishnan, Martin Rötteler and Pranab Sen, 2005,
*On the Power of Random Bases in Fourier Sampling: Hidden Subgroup Problem in the Heisenberg Group.*
http://arxiv.org/abs/quant-ph/0501044

**[Reg2003]** Oded Regev, 2003,
*Quantum Computation and Lattice Problems.*
http://arxiv.org/abs/cs/0304005

**[Reg2003bis]** Oded Regev, 2003,
*New Lattice Based Cryptographic Constructions.*
http://arxiv.org/abs/cs/0309051v1

**[Reg2004]** Oded Regev, 2004,
*A Subexponential Time Algorithm for the Dihedral Hidden Subgroup Problem with Polynomial Space.*
http://arxiv.org/abs/quant-ph/0406151

**[Röt2008]** Martin Rötteler, 2008,
*Quantum algorithms for highly non-linear Boolean functions.*
to appear in Proceedings of the 21st Annual ACM-SIAM Symposium on Discrete Algorithms